\documentclass[preprint]{aastex631}
\usepackage{rotating}
\usepackage{comment}
\usepackage[maxfloats=256]{morefloats}
\accepted{}
\shorttitle{Evolution of stars in the range 7-15 $\rm M_\odot$}
\shortauthors{Limongi et al.}
\graphicspath{{./}{figures/}}

\begin{document}
%\title{Evolution and final fate of solar metallicity stars in the mass range 7-15 $\rm M_\odot$: the transition through AGB, SAGB and Massive Stars}
\title{Evolution and final fate of solar metallicity stars in the mass range 7-15 $\rm M_\odot$. I. The transition from AGB to SAGB stars, Electron Capture and Core Collapse Supernovae progenitors}
\correspondingauthor{}
\email{marco.limongi@inaf.it}

\author[0000-0002-3164-9131]{Marco Limongi}
\affiliation{Istituto Nazionale di Astrofisica - Osservatorio Astronomico di Roma, Via Frascati 33, I-00040, Monteporzio Catone, Italy}
\affiliation{Kavli Institute for the Physics and Mathematics of the Universe (WPI), The University of Tokyo Institutes for Advanced Study, The University of Tokyo, Kashiwa, Chiba 277-8583, Japan}
\affiliation{INFN. Sezione di Perugia, via A. Pascoli s/n, I-06125 Perugia, Italy}

\author[0000-0003-0390-8770]{Lorenzo Roberti}
\affiliation{Konkoly Observatory, Research Centre for Astronomy and Earth Sciences, E\"otv\"os Lor\'and Research Network (ELKH), Konkoly Thege Mikl\'{o}s \'{u}t 15-17, H-1121 Budapest, Hungary}
\affiliation{CSFK, MTA Centre of Excellence, Budapest, Konkoly Thege Mikl\'{o}s \'{u}t 15-17, H-1121, Hungary}
\affiliation{Istituto Nazionale di Astrofisica - Osservatorio Astronomico di Roma, Via Frascati 33, I-00040, Monteporzio Catone, Italy}

\author[0000-0002-3164-9131]{Alessandro Chieffi}
\affiliation{Istituto Nazionale di Astrofisica - Istituto di Astrofisica e Planetologia Spaziali, Via Fosso del Cavaliere 100, I-00133, Roma, Italy}
\affiliation{Monash Centre for Astrophysics (MoCA), School of Mathematical Sciences, Monash University, Victoria 3800, Australia}
\affiliation{INFN. Sezione di Perugia, via A. Pascoli s/n, I-06125 Perugia, Italy}

\author[0000-0001-9553-0685]{Ken'ichi Nomoto}
\affiliation{Kavli Institute for the Physics and Mathematics of the Universe (WPI), The University of Tokyo Institutes for Advanced Study, The University of Tokyo, Kashiwa, Chiba 277-8583, Japan}

\begin{abstract}

According to a standard initial mass function, stars in the range $\rm
7-12~M_\odot$ constitute $\sim 50\%$ (by number) of the stars more
massive than $\sim 7~M_\odot$, but, in spite of this, their
evolutionary properties, and in particular their final fate, are still
scarcely studied. In this paper we present a detailed study of the
evolutionary properties of solar metallicity, non rotating stars in
the range $\rm 7-15~M_\odot$, from the pre main sequence phase up to the
presupernova stage or up to an advanced stage of the thermally pulsing
phase, depending on the initial mass. We find that (1) the $\rm
7.00~M_\odot$ develops a degenerate CO core and evolves as a classical
AGB star in the sense that it does not ignite the C burning reactions;
(2) stars with the initial mass $\rm M\geq 9.22~M_\odot$ end
their life as core collapse supernovae; (3) stars in the range $\rm
7.50\leq M/M_\odot\leq 9.20$ develop a degenerate ONeMg core and evolve
through the thermally pulsing SAGB phase; 4) stars in the mass range
$\rm 7.50\leq M/M_\odot\leq 8.00$ end their life as hybrid CO/ONeMg- or
ONeMg-WD; (5) stars with the initial mass in the range $\rm 8.50\leq
M/M_\odot\leq 9.20$ may potentially explode as electron capture supernovae.
%most likely achieve the central densities in
%excess of the threshold value for the activation of the electron
%capture on $\rm ^{20}Ne$ before losing the entire H-rich envelope and
%therefore may potentially explode as electron capture supernovae.

\end{abstract}

\bigskip
\bigskip
\bigskip
\bigskip

\section{Introduction}\label{sec:intro}
In the general picture of stellar evolution, stars with
$\rm M\lesssim 7~M_\odot$ evolve along the thermally pulsing
asymptotic giant branch (TP-AGB) phase and end their life as CO White
Dwarfs (CO-WDs). On the contrary, stars with $\rm \gtrsim 12~M_\odot$,
the so-called massive stars (MS), evolve through all the major stable
nuclear burning stages and eventually explode as core collapse
supernovae (CCSNe) leaving a compact remnant, i.e. either a neutron
star or a black hole. Stars between $\rm \sim 7~M_\odot$ and
$\rm \sim 12~M_\odot$ have a much more complex evolution. The lower
masses ignite C off-center in an electron degenerate environment and
their ability in removing the degeneracy totally relies on the
capability of the thermally unstable zones (the convective zones) to
heat the layers below the main burning front. The higher masses, on
the contrary, ignite C burning centrally. After the C burning phase,
the lower mass stars ($\rm M\lesssim 10~M_\odot$) develop an inert ONeMg
electron degenerate core and enter the thermally pulsing phase along
the AGB. These stars are generally referred to as super-AGB (SAGB)
stars. The final fate of the SAGBs depends on the competition between
the core growth and the mass loss \citep{Nomoto84}. If mass loss dominates, the
envelope is completely lost and the result is a ONeMg-WD. On the
contrary, if the core grows in mass enough to achieve central
densities close to the threshold density for electron capture (EC)
$\rm ^{20}Ne(e^{-},\nu)^{20}Fe$, deleptonization and thermonuclear
instability develops to induce the electron capture supernova (ECSN).  
The final fate of such stars depends on the competition between the
energy released by the nuclear burning front and the loss of pressure
due to the deleptonization occurring in the central zones. If the
energy released by nuclear burning prevails, degeneracy is removed and
the star explodes as a thermonuclear ECSN \citep{Miyaji+80,NomotoKondo91,Isern+91,Jones+16a,NomotoLeung2017}. 
If, on the contrary, the deleptonization dominates, the collapse cannot be
halted and the star collapses into a neutron star (core collapse ECSN) \citep{Miyaji+80,MiyajiNomoto87,Nomoto87,Kitaura+06,Fischer+10,Jones+16a,Radice+17,Zha+19,Zha+22}. Which
outcome is realized from ECSNe depends on both the details of the
modeling of the presupernova evolution and explosion.

Stars in the mass range $\rm \sim 10-12~M_\odot$ ignite Ne-burning
off-center that develops in a similar fashion as the off-center C burning \citep{Nomoto84}.  The final fate of these stars depends, once again, on the behavior of the off-center Ne burning \citep{Nomoto88}. If the Ne-burning is ignited far enough
off-center, the contracting core may achieve densities sufficiently
high for the URCA processes to be activated until the conditions for
an ECSN are reached before the Ne burning front is able to reach the
center. If, on the contrary the Ne burning front reaches the center
before the activation of the URCA processes, O-burning first and
Si-burning later are ignited off-center leading ultimately to a CCSN
\citep{jones+13,jones+14,Nomoto14,WH2015}.

%Due to the low entropy, electron captures in these stars are efficient in reducing the electron fraction producing a chemical composition enriched in isotopes in general more neutron rich than their more massive counterparts. These stars are though to eventually produce a failed massive star explode as CCSNe \citep{WH2015}.

Stars in the range $\rm 7-12~M_\odot$ constitute roughly $50\%$ of the
stars (by number) more massive than $\sim 7~M_\odot$ according to a
standard initial mass function (IMF). This means that a proper
knowledge of how they evolve and die, is crucial for many
astrophysical subjects. Determining the mass boundaries between stars
that form CO-WD, ONeMg-WD, ECSNe and CCSNe is mandatory to understand
the relative frequencies among these objects. If a substantial
fraction of stars does indeed explode as ECSNe, then they may
contribute significantly to the overall SN rate and also to the
population of neutron stars. ECSNe have also been proposed as
potential sites for the r-process \citep{Ning+07}, which is
responsible for the synthesis of roughly half of the nuclei above the
Fe group.  The ECSN has been proposed for SN 1054 that formed the Crab
Nebula in view of the similarity of low explosion energies and small
amount of heavy elements between the Crab Nebula and the ejecta of the
ECSN \citep{Nomoto+1982}.  ECSNe may explain the observations of
sub-luminous type II plateau supernovae (SNIIP) with a low amount of
$\rm ^{56}Ni$ ejected \citep{Smartt09}.
Recently, \cite{Hiramatsu+21} observed SN II 2018zd and found its
several observed features can be well-explained by an ECSN but not by
an Fe-core collapse SN.
In the context of the chemical evolution of the galaxies, because of
the shape of the IMF, these stars should contribute significantly to
the production of some specific isotopes. Also, the chemical yields
produced by stars in this mass interval are either completely ignored
or obtained by means of an arbitrary interpolation between the yields
produced by the AGB stars and those produced by the MS. In both cases
a significant error can be made. Moreover, a large fraction of massive
stars is part of binary or multiple systems \citep{DK13}. Binary
interactions crucially affect the amount of mass lost by these stars
and hence even more massive progenitors may contribute to produce
ECSNe \citep[see, e.g,][and references
therein]{Brinkman+23}. Unfortunately, the binary scenario remains so
far mostly unexplored.

Despite their astrophysical relevance, not many models in the range $\rm 7-12~M_\odot$ that cover the full evolution are available at present.
In particular, in none of the papers found in literature on this subject \citep[e.g.][]{ritossa+99,Siess10,Ventura+11,Lau+12,Karakas+12,Taka+13,gilpons+13,Ventura+13,jones+13,jones+14,doherty+15,WH2015,Taka+19,Zha+19}
a homogenous, detailed and comprehensive study of the full evolution of stars in the mass range $\rm 7-12~M_\odot$, with a rather fine mass grid, is presented.
In some cases, as it is reported in a number of recent papers, the TP phase cannot be followed due to numerical problems and the core growth is treated either parametrically, assuming an arbitrary accretion rate, or it is even neglected \citep{WH2015,Taka+19}. Moreover, the full evolution of these stars with rotation has never been computed, as far as we know. The main reason for the paucity of models in this mass range is that the computation of their evolution is extremely challenging from a numerical point of view and in general it also requires an enormous amount of computer time and memory. As already mentioned above, in these stars, depending on the initial mass, C-, Ne-, O- and Si-burning ignite off-center and the burning front propagates inward in mass accompanied by a number of convective shells. A proper treatment of the heat transfer from the burning front to the underlying layers requires an extremely fine spatial resolution of the order of a few km. The full coupling of convective mixing and nuclear burning, possibly coupled to the structure equations, is also necessary to properly follow these stages and avoid numerical instabilities. The adoption of a quite extended nuclear network, including at least 100 or more nuclear species fully coupled to the stellar evolution, is also needed to properly compute the energy generation and to trace the abundance of the various isotopes and in particular the electron fraction. A very large number of thermal pulses are expected for SAGB stars, ranging from several tens to thousands. Typically, the thickness of the inter-shell region, i.e. between the He- and the H-shell, is of the order of $\rm 10^{-4}-10^{-5}~M_\odot$. Resolving all these phenomena, both in space and in time, is extremely challenging and requires in general several thousands of zones per model and hundreds of thousands, if not millions, of models to cover the entire evolution. Coupling an extended network to these kinds of models implies the need of an enormous amount of computer time to compute a full single evolution.

This is the first paper of a series in which we aim to study in detail the evolutionary properties of stars in the transition from AGB stars to CCSN progenitors and how they change with the initial metallicity and rotation velocity. In this paper we present detailed evolution of solar metallicity, non rotating, stars in the range $\rm 7-15~M_\odot$ from the pre-main sequence phase up to the presupernova stage or up to an advanced stage of the thermally pulsing phase, depending on the initial mass. The main goal is to (1) study how the evolutionary properties of these stars change in the transition from AGB to Massive Stars, (2) determine the limiting masses between AGB and SAGB stars and between SAGB stars and Massive Stars, (3) determine their final fate and ultimately (4) identify the limiting masses that mark the transitions from the various final outcomes, i.e., CO-WD, ONe-WD, ECSNe and CCSNe.

\section{Stellar evolution code and nuclear network}\label{code}

In the last 20 years we developed and continuously improved our stellar evolution code {\scshape{franec}} \cite{cl13,lc18}. One of the strengths of the code is that it can automatically manage nuclear networks including an arbitrarily extended list of isotopes and associated reactions. Another feature that makes this code unique in the panorama of the stellar evolution codes is that all the equations describing a) the physical structure of the star, b) the chemical evolution due to the nuclear reactions and c) the chemical mixing due to a variety of instabilities (thermal and mechanical)  are coupled together in a single system of equations and solved simultaneously. Last, but not least, {\scshape{franec}} includes the treatment of the stellar rotation, the transport of the angular momentum and the rotation induced mixing of the chemicals. The multi-threads/parallel solver algorithms used to invert the large sparse matrices, needed to solve the system of equations, are the most sophisticated and fast presently available. All the features mentioned above make this code extremely robust and fast from a numerical point of view and suitable to study essentially all the evolutionary phases of stars, even the most challenging ones like the advanced burning stages of massive stars, the off-center C-, Ne-, O- and Si-burning, as well as the thermal pulses in AGB and SAGB stars.

All the models presented in this paper have been computed by means of the latest version of the {\scshape{franec}} code. This version is essentially the one adopted in \cite{cl13} and \cite{lc18}, with the following updates/differences. 

The induced overshooting occurring during the core He burning phase, because of the transformation of He into C and O, is properly
taken into account following \cite{Cast+85}. During the late stages of core He burning the occurrence of the breathing pulses has been 
properly inhibited as described in \cite{Caputo+89} and \cite{CS89}. 

The equation of state adopted is the one provided by F.X. Timmes and available in his web page (https://cococubed.com). It takes into account all stages of Saha ionization, plus a simple density ionization model, for elements between H and Zn, for all the degree of degeneracy and relativity. In addition to that, electron positron pairs and Coulomb corrections are also taken into account. 

%\begin{figure*}[ht!]
%\epsscale{1.0}
%\plotone{comp_model_alfetta.eps}
%\caption{Comaprison of two models of $\rm 15~M_\odot$ computed with the %full network (dashed lines), and with the reduced network (solid lines) (see text). Upper left panel: central abundance in mass fraction of the most abundant isotopes (see the legend on the right). Upper right panel: central temperature versus central density. Lower left panel: convective core mass as a function of time. Lower right panel: central value of the electron fraction as a function of the central temperature.\label{fig:comp_model_alfetta}}
%\end{figure*}

Since in this work we are mainly interested in studying the physical properties of the stars in the transition between AGB and massive stars and since the calculation of these models requires an enormous amount of computer time and memory, we have chosen a nuclear network that includes the minimum number of isotopes but that, at the same time, guarantees the calculation of the nuclear energy with great accuracy. 
%In order to find such a nuclear network as a first step we computed the presupernova evolutions of models with 13, 15, 20 and 25 $\rm M_\odot$, by means of the present version of the {\scshape{franec}} and with the large network adopted in \cite{lc18}. Then, we calculated a series of presupernova evolutions by changing the size of the network, both the number of isotopes and the number of reactions, until we could reproduce the evolutionary results obtained with the larger network (Figure \ref{fig:comp_model_alfetta}). 
The 112 isotopes included in the adopted network are reported in Table \ref{tabnetwork}. 
All these isotopes are coupled among each other by the most efficient reactions due to the weak and strong interactions for a total of about 466 reactions. Among them, we have also taken into account the following 20 URCA processes, due to their crucial role played in some of the models presented in this work:
\begin{eqnarray}
^{27}{\rm Al}(e^{-},\nu)^{27}{\rm Mg}(e^{-},\nu)^{27}{\rm Na} \nonumber \\
^{27}{\rm Na}(\beta^{-},\bar{\nu})^{27}{\rm Mg}(\beta^{-},\bar{\nu})^{27}{\rm Al} \nonumber \\
^{25}{\rm Mg}(e^{-},\nu)^{25}{\rm Na}(e^{-},\nu)^{25}{\rm Ne} \nonumber \\
^{25}{\rm Ne}(\beta^{-},\bar{\nu})^{25}{\rm Na}(\beta^{-},\bar{\nu})^{25}{\rm Mg} \nonumber \\
^{23}{\rm Na}(e^{-},\nu)^{23}{\rm Ne}(e^{-},\nu)^{23}{\rm F} \nonumber \\
^{23}{\rm F}(\beta^{-},\bar{\nu})^{23}{\rm Ne}(\beta^{-},\bar{\nu})^{23}{\rm Na} \nonumber \\
^{24}{\rm Mg}(e^{-},\nu)^{24}{\rm Na}(e^{-},\nu)^{24}{\rm Ne} \nonumber \\
^{24}{\rm Ne}(\beta^{-},\bar{\nu})^{24}{\rm Na}(\beta^{-},\bar{\nu})^{24}{\rm Mg} \nonumber \\
^{20}{\rm Ne}(e^{-},\nu)^{20}{\rm F}(e^{-},\nu)^{20}{\rm O} \nonumber \\
^{20}{\rm O}(\beta^{-},\bar{\nu})^{20}{\rm F}(\beta^{-},\bar{\nu})^{20}{\rm Ne} \nonumber
\end{eqnarray}
The energy release/absorption associated with these reactions has been treated as described in \cite{Miyaji+80}. 

The nuclear cross sections have been updated by taking into account the most recent experimental data and theoretical calculations, as described in Roberti et al. 2023 (submitted to ApJS). In particular, since the URCA rates must be carefully calculated \citep{Toki+2013,NomotoLeung2017}, we have adopted for these processes the refined electron capture and beta decay rates provided by \cite{Suzuki+2016}.

The zones unstable for convection are determined according to the Ledoux criterion in the H rich layers and according to the Schwarzschild criterion in all the other cases. 
As it is well known, convective core overshooting during core H burning determines the size of the He core at core H depletion that, in turn, drives all the following evolution of the star \citep{Bressan+81,Bertelli+86,Temaj+23}, unless the He core is further reduced by the mass loss during core He burning \citep{lc06,cl13}. In the present version of the code we use the same prescription used in the previous version adopted in \cite{lc18}, i.e., during core H burning we assume $0.2~H_{\rm P}$ of convective core overshooting by computing $H_{\rm P}=\left( P/\rho g \right)$ at the outer edge of the formal convective core, defined by the stability criterion mentioned above.
Some amount of extra-mixing is assumed at the lower edge of the convective envelope and of the convective shells that form within the electron degenerate CO core and that are associated to C-, NeO- and Si-off-center nuclear burning. The mixing efficiency in these zones are determined by assuming an exponentially decaying convective velocity given by:
$$
v_{\rm conv}(r)=v_{\rm conv}(r_0) {\rm exp}\left(-{r_0-r \over f {\rm H}_{\rm P}(r_0)}\right)
$$
where the subscript $0$ refers to values corresponding to the lower border of the convective zone, $v_{\rm conv}(r_0)$ is computed in the context of the mixing-length theory and the free parameter $f$ is assumed equal to 0.014. The diffusion coefficient is then computed as
$$
D(r)={1 \over 3} v_{\rm conv}(r) {\rm H}_{\rm P}(r)
$$

When a fuel is ignited off-center, because of the temperature
inversion due to the degeneracy of the matter \cite[see, e.g., Figures
 1-3 in][]{Nomoto84}, a convective shell develops and a sharp
discontinuity in the temperature forms at the base of the convective
zone, where burning is occurring. The capability of the burning
front to propagate inward in mass, i.e. as a continuous flame or by
recurrent flashes, as well as its speed, is in general controlled by
the coupling of convective mixing and heat transfer from the hot zones
at the base of the convective shell, where burning is occurring,
to the radiative cooler and inert zones beneath. Generally speaking,
if the heat transfer is efficient enough one expects a burning front
continuously propagating toward progressively more internal zones. On
the contrary, if the heat transfer is not efficient enough, the
propagation of the burning front toward the center, 
e.g., ONe burning, has been found to occur
by compressional heating \citep[Figures 26 and 27 in][]{Nomoto88}
through a sequence of convective shells that form where the fuel is
abundant and disappears as the fuel is locally exhausted. The
treatment of such a phenomenon is not trivial and different approaches and assumptions can be followed \cite[see, e.g.,][and references therein]{Nomoto88,jones+14,WH2015}. In this work we assume that every time a major fuel (C, NeO, Si) is ignited off-center, 
burning propagates as a convectively bounded flame (CBF). More
specifically, once the speed and the width of the burning flame is
assumed, a given amount of energy in the radiative layers below the
convective bound is deposited following the prescription of
\cite{WH2015}.

%Such a physical phenomenon needs a special treatment since it cannot be followed in a self consistent way and it is usually named Convective Bounded Flame (CBF). For the C-, ONe- and Si-CBF we follow the prescriptions provided by \cite{WH2015}. \cite{Nomoto88} discussed the competition between the two above mentioned phenomena finding that the heat conduction dominates in C burning while in ONe burning the timescale of the inward propagation of the burning front is shorter than the heat conduction \citep[Figures 26 and 27 in][]{Nomoto88}.
%  See discusion in section 6.3, p.32 of {Nomoto88}.

We adopt the commonly used mass loss rate provided by \cite{bloecker95} (equation 17 with $\eta=0.05$) during the AGB and SAGB phases \citep[see, e.g.][]{jones+13,doherty+15,doherty+17}. This mass loss rate adds to the other prescriptions already adopted in \cite{lc18}.

The initial composition adopted for the solar metallicity is the one provided by \cite{asplund+09}, which corresponds to a total metallicity of $\rm Z=1.345\cdot 10^{-2}$. The adopted initial He mass fraction is 0.265.

\begin{deluxetable}{lrrlrr}
\tablewidth{0pt}
\tablecaption{Nuclear network adopted in the present calculations\label{tabnetwork}}
\tablehead{
\colhead{Element} & \colhead{$\rm A_{\rm min}$} & \colhead{$\rm A_{\rm max}$} &
\colhead{Element} & \colhead{$\rm A_{\rm min}$} & \colhead{$\rm A_{\rm max}$}
}
\startdata
H........  &   1   &   3 &  P........  &  29   &  33 \\  
He.......  &   3   &   4 &  S........  &  31   &  35 \\  
Li.......  &   7   &   7 &  Cl.......  &  33   &  37 \\  
Be.......  &   7   &   7 &  Ar.......  &  36   &  38 \\  
C........  &  12   &  13 &  K........  &  39   &  39 \\  
N........  &  13   &  15 &  Ca.......  &  40   &  44 \\  
O........  &  15   &  20 &  Sc.......  &  43   &  45 \\  
F........  &  17   &  23 &  Ti.......  &  44   &  50 \\  
Ne.......  &  20   &  25 &  V........  &  47   &  51 \\  
Na.......  &  23   &  27 &  Cr.......  &  48   &  54 \\  
Mg.......  &  24   &  27 &  Mn.......  &  51   &  55 \\  
Al.......  &  26   &  27 &  Fe.......  &  52   &  58 \\  
Si.......  &  28   &  30 &  Co.......  &  55   &  59 \\  
\enddata                                              
\end{deluxetable}

\section{Results}

We computed the evolution of solar metallicity, non rotating, stars with the initial mass between 7 and $\rm 15~M_\odot$ from the pre-main sequence phase up to the presupernova stage or up to an advanced stage during the thermally pulsing phase, depending on the initial mass.
The main evolutionary properties of all the computed models are reported in Table \ref{tab_main_prop}, for the stars with the initial mass between 7 and $\rm 9.20~M_\odot$, and in Table \ref{tab_main_prop2} for stars with the initial mass between 9.22 and $\rm 15~M_\odot$. The various entries have the following (in most cases usual) meaning: $\rm M_{CC}$ is the maximum size of the convective core in units of $\rm M_\odot$; $\rm t$ is the evolutionary time in units of yr; $\rm ^{12}C$ is the central carbon mass fraction (this entry is reported only for the He burning phase and refers to the value at core He depletion); $\rm M_{Fe}$, $\rm M_{SiS}$, $\rm M_{ONe}$, $\rm M_{CO}$, $\rm M_{He}$, $\rm M_{CE}$ and $\rm M_{tot}$ are the iron core mass, the SiS (O depleted) core mass , the ONe core mass, the CO core mass, the He core mass, the convective envelope mass and the total mass, respectively, in units of $\rm M_\odot$; $\rm \psi_{c}$, $\rm T_{c}~(K)$, $\rm \rho_c~(g~cm^{-3})$ and $\rm Y_{e,c}$ are the central values of the degeneracy parameter, the temperature, the density and the electron fraction; $\rm T_{ign}~(K)$, $\rm \rho_{ign}~(g~cm^{-3})$, $\rm \psi_{ign}$ and $\rm M_{ign}~(\rm M_\odot)$ are the temperature, density, degeneracy parameter and mass coordinate corresponding to an off-center nuclear ignition; $\rm M_C$, reported among the quantities at the end of the second dredge up in Table \ref{tab_main_prop}, and at Neon ignition in Table \ref{tab_main_prop2}, refers to the mass coordinate, in units of $\rm M_\odot$, marking the central zone where the carbon mass fraction exceeds 0.01, this quantity being relevant for those stars that form hybrid CO/ONeMg cores, i.e., cores mainly composed by ONeMg but with a central region still rich in C.
% $\rm X_1$, $\rm X_2$, $\rm X_3$ and $\rm X_4$, reported in Table \ref{tab_main_prop2} are the central mass fractions of the most abundance nuclear species.

\subsection{Evolution during core H- and core He-burning}

\begin{figure*}[ht!]
\begin{center}
\includegraphics[width=.45\linewidth]{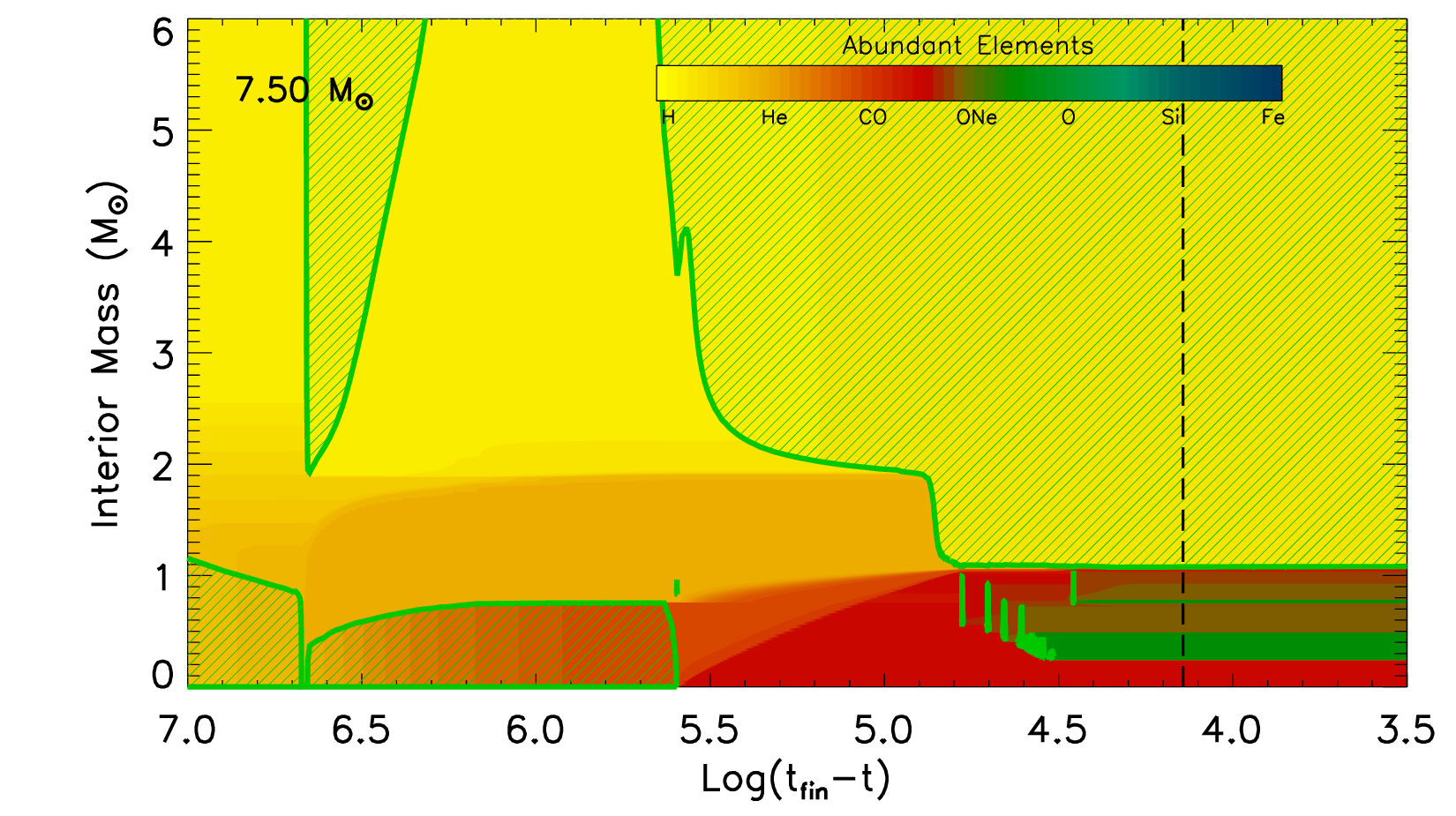}\quad\includegraphics[width=.45\linewidth]{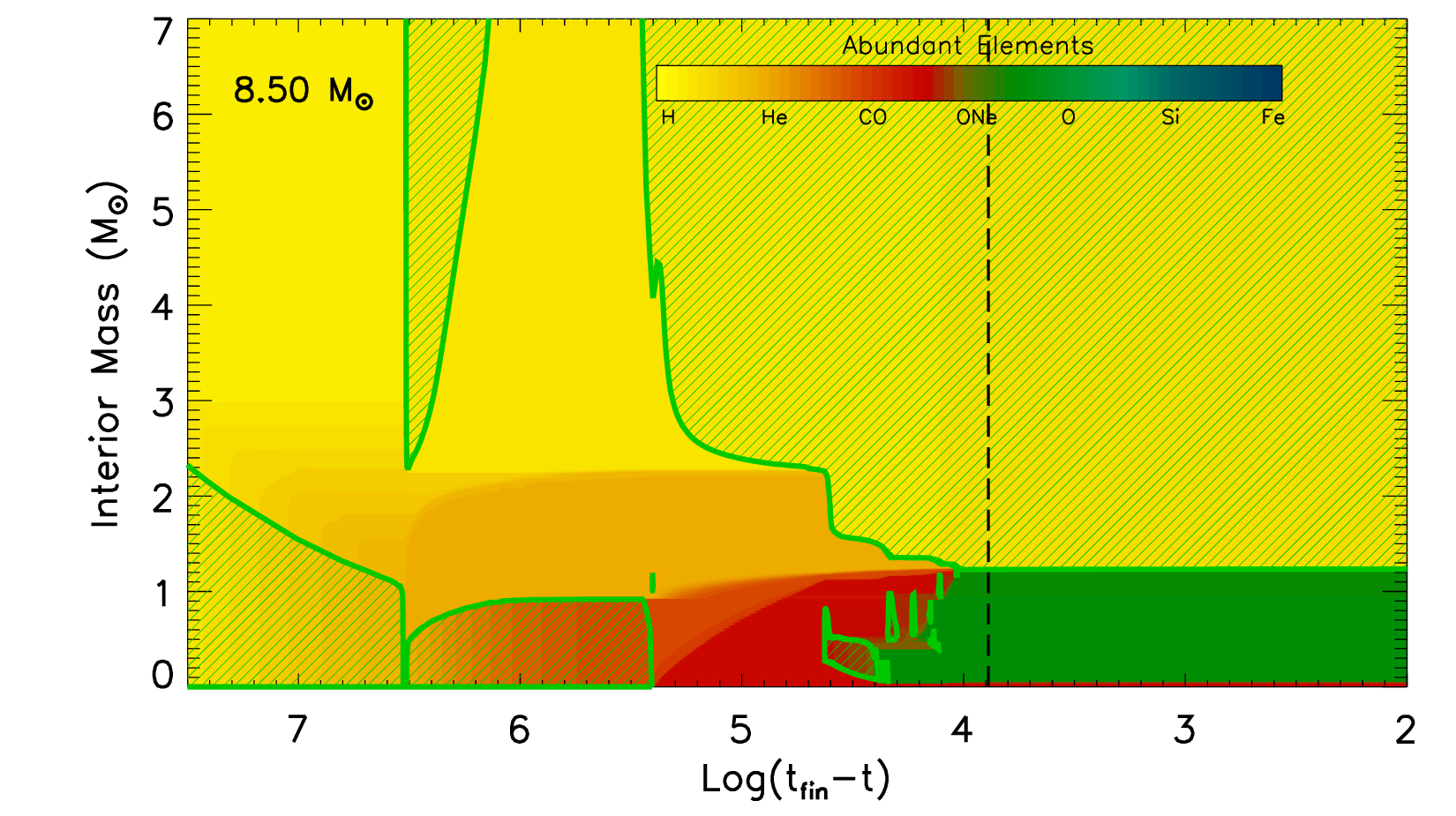}
\\[\baselineskip]% adds vertical line spacing
\includegraphics[width=.45\linewidth]{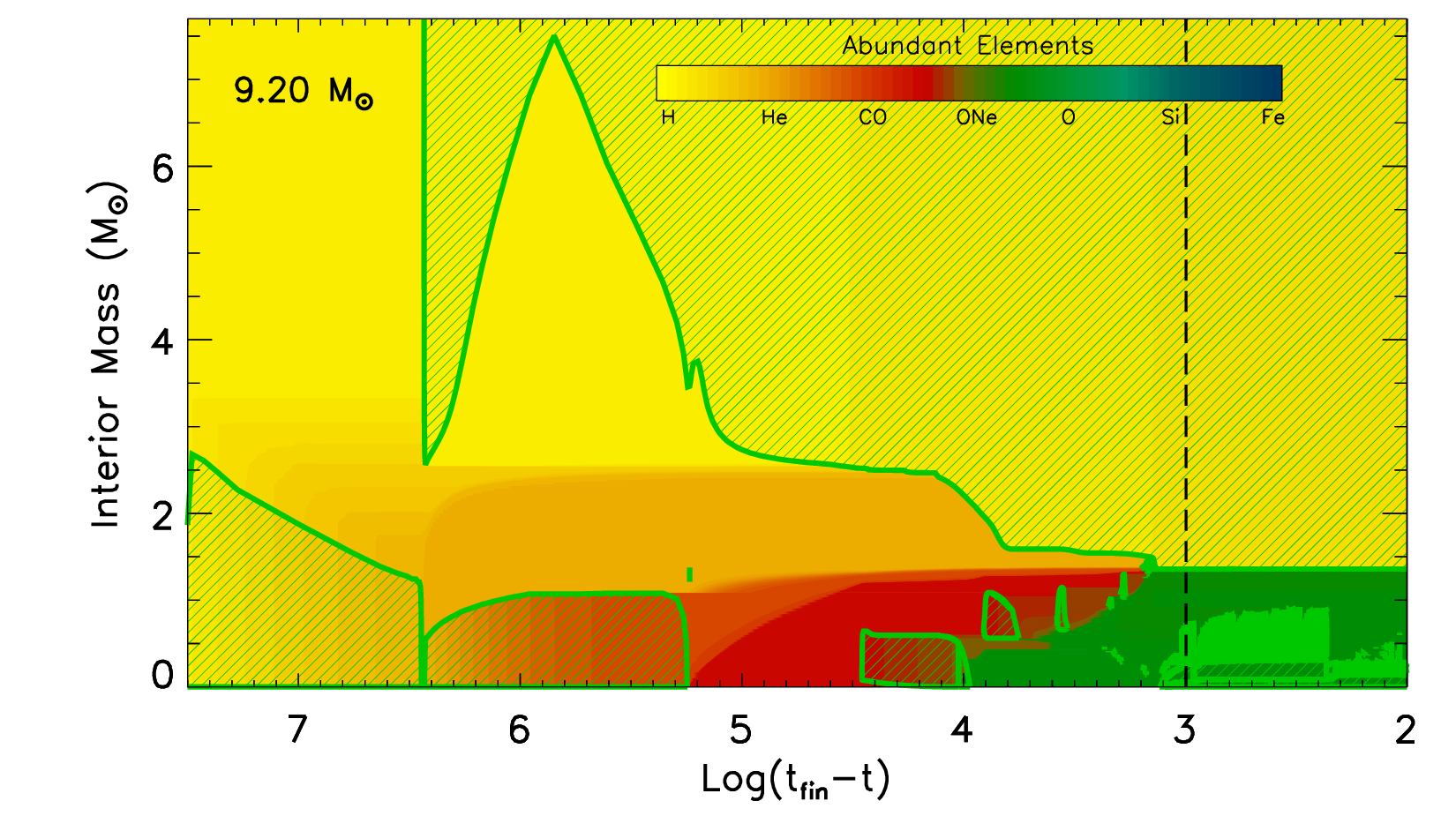}\quad\includegraphics[width=.45\linewidth]{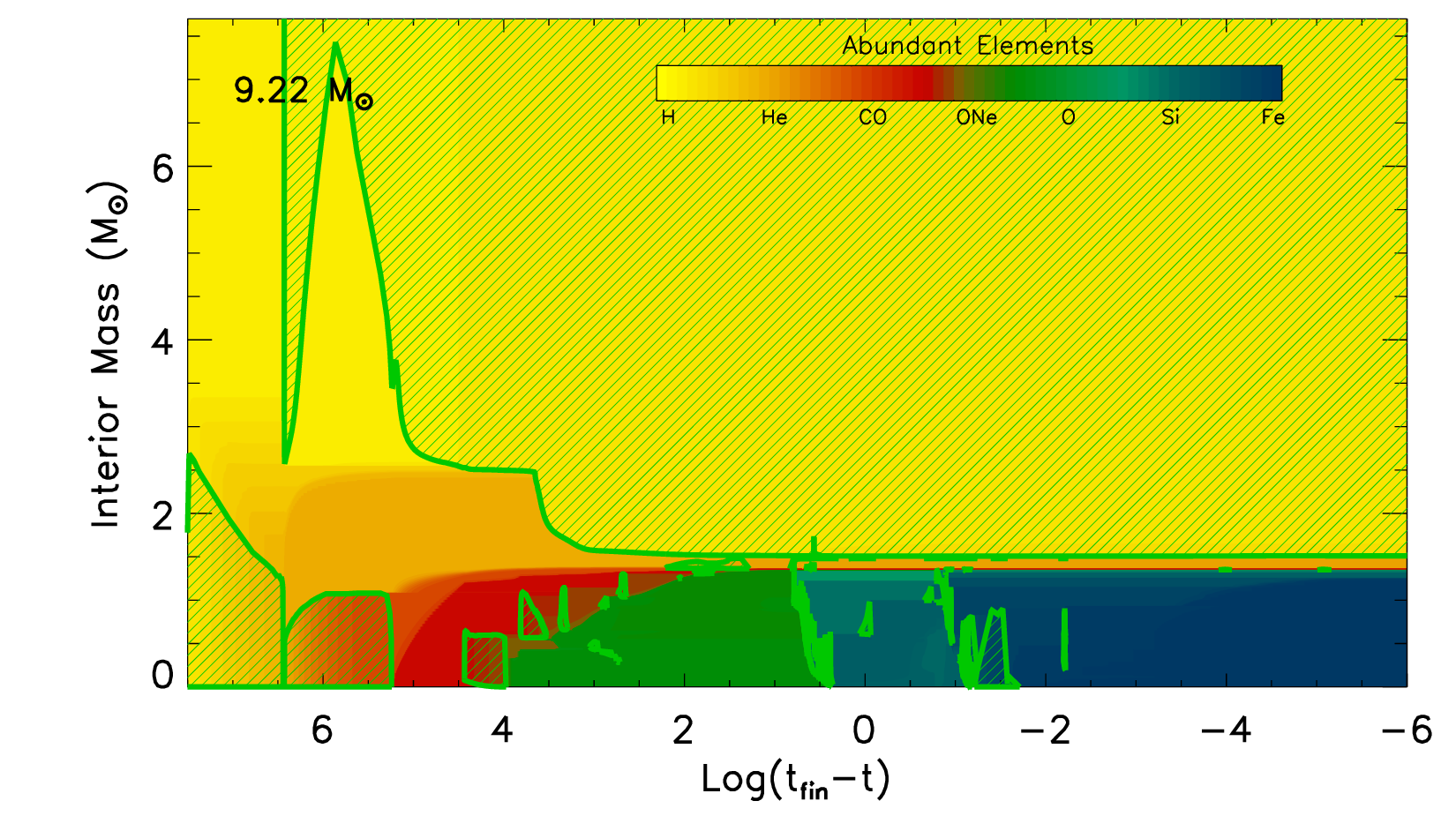}
\\[\baselineskip]% adds vertical line spacing
\includegraphics[width=.45\linewidth]{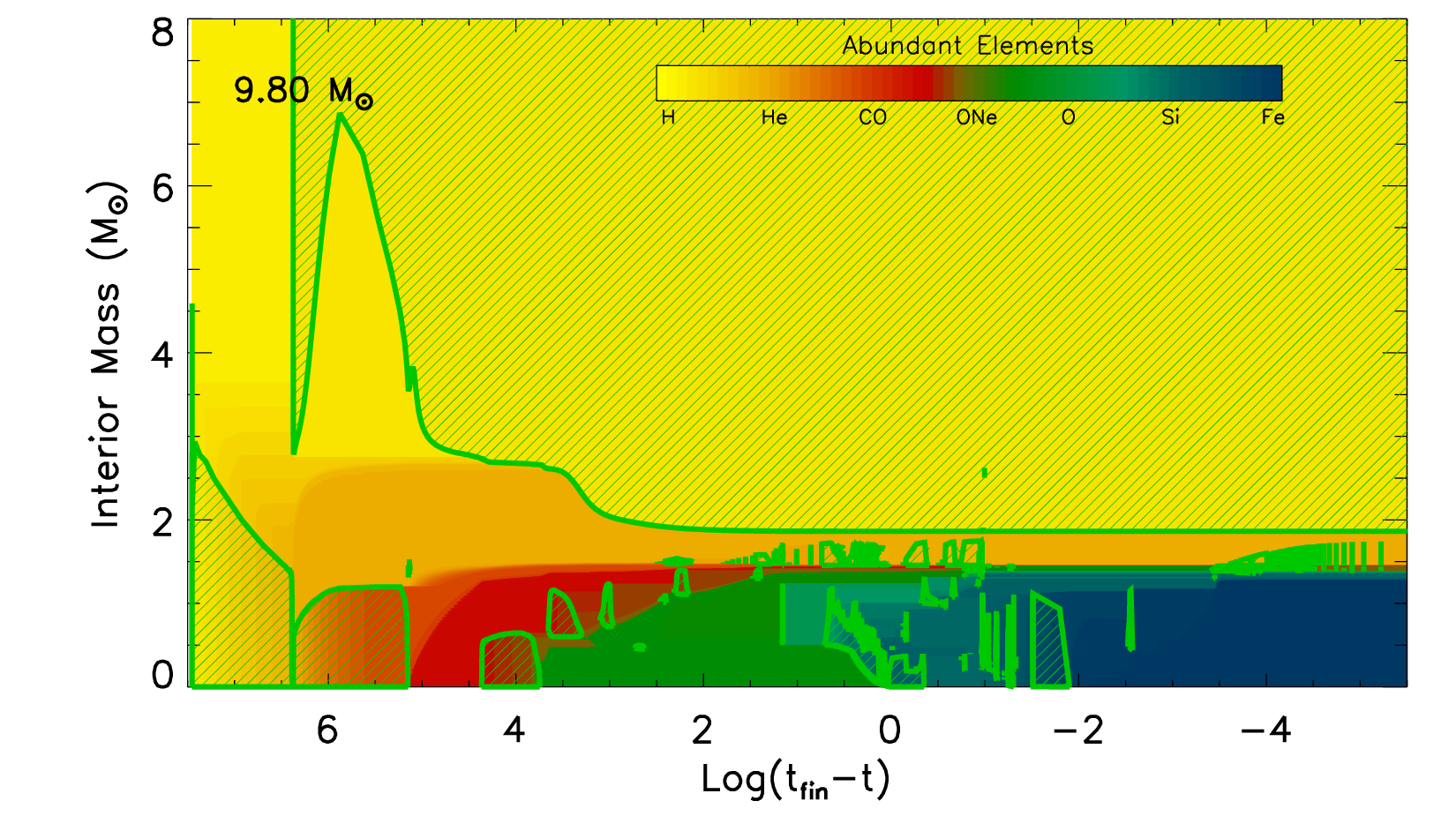}\quad\includegraphics[width=.45\linewidth]{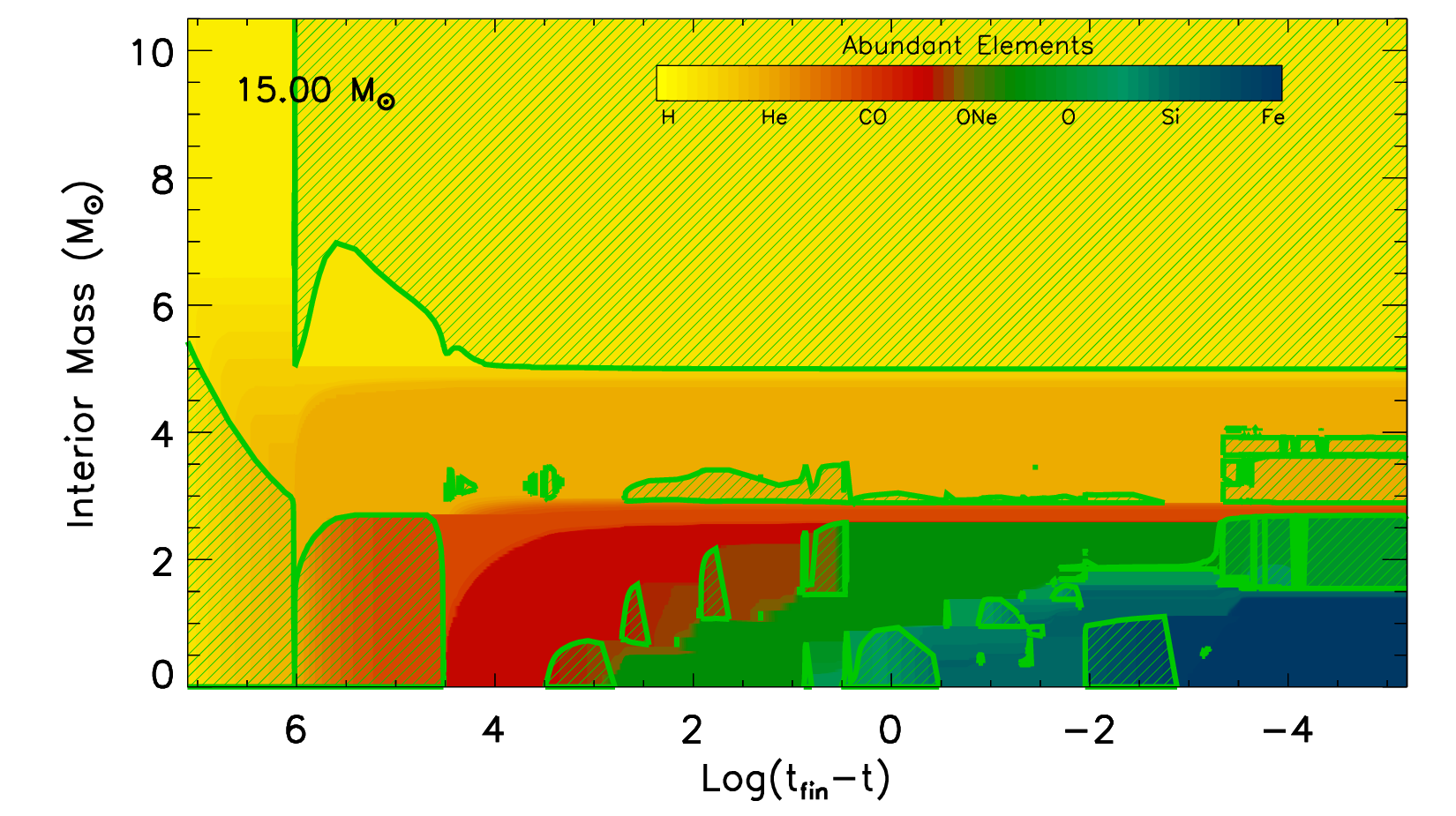}
\end{center}
\caption{Convective (green shaded areas) and chemical (color codes reported in the color bar) internal history of selected models from the main sequence phase up to the an advanced stage of TP-SAGB phase (upper and middle left panels) or up to the the onset of the iron core collapse prior to the core collapse supernova explosion (middle right and lower panels). The dashed line in the upper and middle left panels marks the onset of the thermally pulsing phase. In the x-axis is reported the logarithm of the time till the end of the evolution ($\rm t_{\rm fin}-t$) in units of yr.\label{kiptot}}
\end{figure*}

\begin{figure*}[ht!]
\epsscale{0.8}
\plotone{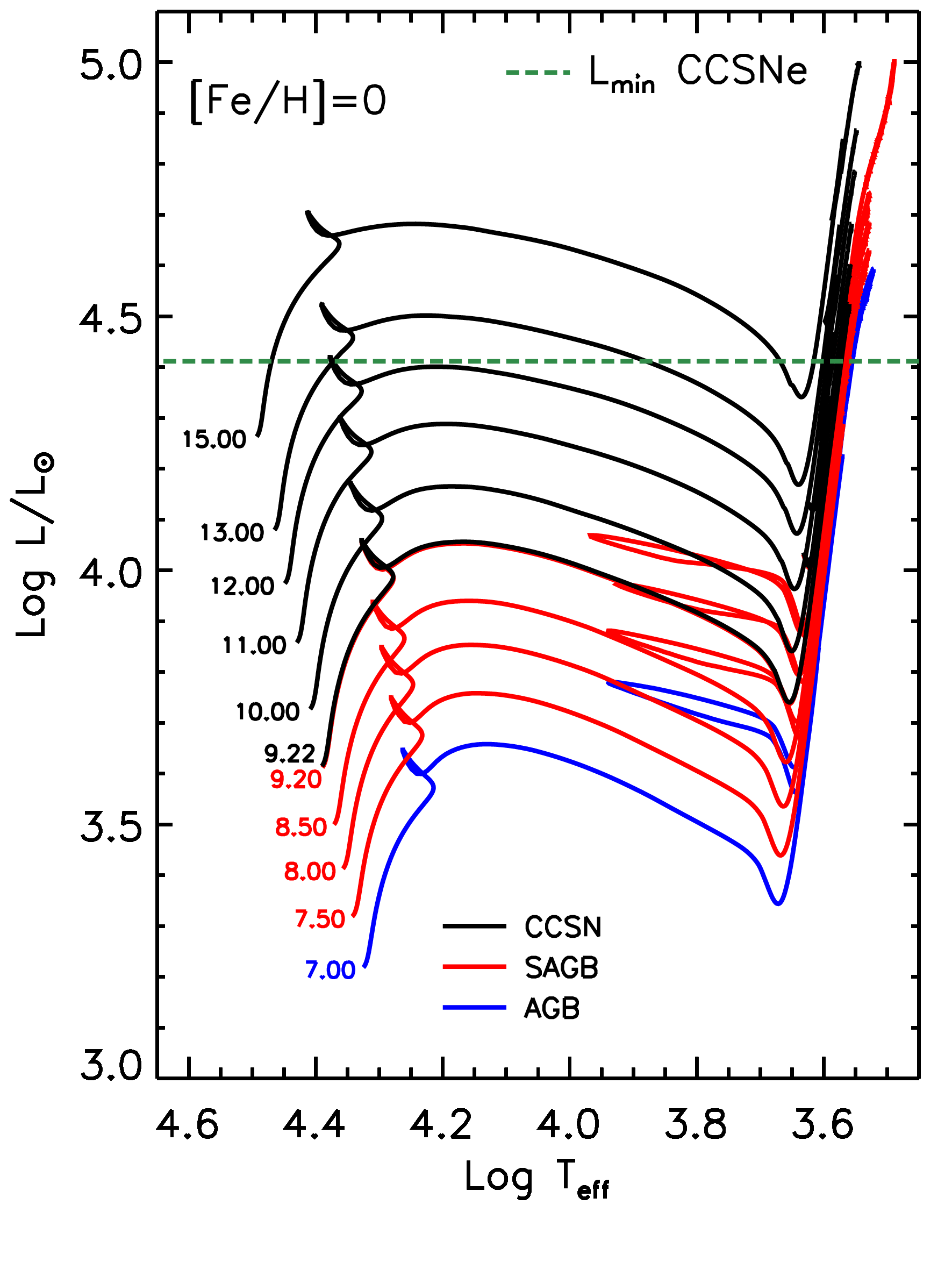}
\caption{Evolutionary path in the HR diagram of selected models. The blue line refers to AGB star, i.e., the one that develops a degenerate CO core, does not ignite C burning and enters the thermally pulsing phase. The red lines refer to SAGB stars, i.e., those that ignite C burning and then do not ignite Ne burning; these stars develop a degenerate ONeMg core and enter the thermally pulsing phase. The black lines refer to those stars that ignite Ne burning and eventually explode as CCSNe. The horizontal green dashed line marks the final luminosity of the lowest mass that explodes as CCSN, i.e., the expected minimum luminosity of CCSNe \citep{Smartt15}.\label{fig:hrtot}}
\end{figure*}

The evolution of all the computed models during core H burning is characterized, as usual, by the formation of a convective core that progressively recedes in mass. As a consequence at core H depletion a He core is formed surrounded by a zone with a gradient of chemical composition (Figure \ref{kiptot}). As H burning shifts progressively in a shell, all the stars move to the red side of the HR diagram (Figure \ref{fig:hrtot}) and become red giants. During this phase of the evolution, the surface temperature decreases, inducing the formation of a convective envelope that penetrates progressively in mass. When the convective envelope reaches the region of variable composition left by the receding H convective core, a dredge up to the surface of the products of the core H burning begins (first dredge up, see Figure \ref{kiptot}). 

In all the models, the core He burning starts after the first dredge up, i.e. when the star is a red giant, and develops in a convective core that, at variance with core H burning, increases progressively in mass (Figures \ref{kiptot}). Such an increase, due to the increase of the opacity as a result of the conversion of He into C and O, produces a sharp discontinuity in the radiative gradient at the edge of the convective core that drives the so called induced overshooting, i.e., the zone homogenized by convection extends beyond the formal border of the convective core, until the neutrality of the radiative and the adiabatic gradients is realized \citep{Cast+85}. During the late stages of core He burning, i.e. when the central He mass fraction decreases below $\sim 0.1$, the fresh He engulfed by the increasing convective core becomes comparable to the central He mass fraction and this produces a burst of nuclear energy that, in turn, drives a so called breathing pulse, i.e., a progressive increase of both the convective core and the central He mass fraction. This occurs until the He ingested does not produce any more a substantial increase of the nuclear energy \citep{Cast+85}. Since the occurrence of the breathing pulses is still highly uncertain, they have been suppressed as already mentioned in section \ref{code}.

\begin{figure*}[ht!]
\epsscale{0.8}
\plotone{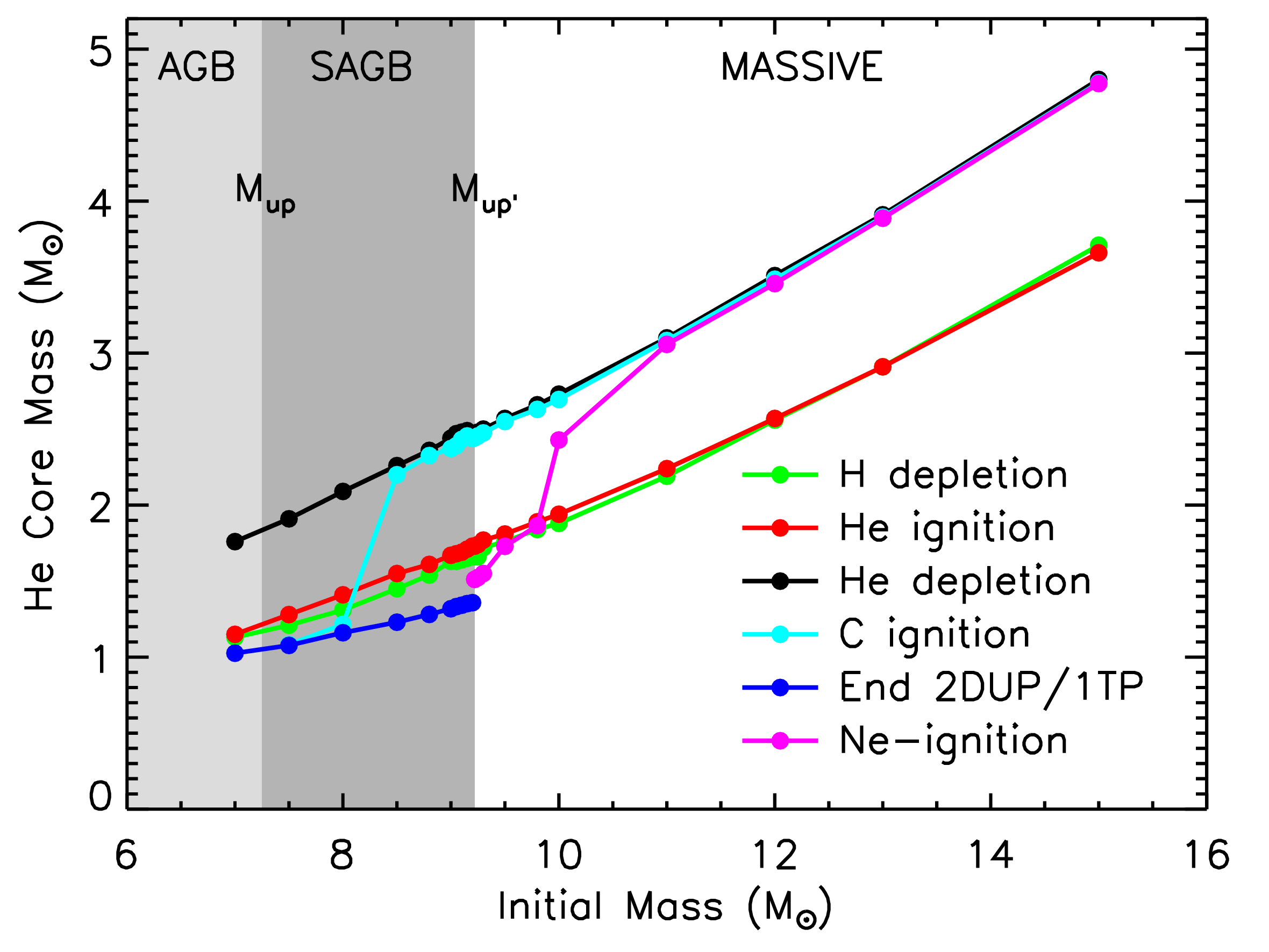}
\caption{He core mass at various evolutionary stages (see legend)\label{fig:mcore_he}}
\end{figure*}

\begin{figure*}[ht!]
\epsscale{0.8}
\plotone{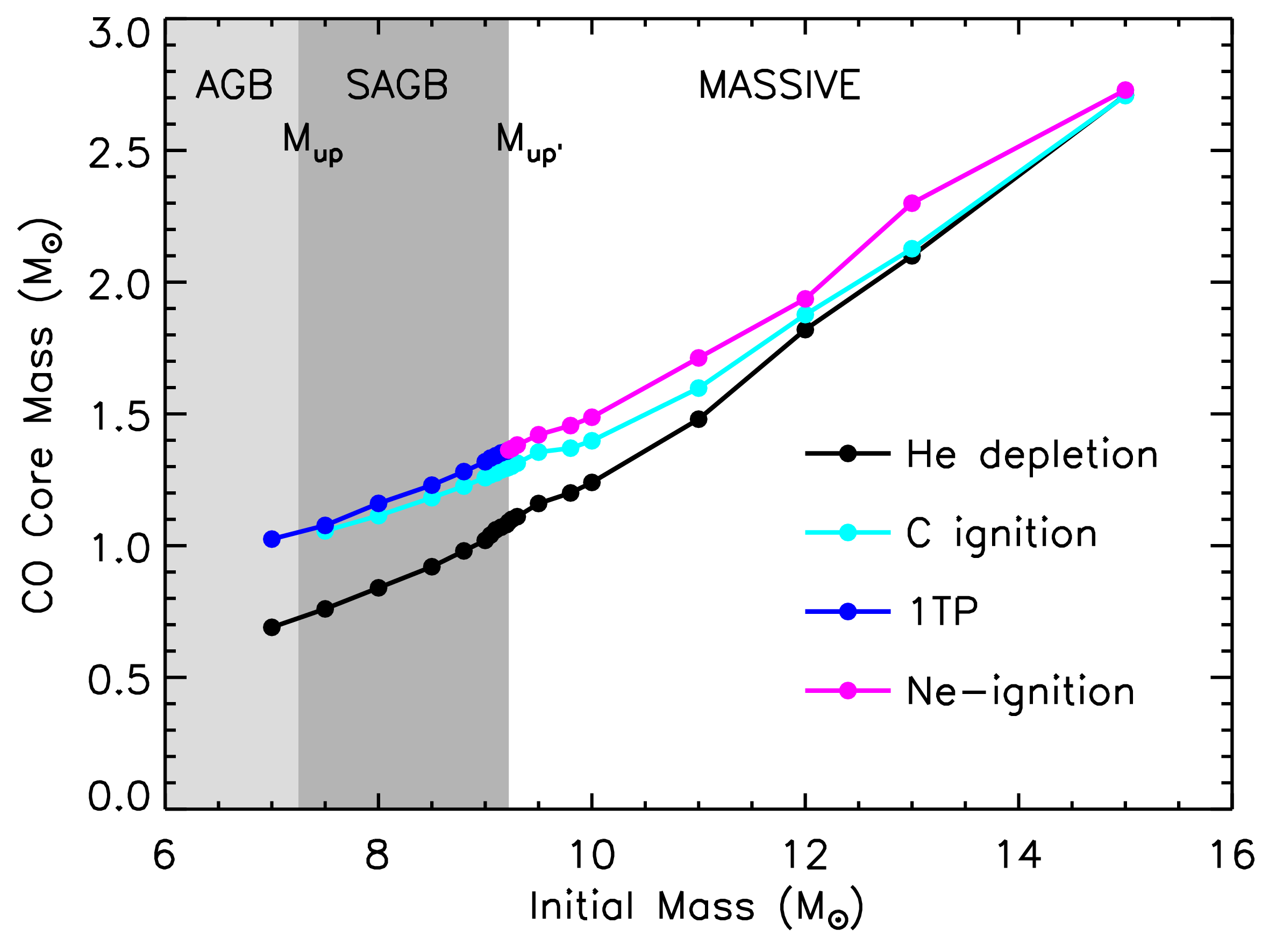}
\caption{CO core mass at various evolutionary stages (see legend)\label{fig:mcore_co}}
\end{figure*}

\begin{figure*}[ht!]
\epsscale{0.8}
\plotone{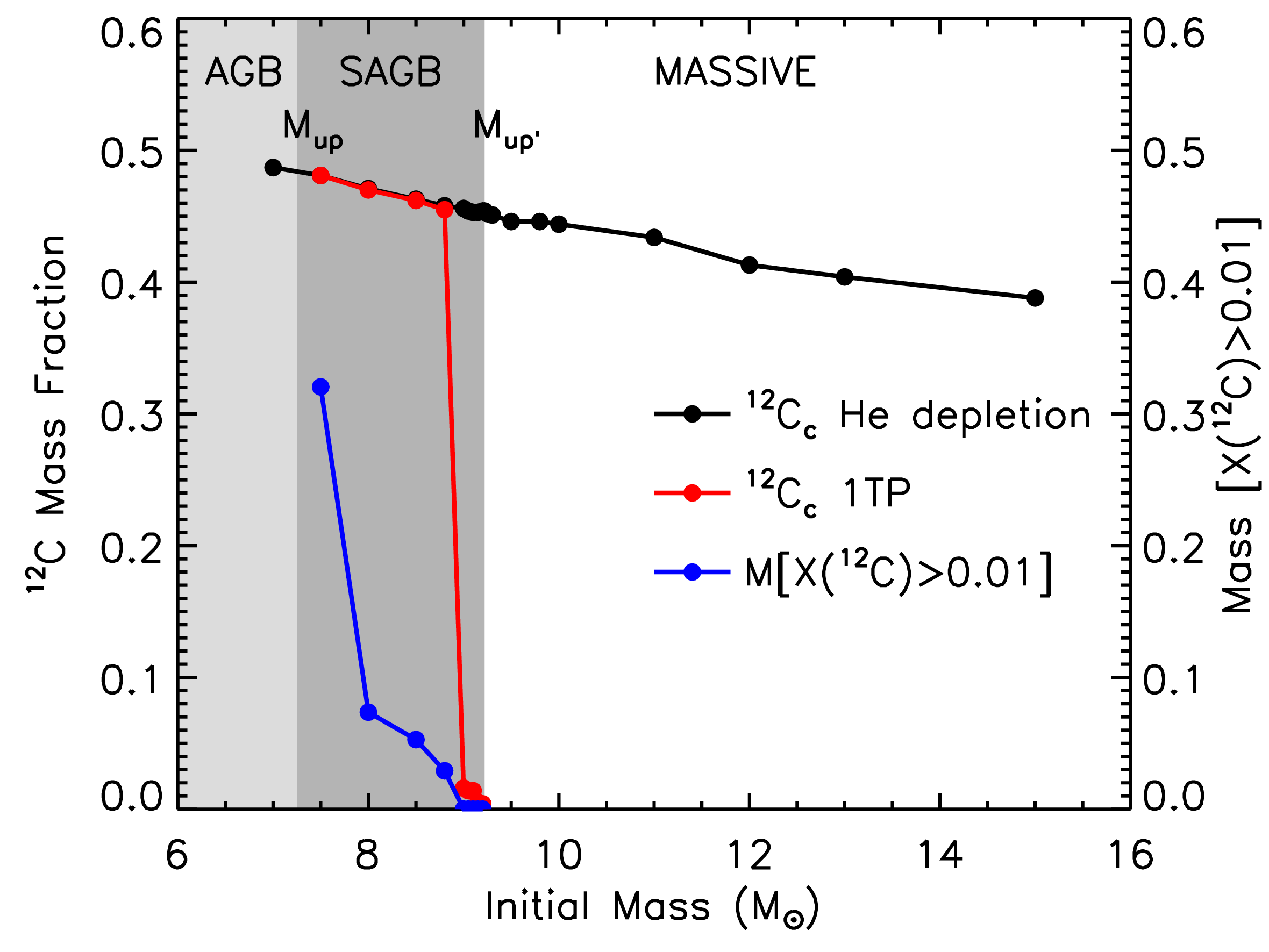}
\caption{Central $\rm ^{12}C$ mass fraction at core He depletion (black line and dots) and at first thermal pulse (red line and dots). The blue line refers to the mass of the central zone where the $\rm ^{12}C$ mass fraction is larger than 0.01. The Figure shows that the models with the initial mass between 7.5 and 8.8 $\rm M_\odot$ form a hybrid degenerate CO core.\label{fig:c12}}
\end{figure*}

%The number and the strength of the breathing pulses depend strongly on the numerics and eventually determine, in a direct way, the carbon abundance at core He depletion, that in turn drives the following evolution of the CO core \citep{cl20}. In absence of a specific treatment of the breathing pulses, the carbon abundance at core He depletion obtained for a grid of models shows, in general, a scattered behavior that is not real but simply the result of the numerical treatment. Since such an artificial scatter in the C abundance left by core He burning may produce unpredictable and non monotonic evolutionary properties during the following phases, the late stages of core He burning are treated in a very accurate way in order to obtained a monotonic and continuous decrease of the central He mass fraction with time. 
During core He burning, models with mass lower than $\rm 9.20~M_\odot$ perform an extended blue loop in the HR diagram (Figure \ref{fig:hrtot}) that is accompanied by a temporary recession and disappearance of the convective envelope when the stars cross the blue side of the HR diagram (Figure \ref{kiptot}). The H shell during core He burning is active and advances in mass increasing progressively the size of the He core. At core He depletion all the models are red giants and their He core mass is increased by $\sim 50\%$, with respect to the one at core H depletion (Figure \ref{fig:mcore_he}). The CO core at core He depletion increases with the initial mass (see Figure \ref{fig:mcore_co}, black line, Table \ref{tab_main_prop} and Table \ref{tab_main_prop2}), ranging between $\rm \sim 0.5~M_\odot$ and $\rm \sim 2.6~M_\odot$, while the $\rm ^{12}C$ mass fraction left by the core He burning decreases smoothly with the initial mass and ranges between $\sim 0.5$ and $\sim 0.38$ (see Figure \ref{fig:c12}, black line, Tables \ref{tab_main_prop} and \ref{tab_main_prop2}).

\subsection{Evolution after core He depletion}\label{hedepletion}

The evolution after core He depletion is characterized by the following processes: (1) the shift of the He burning shell, outward in mass, that progressively increases the size of the CO core and switches off the H burning shell;
(2) the substantial energy loss from the central zones due to the neutrino emission; (3) the energy deposition in the more external zones of the CO core due to compressional heating induced by the advancing of the He burning shell;
(4) the progressive penetration of the convective envelope that eventually may produce the second dredge-up (2nd-dup). The interplay and timing of these processes and the final fate of the star depend on the CO core mass at core He depletion that, in turn, depends on the He core mass at core H depletion and ultimately on the initial mass.

\begin{figure*}[ht!]
\epsscale{0.8}
\plotone{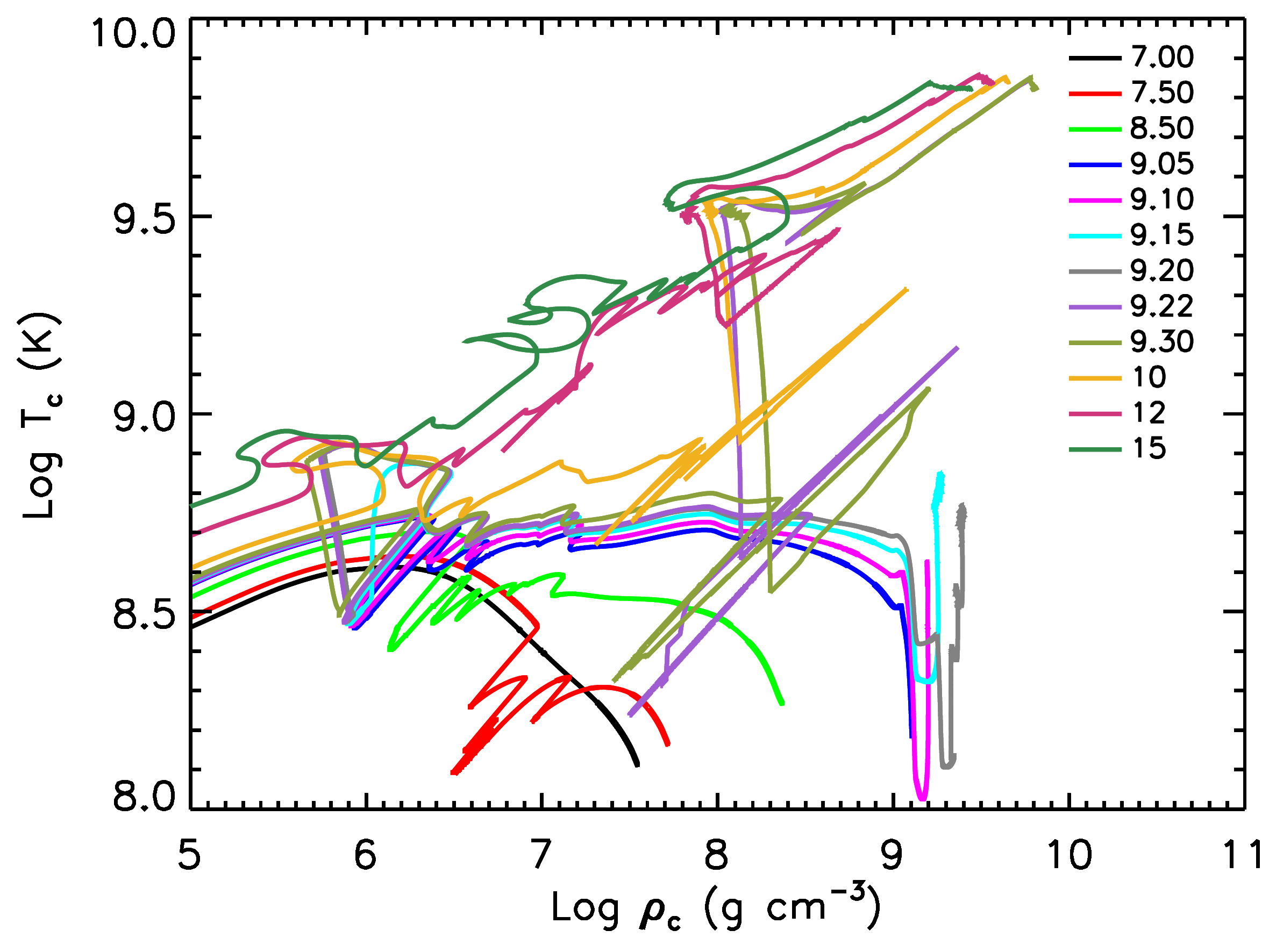}
\caption{Evolution of the central temperature and density of all the computed models.\label{fig:tcrocall}}
\end{figure*}

\subsection{Evolution toward the TP-AGB phase. Stars with initial mass $\rm M\leq 7.0~M_\odot$ $\rm (M_{CO}\leq 0.69~M_\odot)$}\label{sec:agb}
In stars with the initial mass $\rm M\leq 7.0~M_\odot$ (that form CO cores at core He exhaustion with $\rm M_{CO}\leq 0.69~M_\odot$), the maximum temperature within the CO core does not reach the threshold value for the C ignition, the CO core becomes progressively more and more degenerate and eventually these stars enter the thermally pulsing asymptotic giant branch (TP-AGB) phase. We followed the evolution of the $\rm 7~M_\odot$ model along 27 TPs. The evolution of solar metallicity AGB stars with mass around  $\rm 7~M_\odot$ has been already discussed in detail in literature (see section \ref{sec:intro}), therefore we will not address them here but we report in Table \ref{tab_main_prop} the main properties of the model at various key times during the evolution. Figures \ref{fig:hrtot} and \ref{fig:tcrocall} show the evolutionary path of the model in the HR diagram and in the central temperature-central density plane.
%A global look to the interior properties of the star during its entire evolution is given in Figures \ref{kiptot} (upper left panel). 
It is worth mentioning the occurrence of the second dredge up starting after core He depletion and going to completion before the onset of the TPs. The effect of the 2nd dredge up is that of reducing the He core from $\rm 1.76~M_\odot$ (value at core He depletion) to $\rm 1.01~M_\odot$ (Figure \ref{fig:mcore_he}).  
%Figures \ref{kip2dupTP} and \ref{kip2dupTPzoom} show a detail of the penetration of the convective envelope into the He core. 

\begin{figure*}[ht!]
\begin{center}
\includegraphics[width=.45\linewidth]{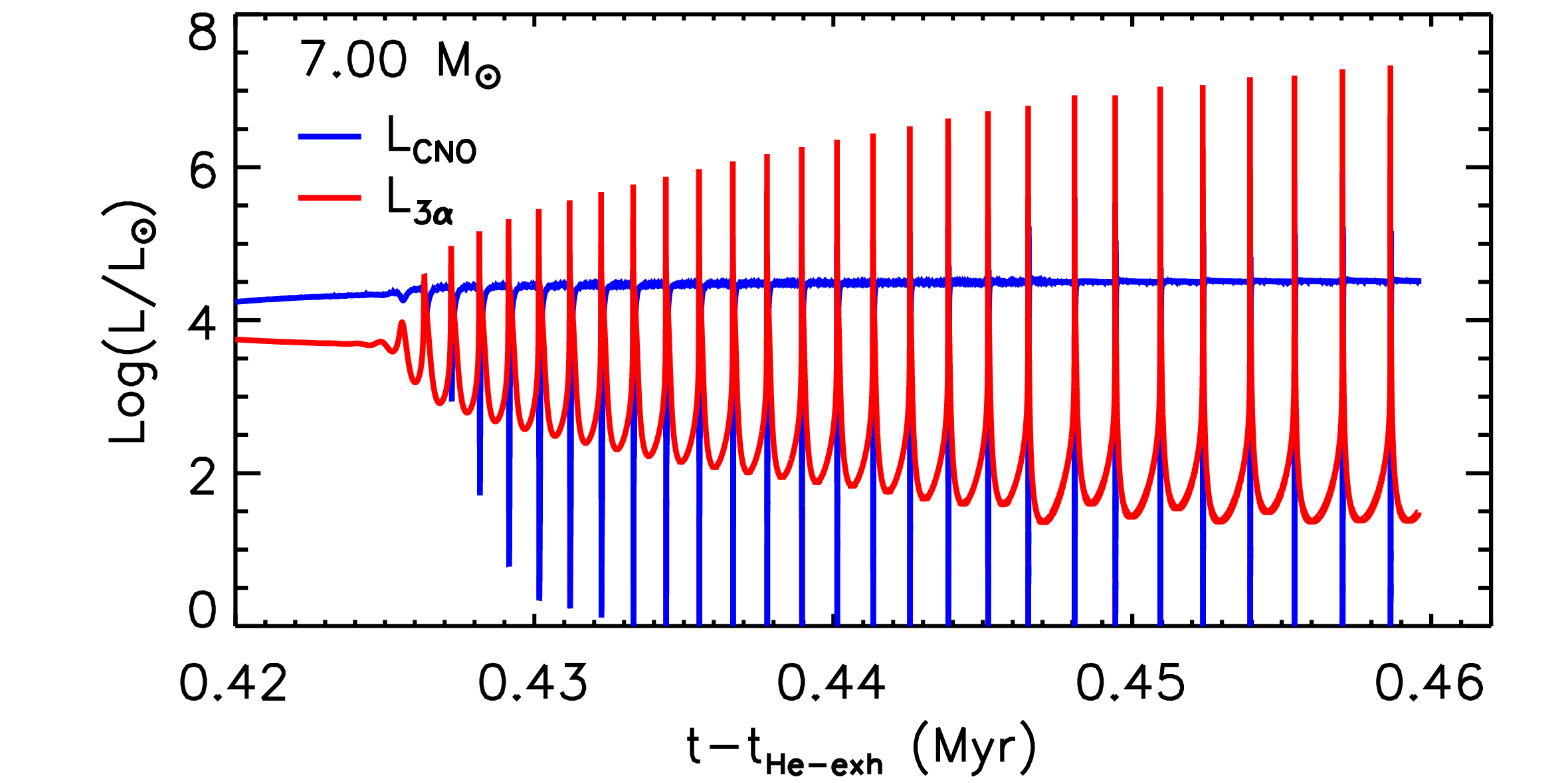}\quad\includegraphics[width=.45\linewidth]{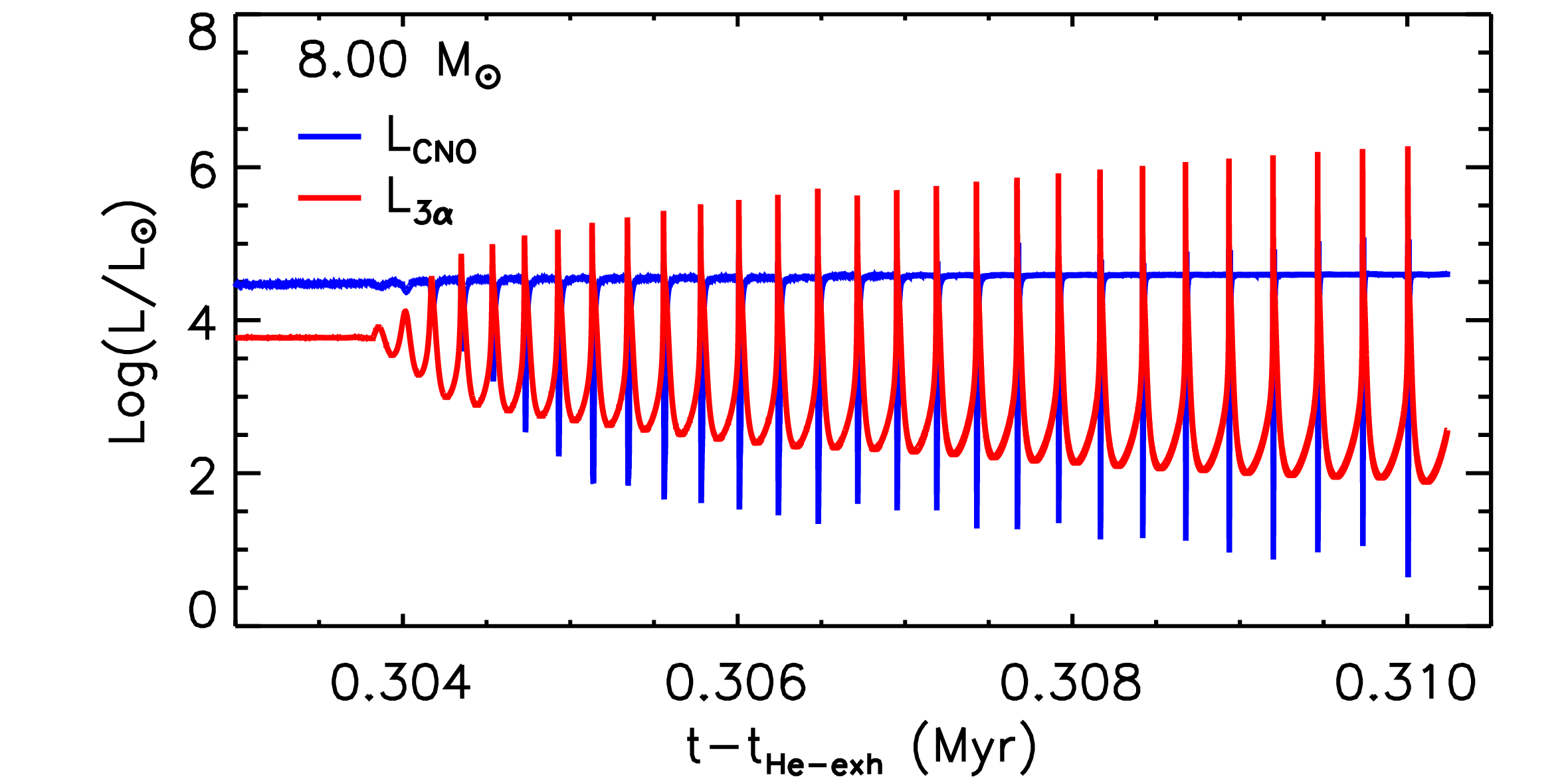}
\\[\baselineskip]% adds vertical line spacing
\includegraphics[width=.45\linewidth]{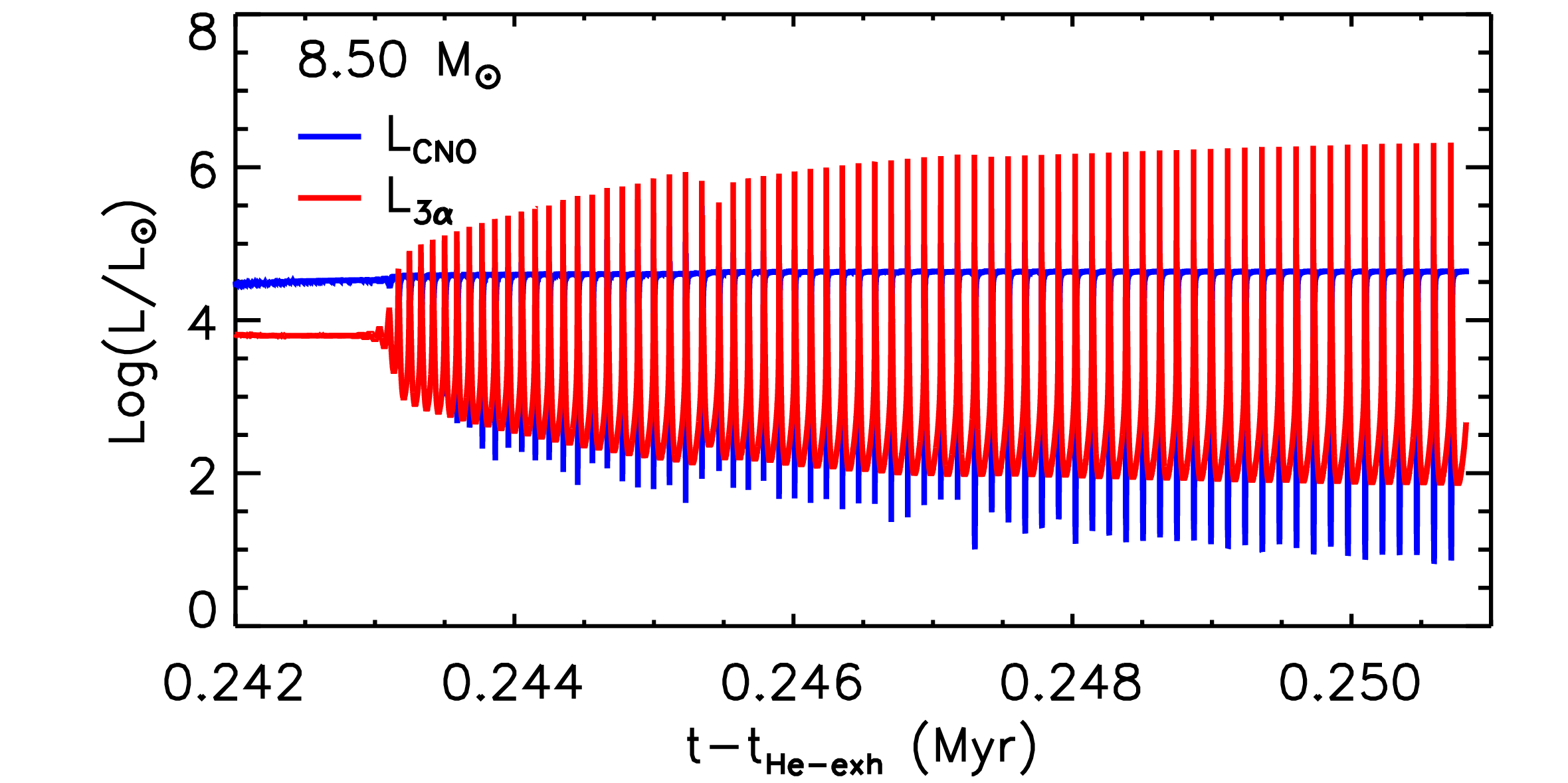}\quad\includegraphics[width=.45\linewidth]{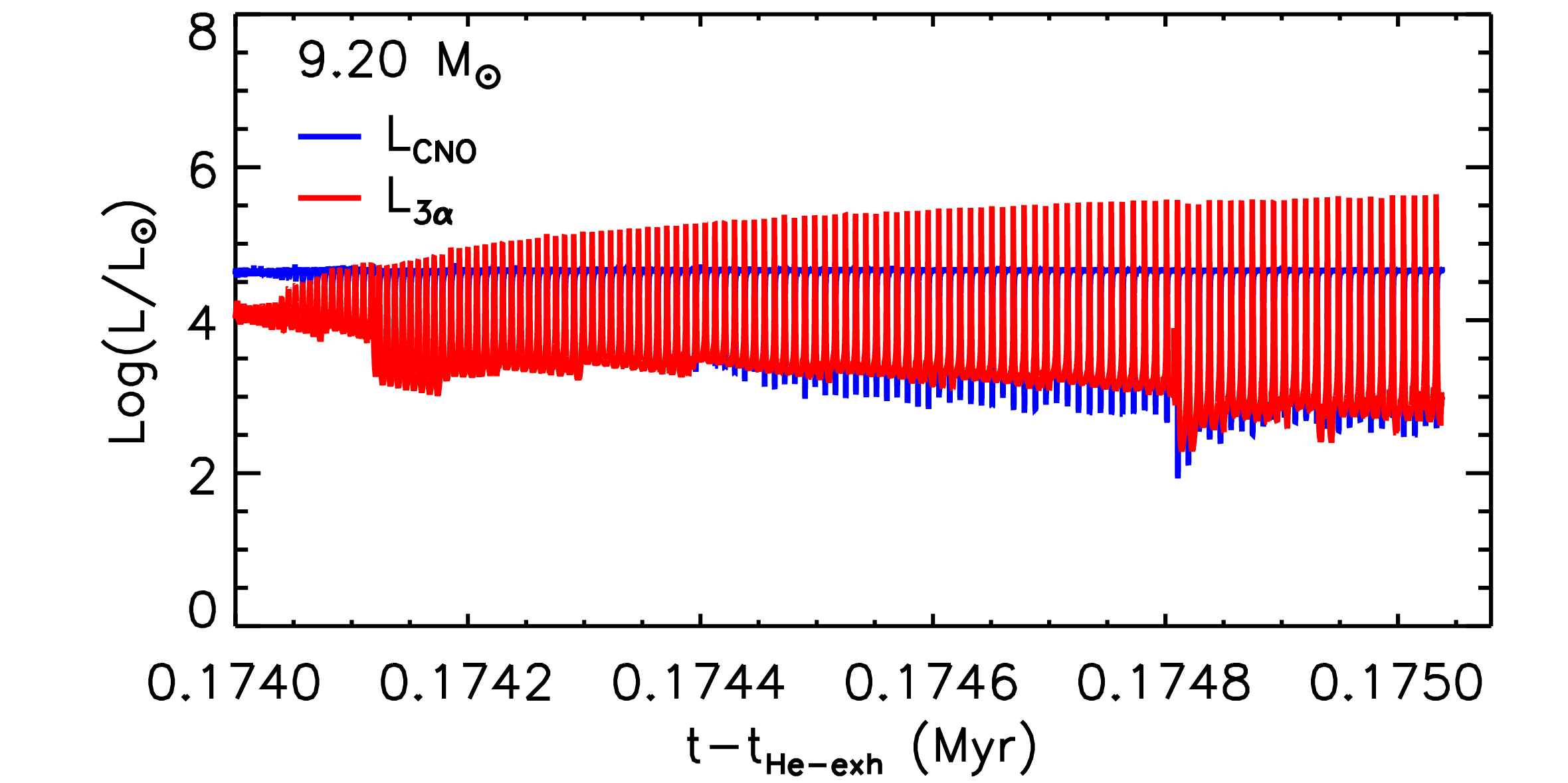}
\end{center}
\caption{Evolution of the H (blue line) and He (red line) luminosities as a function of time during the AGB phase (for the $\rm 7.0~M_\odot$, upper left panel) and during the SAGB phase (for the $\rm 8.0,~8.5,~9.2~M_\odot$ models, upper right panel and lower left and lower right panels, respectively. The time has been reset at the core He exhaustion. \label{l3aall}}
\end{figure*}

\begin{figure*}[ht!]
\begin{center}
\includegraphics[width=.45\linewidth]{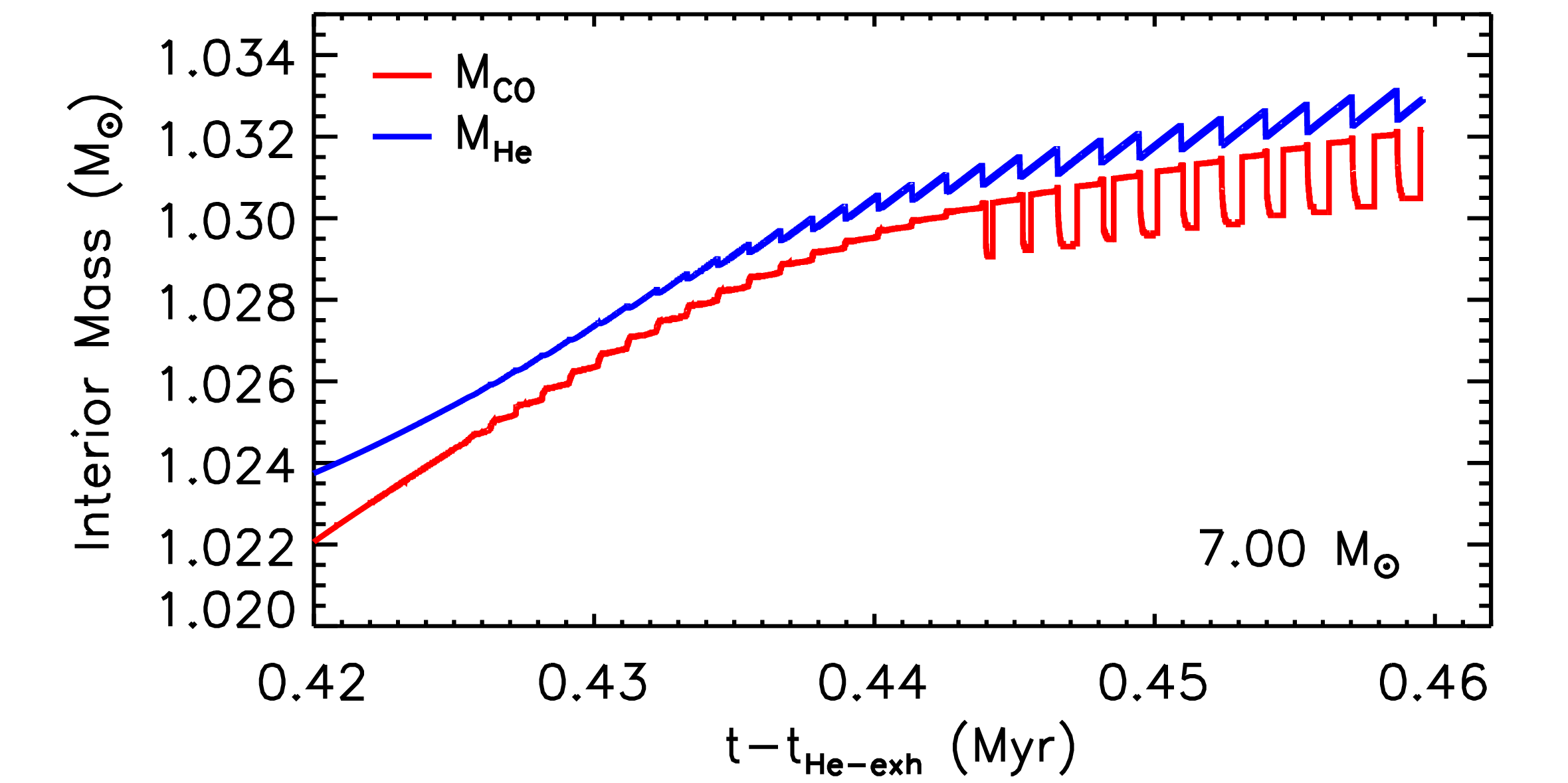}\quad\includegraphics[width=.45\linewidth]{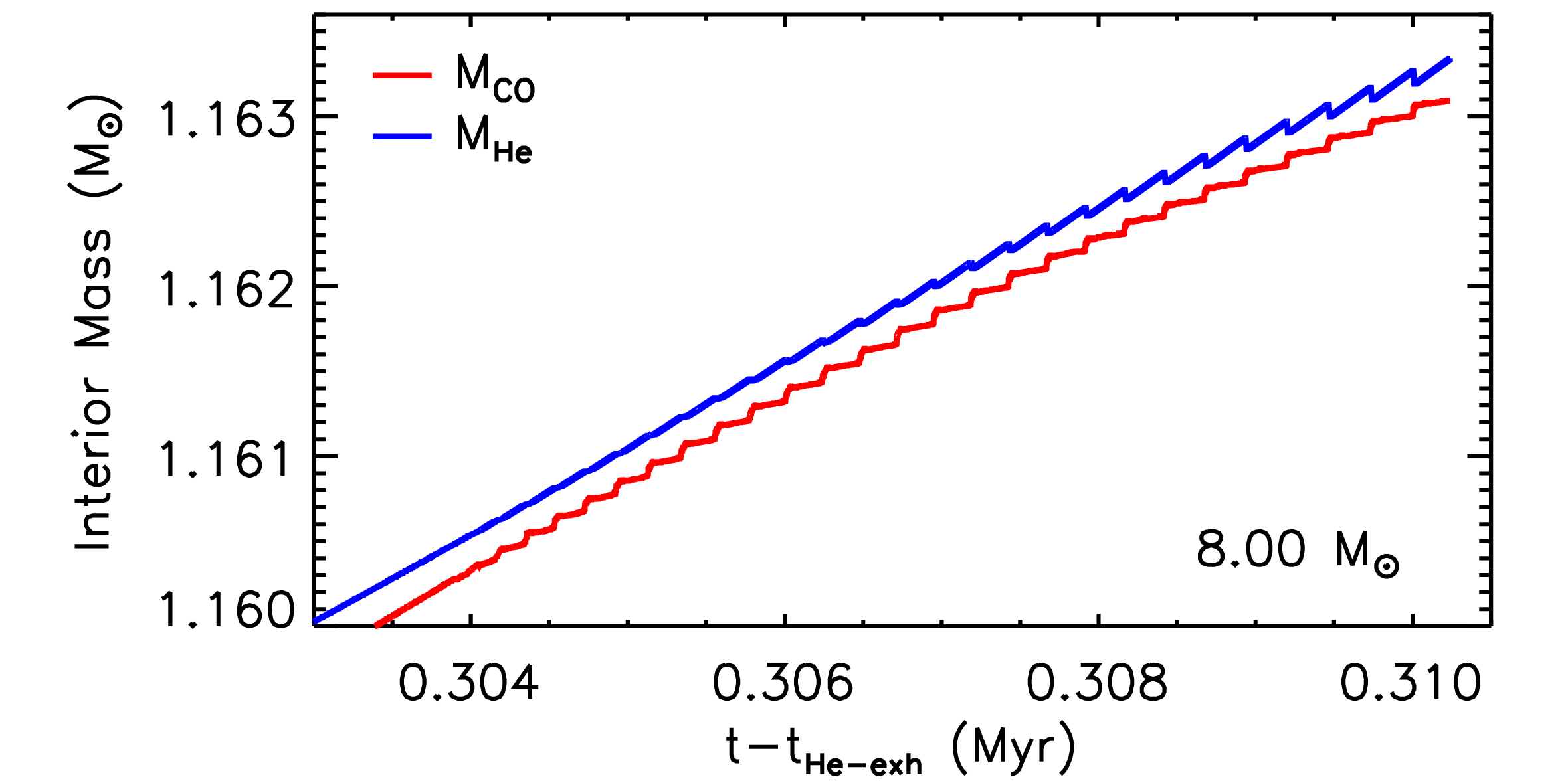}
\\[\baselineskip]% adds vertical line spacing
\includegraphics[width=.45\linewidth]{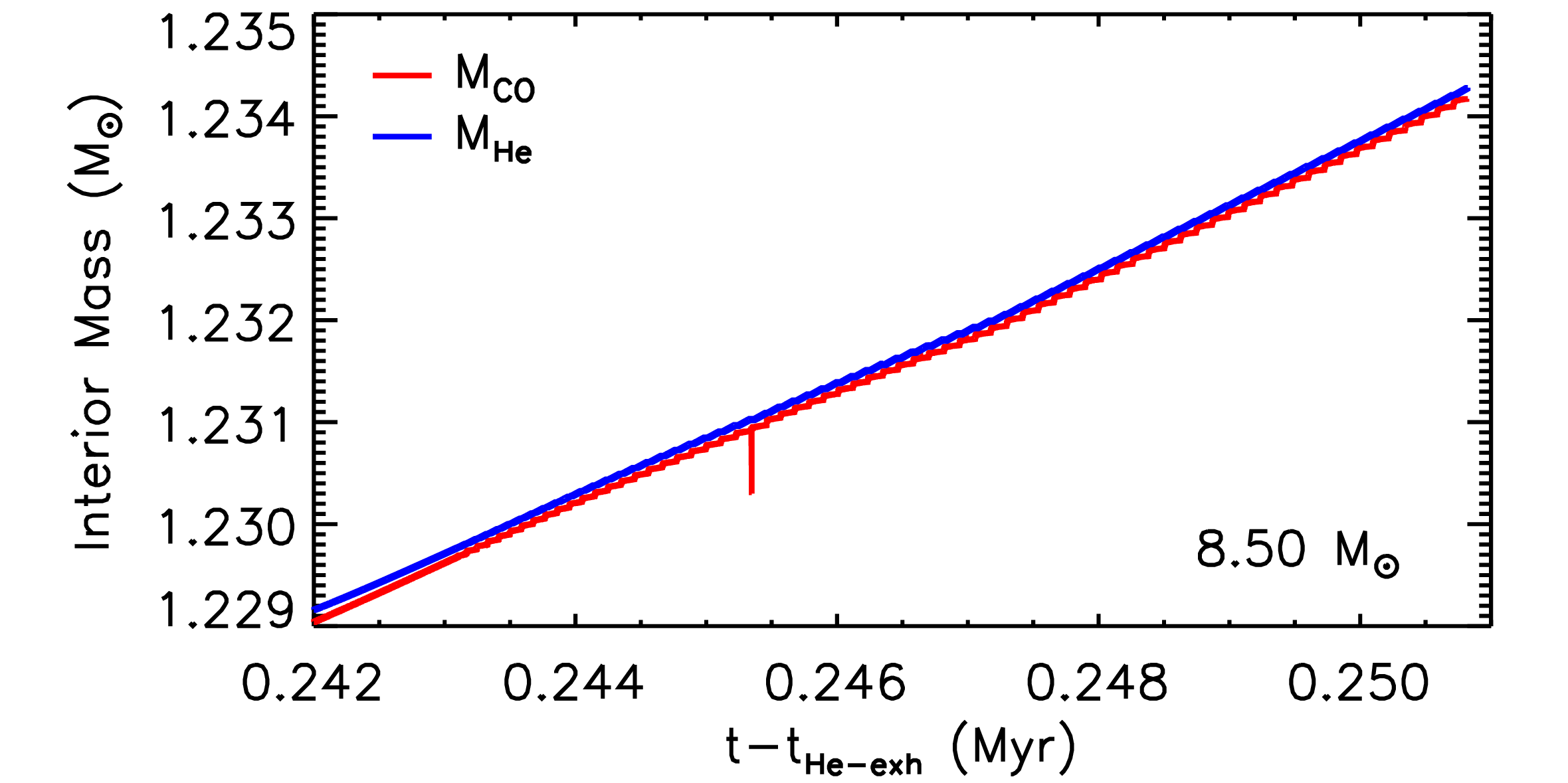}\quad\includegraphics[width=.45\linewidth]{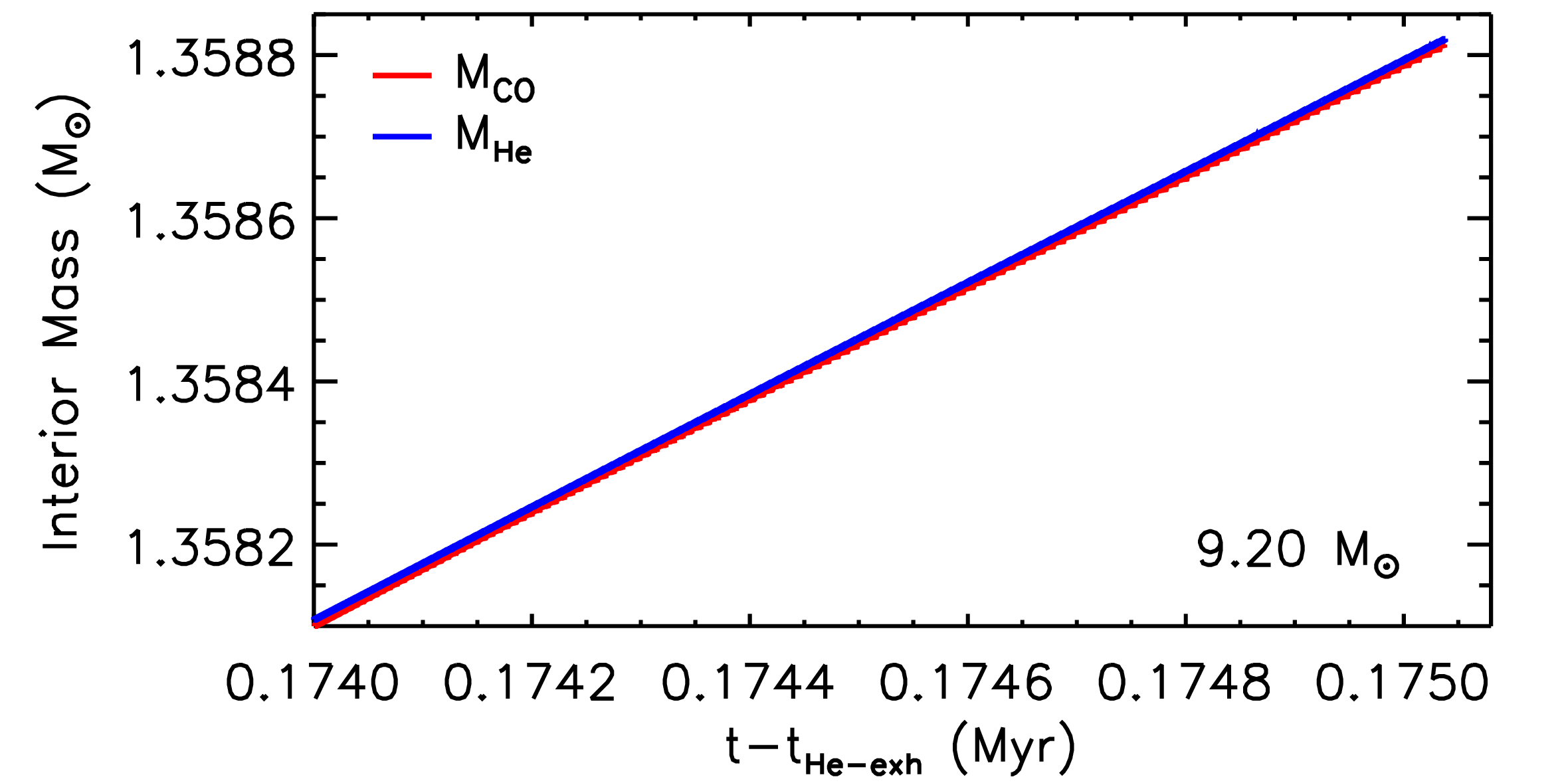}
\end{center}
\caption{Evolution of the He (blue line) and C (red line) core masses as a function of time during the AGB phase (for the $\rm 7.0~M_\odot$, upper left panel) and during the SAGB phase (for the $\rm 8.0,~8.5,~9.2~M_\odot$ models, upper right panel and lower left and lower right panels, respectively. The time has been reset at the core He exhaustion.\label{mcoresall}}
\end{figure*}

\begin{figure*}[ht!]
\begin{center}
\includegraphics[width=.45\linewidth]{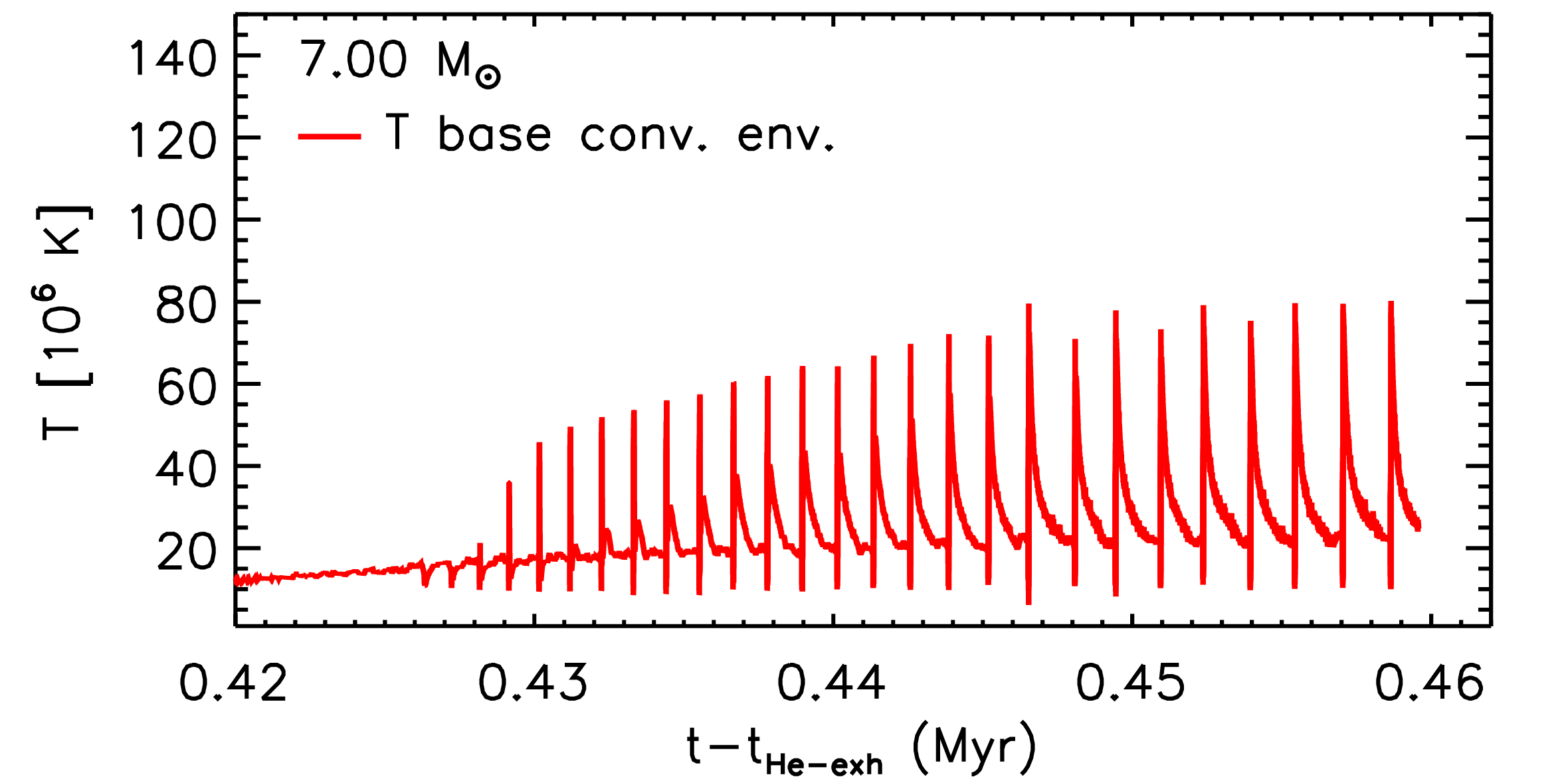}\quad\includegraphics[width=.45\linewidth]{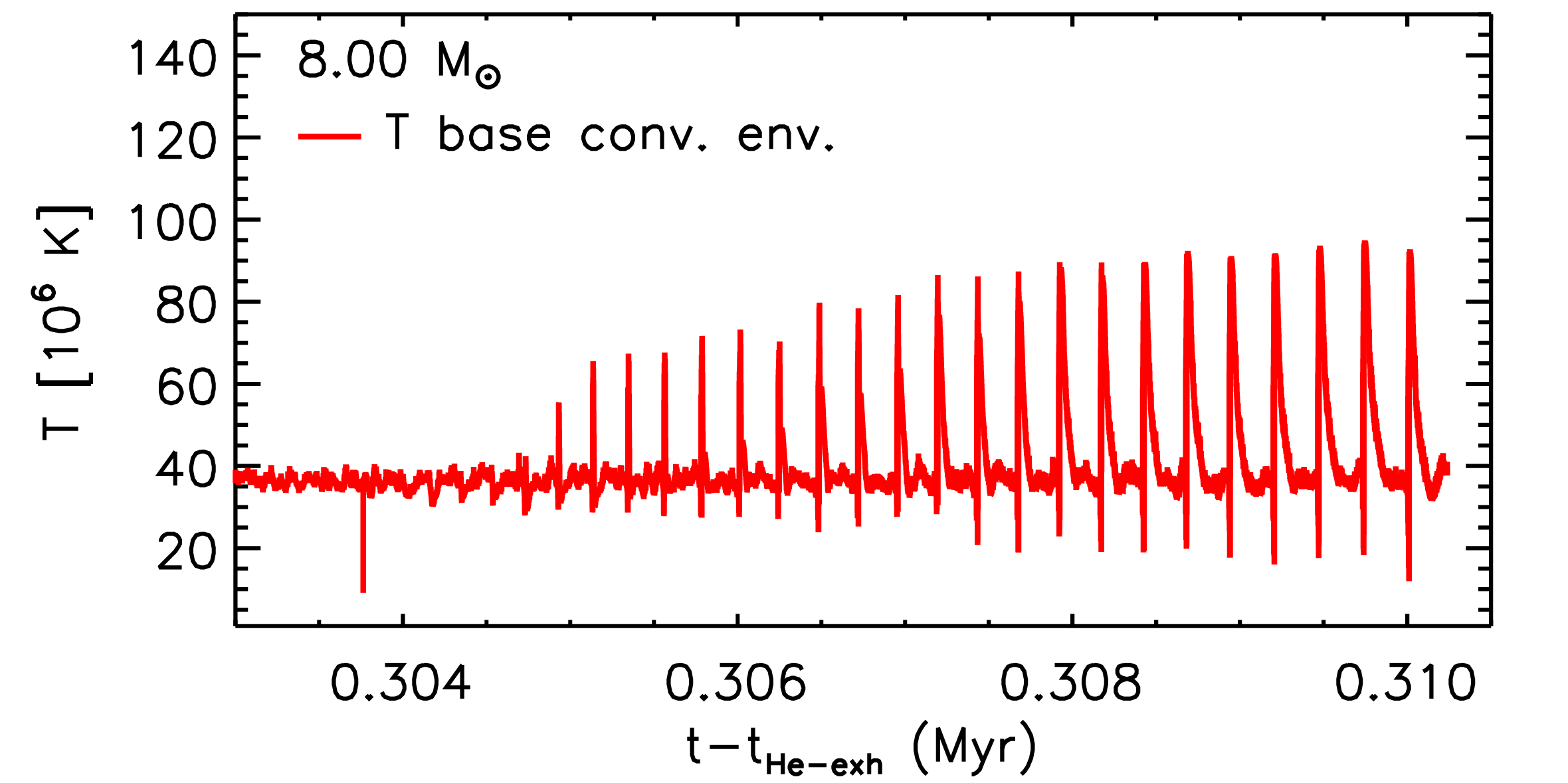}
\\[\baselineskip]% adds vertical line spacing
\includegraphics[width=.45\linewidth]{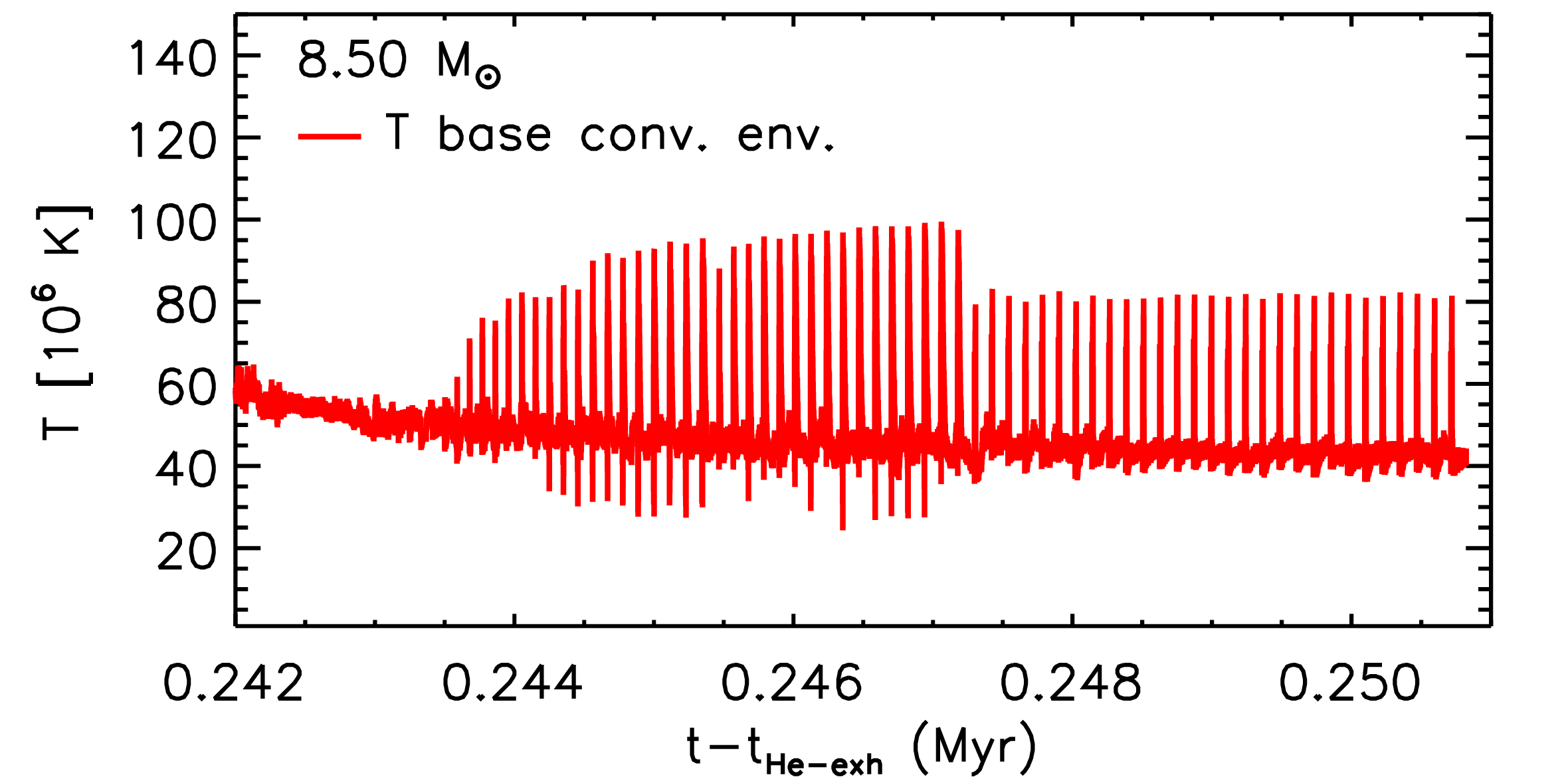}\quad\includegraphics[width=.45\linewidth]{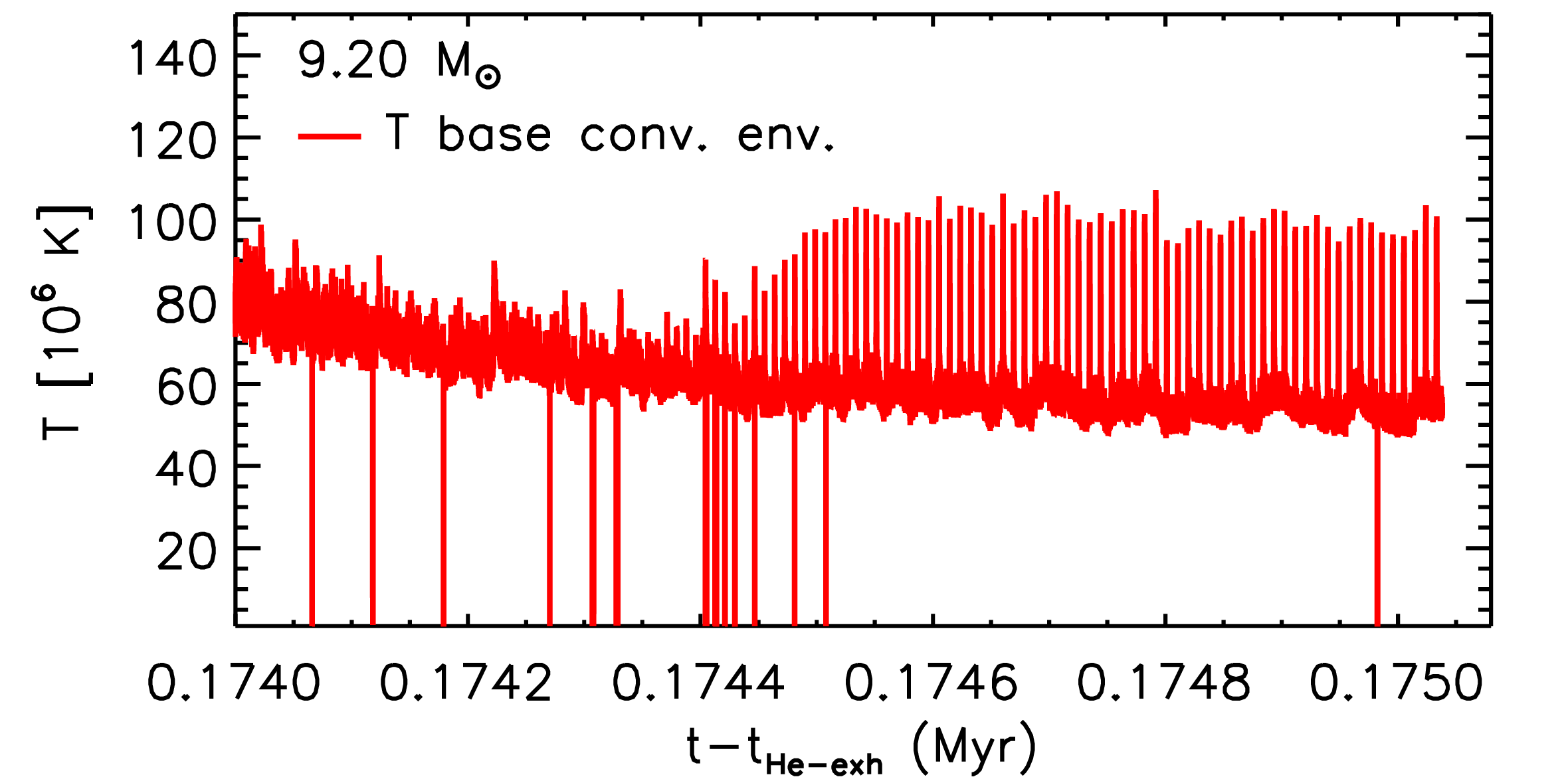}
\end{center}
\caption{Evolution of the temperature at the base of the convective envelope as a function of time during the AGB phase (for the $\rm 7.0~M_\odot$, upper left panel) and during the SAGB phase (for the $\rm 8.0,~8.5,~9.2~M_\odot$ models, upper right panel and lower left and lower right panels, respectively. The time has been reset at the core He exhaustion.\label{tceall}}
\end{figure*}

\begin{figure*}[ht!]
\begin{center}
\includegraphics[width=.45\linewidth]{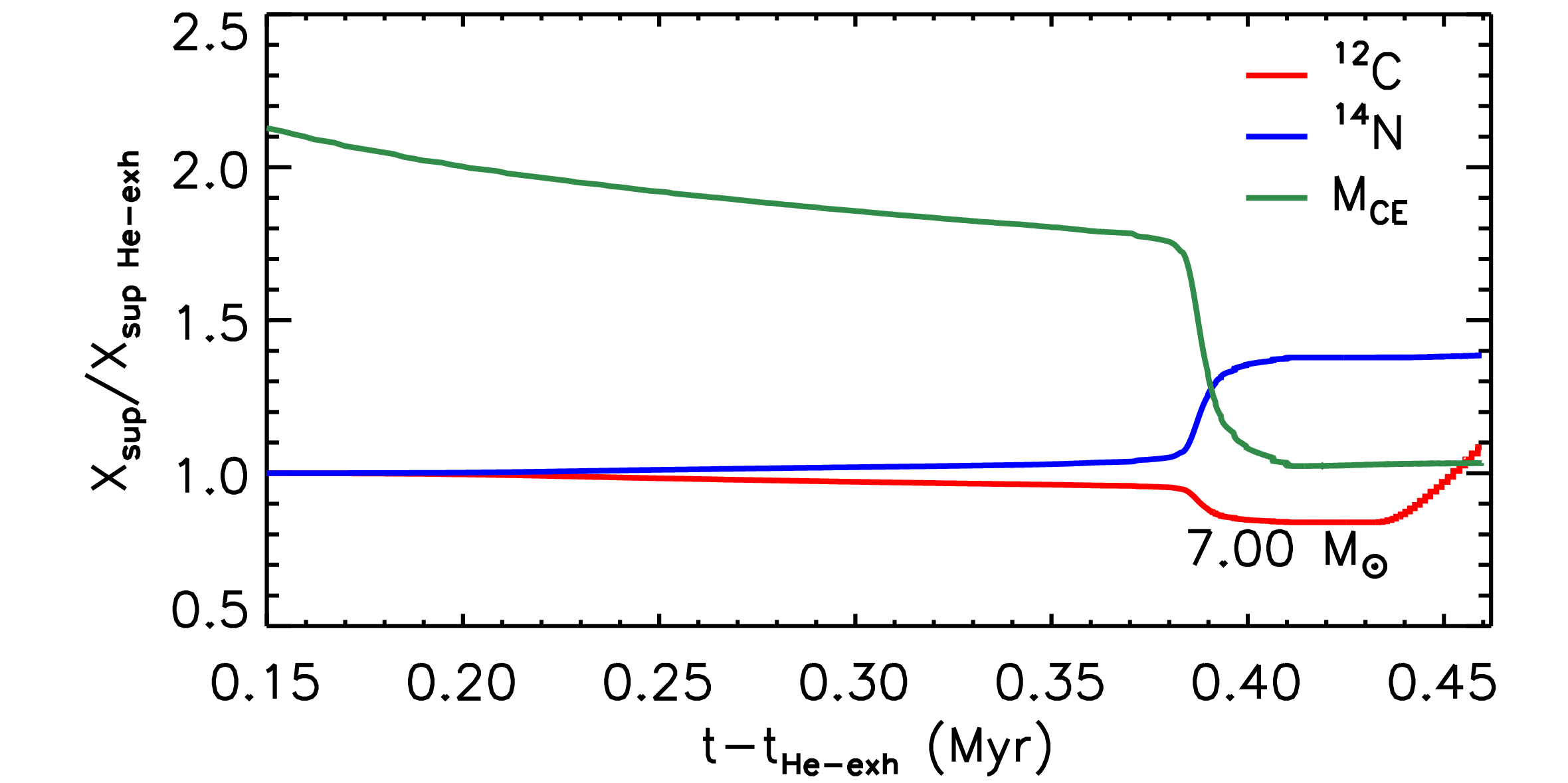}\quad\includegraphics[width=.45\linewidth]{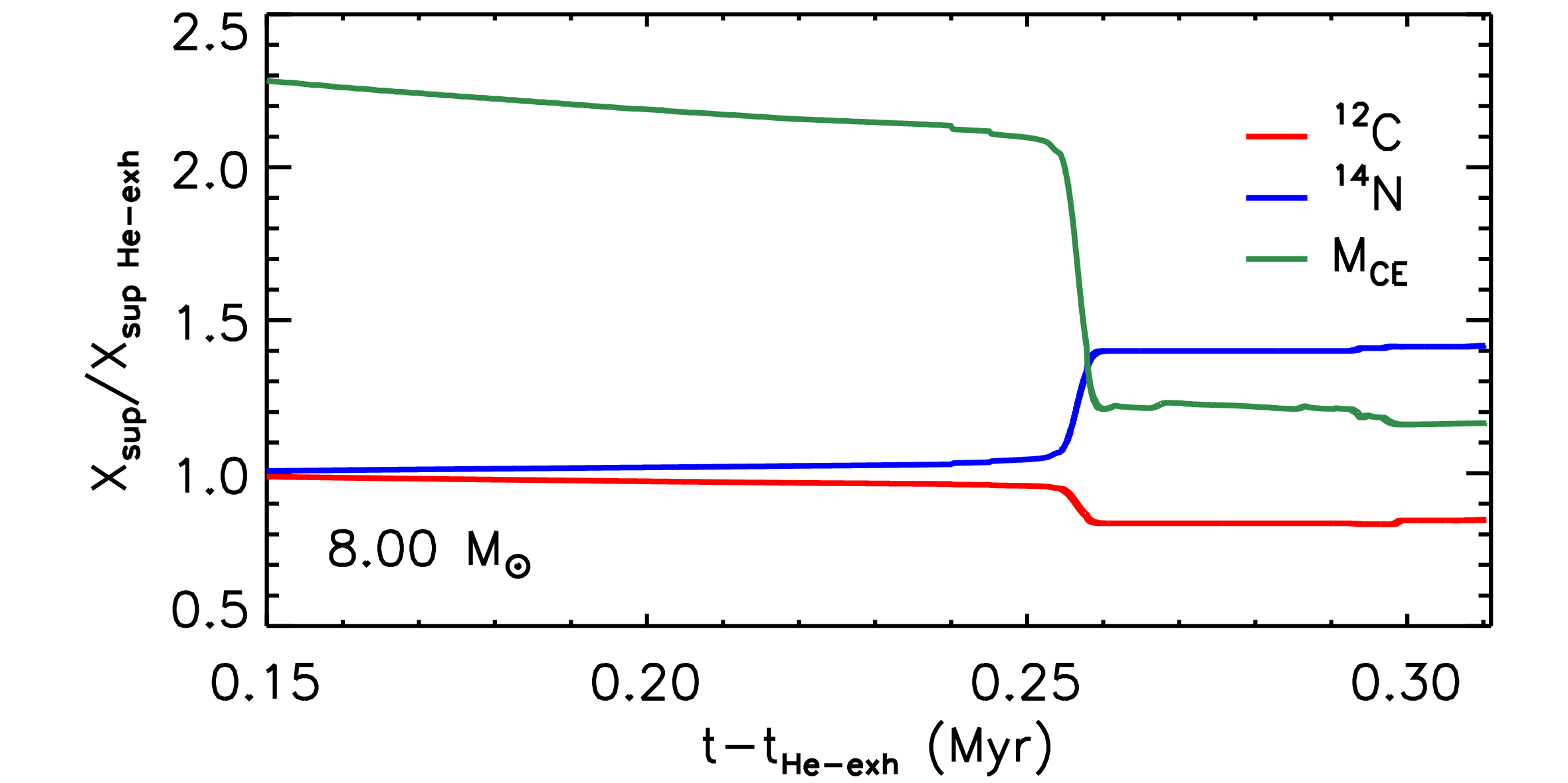}
\\[\baselineskip]% adds vertical line spacing
\includegraphics[width=.45\linewidth]{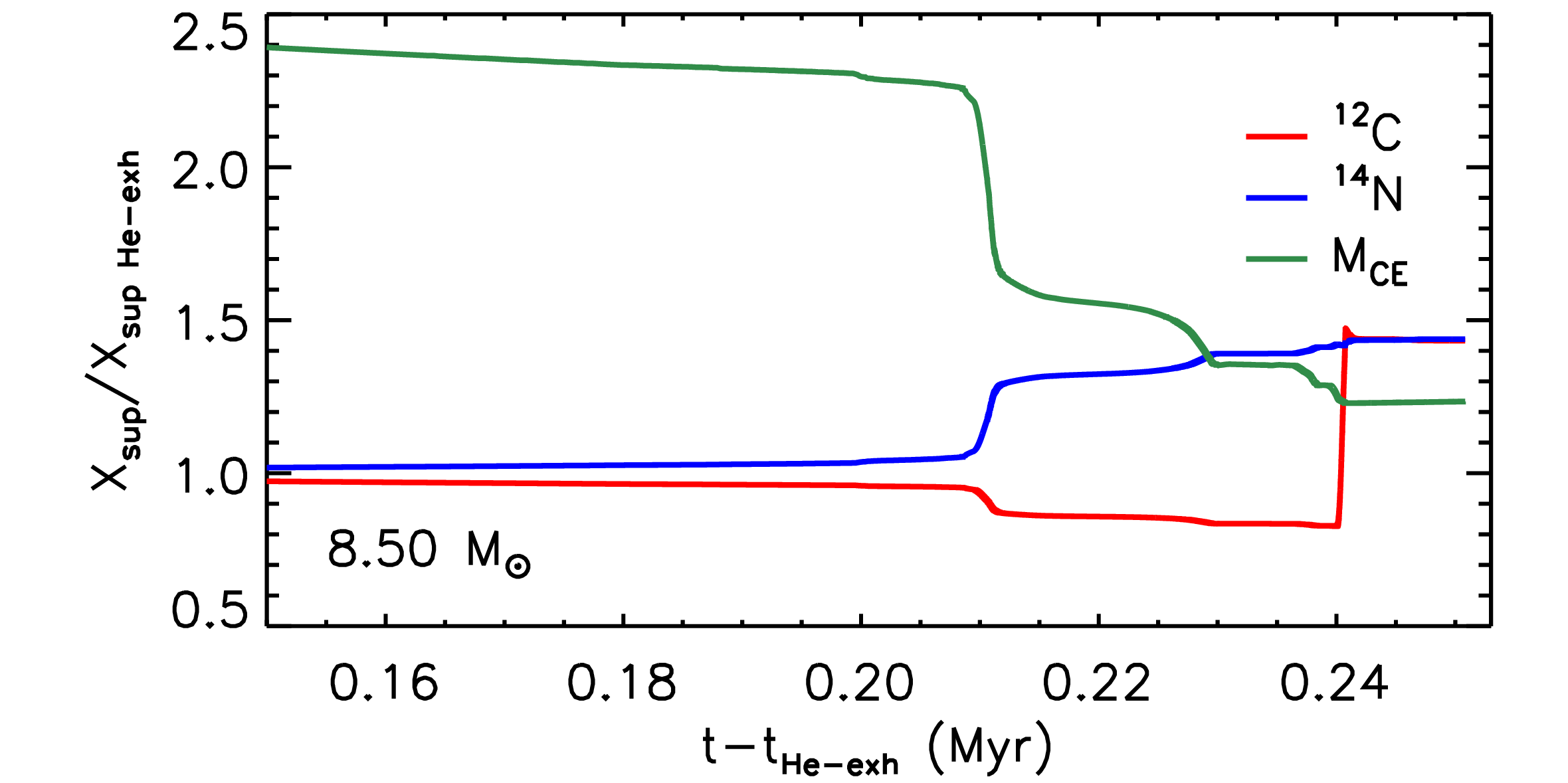}\quad\includegraphics[width=.45\linewidth]{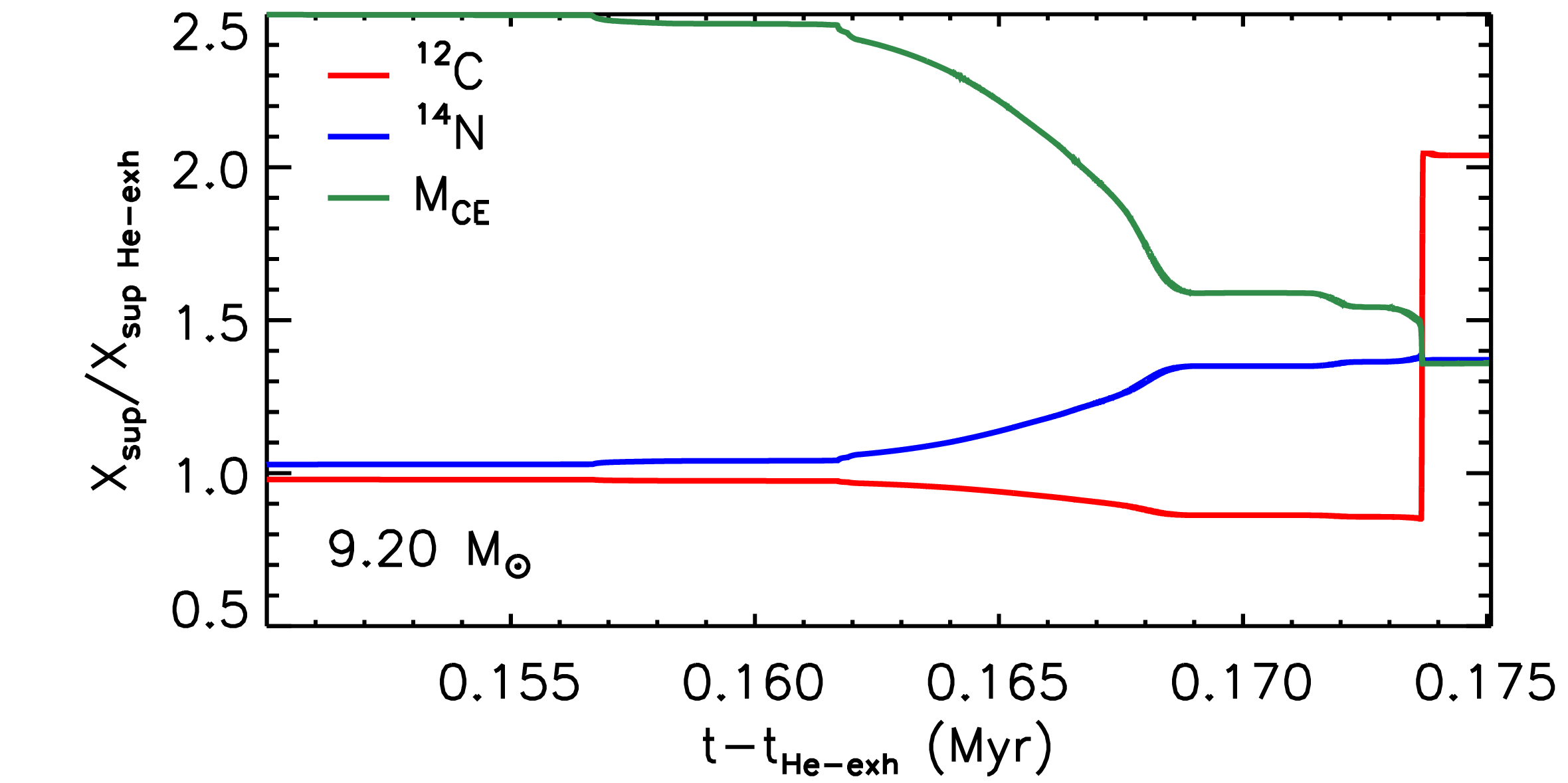}
\end{center}
\caption{Evolution of the surface $\rm ^{12}C$ (red line) and $\rm ^{14}N$ (red line) mass fraction and of the mass coordinate of the bottom of the convective envelope (green line) as a function of time during the AGB phase (for the $\rm 7.0~M_\odot$, upper left panel) and during the SAGB phase (for the $\rm 8.0,~8.5,~9.2~M_\odot$ models, upper right panel and lower left and lower right panels, respectively. The time has been reset at the core He exhaustion.\label{c12supall}}
\end{figure*}

\begin{figure*}[ht!]
\begin{center}
\includegraphics[width=.45\linewidth]{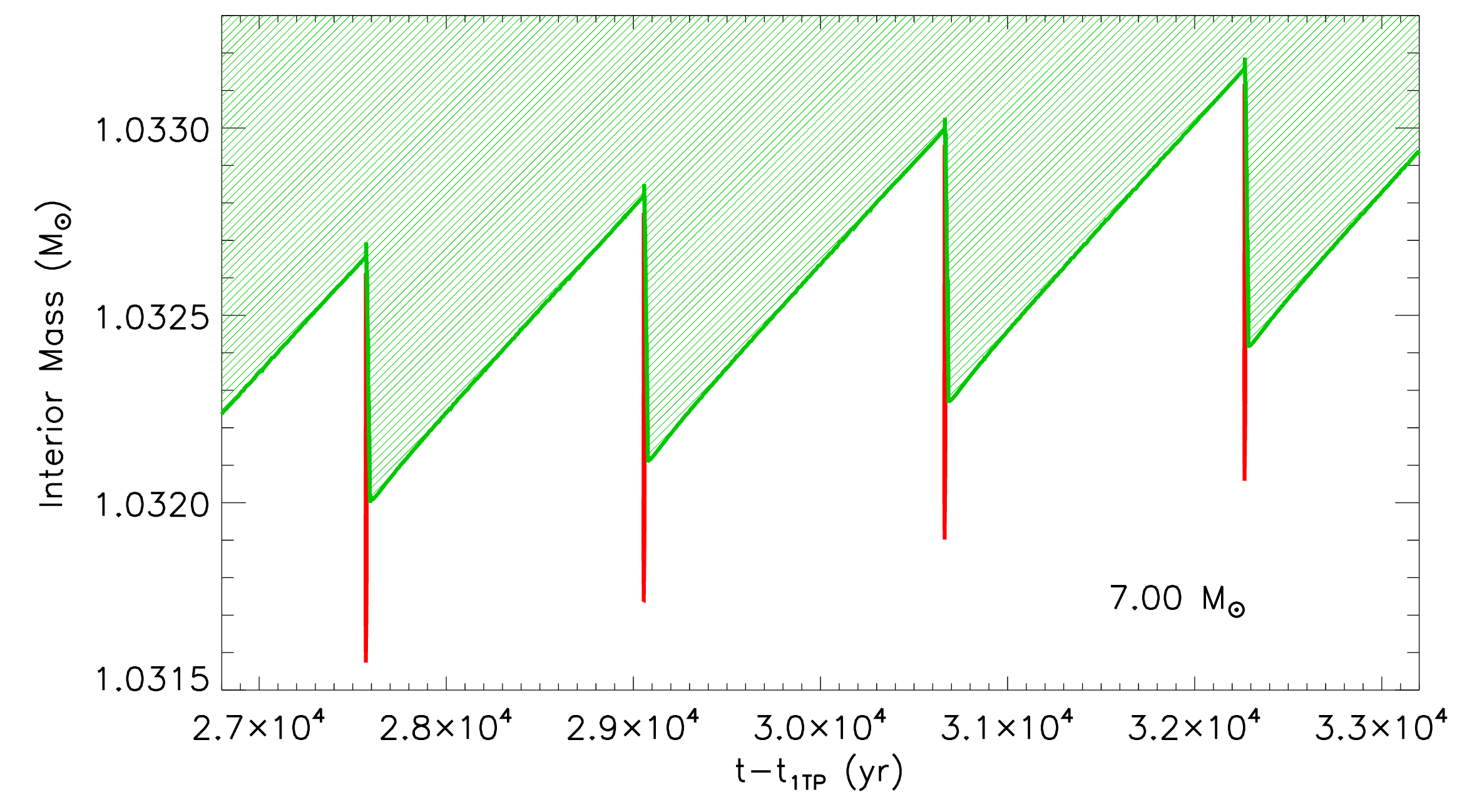}\quad\includegraphics[width=.45\linewidth]{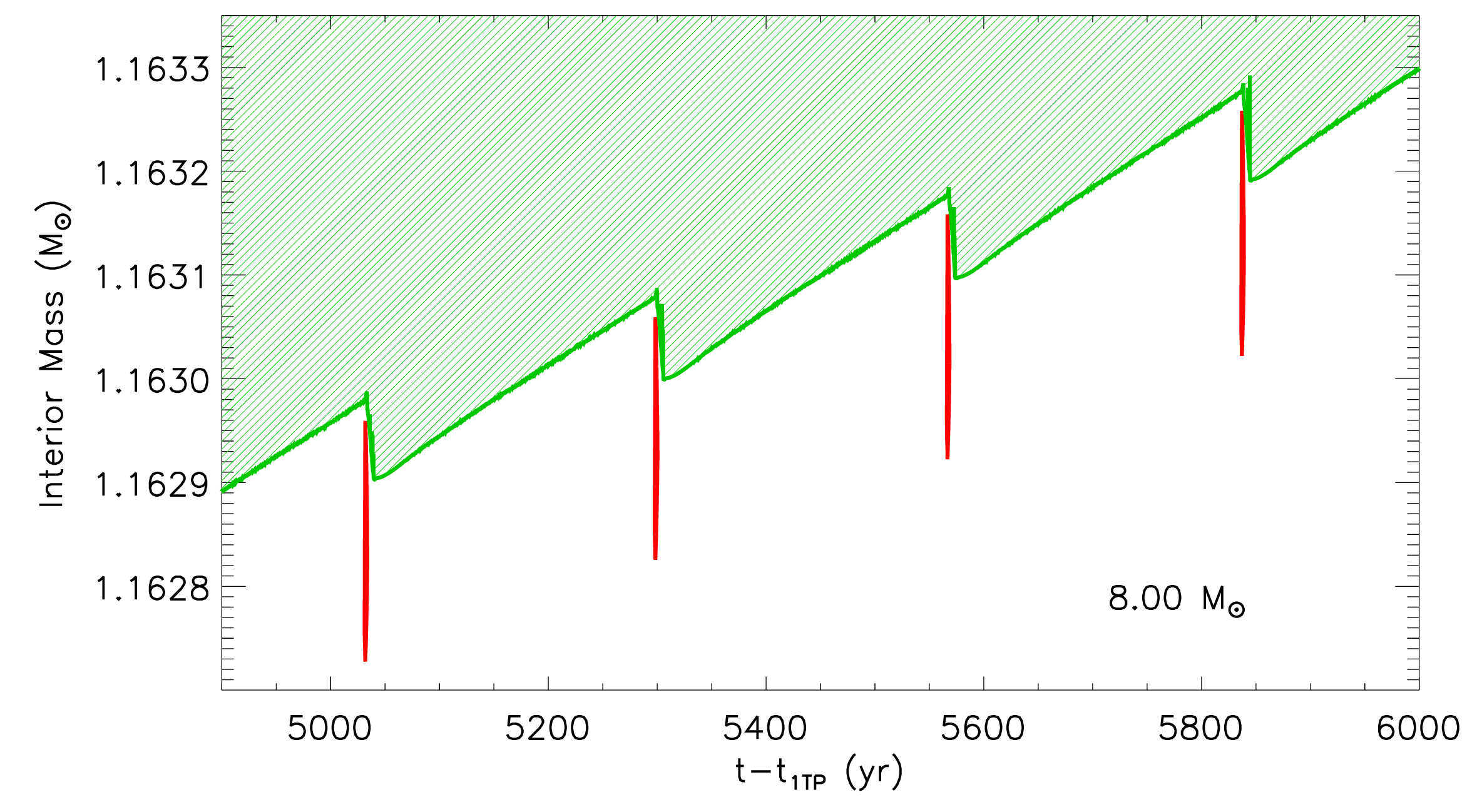}
\\[\baselineskip]% adds vertical line spacing
\includegraphics[width=.45\linewidth]{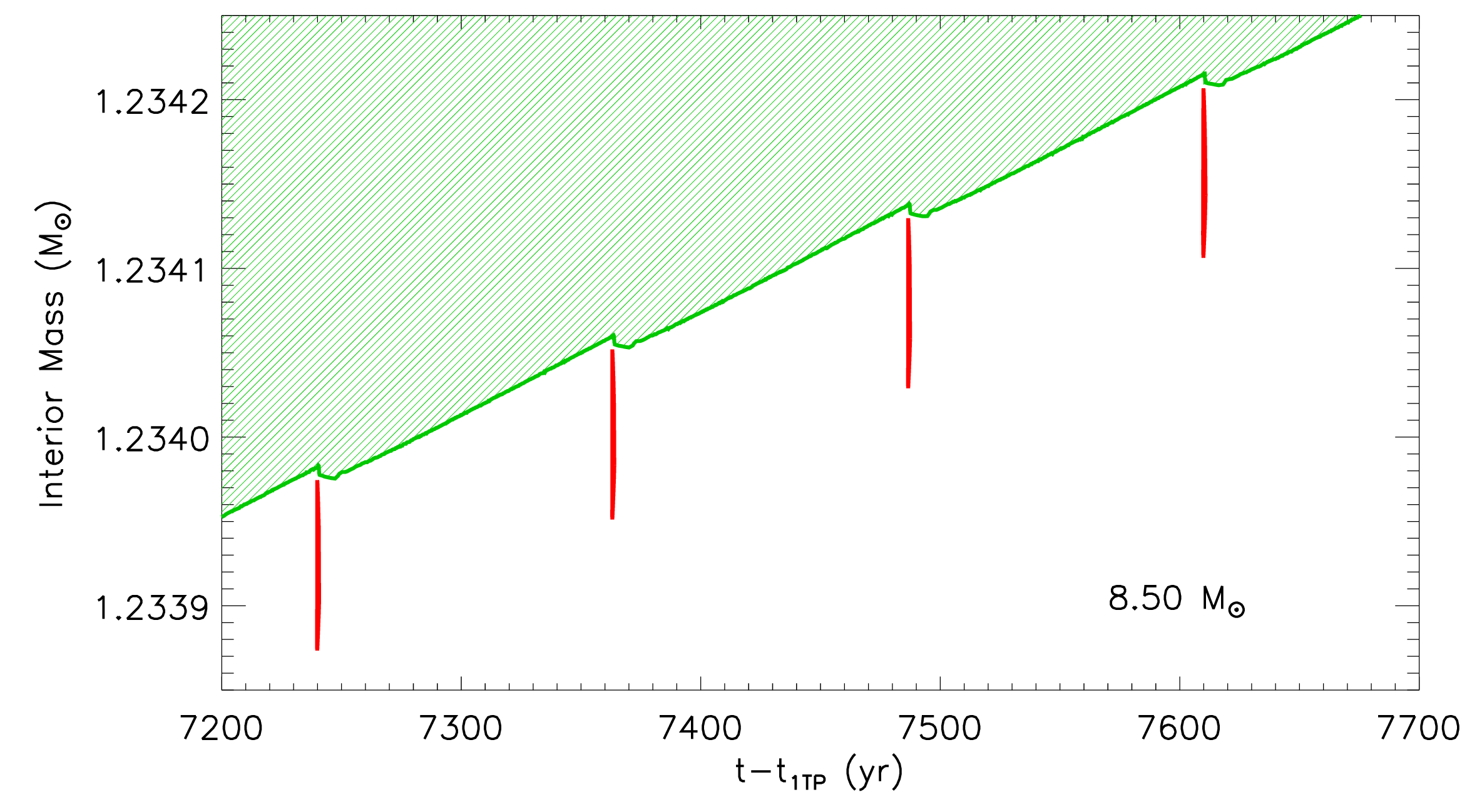}\quad\includegraphics[width=.45\linewidth]{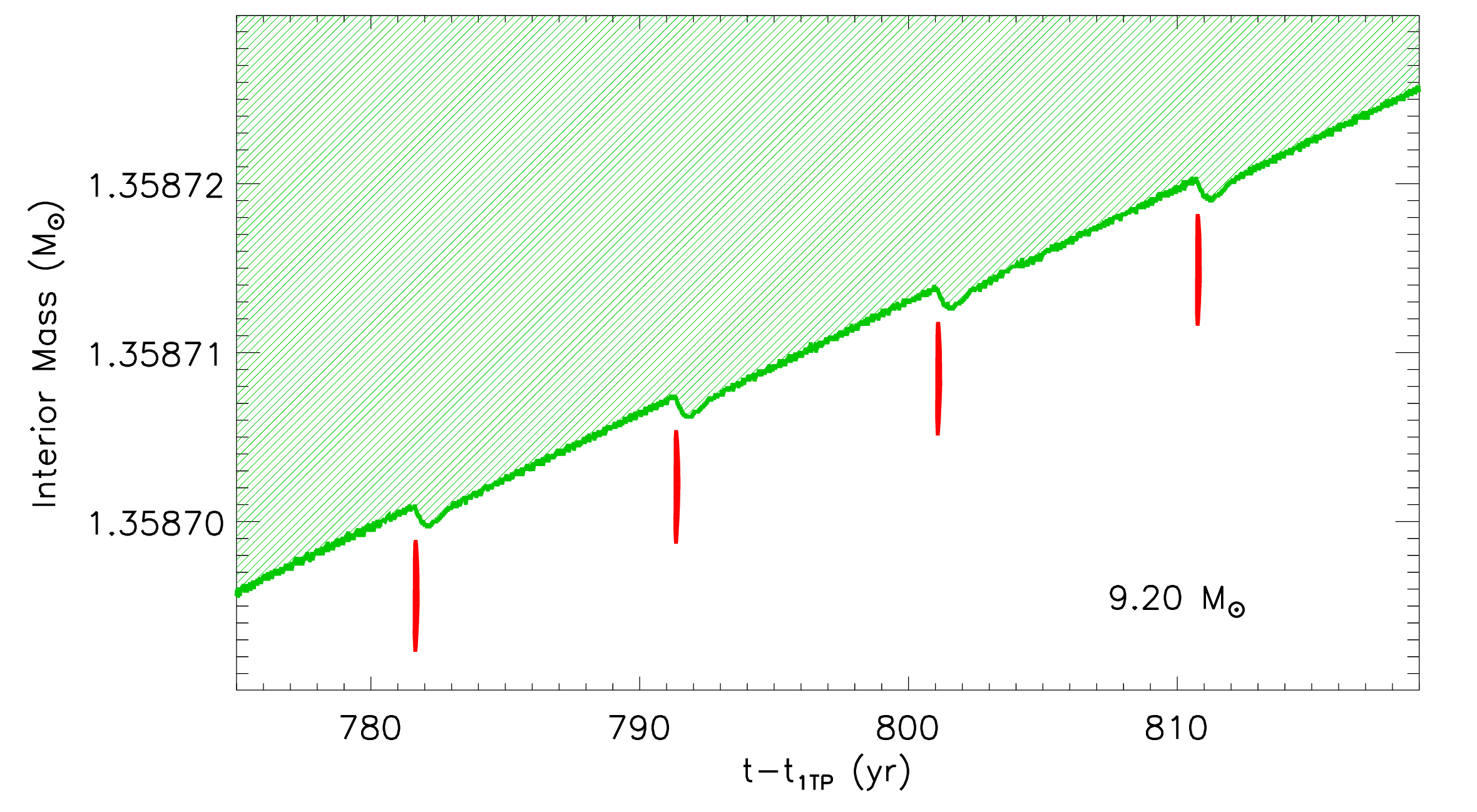}
\end{center}
\caption{Evolution of the convective envelope (green shaded area) and of the He convective shell (red line) as a function of time during the last 4 thermal pulses for the AGB $\rm 7.0~M_\odot$ model (upper left panel) and for the $\rm 8.0,~8.5,~9.2~M_\odot$ SAGB models, (upper right panel and lower left and lower right panels, respectively). The time has been reset at the core He exhaustion.\label{tpzoomalllin}}
\end{figure*}

The main properties of the thermal pulse phase are reported in Table \ref{tpprop700} and in the upper left panels of Figures  \ref{l3aall}, \ref{mcoresall}, \ref{tceall} and \ref{c12supall}. The values reported in the Table as well as the behavior of the various quantities shown in the Figures are consistent with what has been found in literature (see section \ref{sec:intro}). Let us note, only, that the maximum temperature at the base of the convective envelope increases progressively from $\rm \sim 40~MK$, at the beginning of the TP phase, to a pleteau value corresponding to $\rm \sim 80~MK$ after the first $\sim 16$ TPs (upper left panel of Figure \ref{tceall}). By the way, \cite{Nomoto1972} investigated in their Figures 2 and 3 how the temperature at the base of the convective envelope and the depth of mixing depends on the luminosity and the mass of the CO core. The third dredge-up, due to the penetration of the convective envelope into the He core during the quiescent shell He burning phase, i.e., when the H-burning shell is switched off, occurs after few TPs and induces from one side a progressive reduction of the rate at which the CO core increases (upper left panel of Figure \ref{mcoresall}) and at the same time a progressive enrichment of the surface carbon abundance (upper left panel of Figure \ref{c12supall}). Note, however, that such an enhancement is very mild, in fact the surface carbon abundance has increased, at the end of this phase, by a factor of $\rm \sim 1.1$ compared to the value at core He depletion. For this reason, we decided to not take into account carbon enhanced opacity tables. By the way, let us remind that, as already mentioned in section \ref{code}, we assume some amount of extra-mixing at the base of the convective envelope and therefore this is applied also during the third dredge up.
%note that some amount of extramixing is assumed at the inner border of the convective envelope also in this case, as it is already discussed in section \ref{code}. 
Figure \ref{tpzoomalllin} shows the behavior of the convective zones during the last thermal pulses before the stop of the calculation, where it can be appreciated the size of the He convective shell that forms after the He shell ignition and the efficiency of the 3rd dredge-up, in particular the quantity $\rm \lambda=\Delta M_{\rm dredge}/\Delta M_{H}\sim 0.78$, where $\rm \Delta M_{H}$ is the increase of the core mass during the interpulse phase and $\rm \Delta M_{\rm dredge}$ is the maximum penetration of the convective envelope following the pulse
\citep[see, e.g., Figure 5 in][]{doherty+17}

\subsection{Evolution during C burning. Stars with initial mass $\rm M\geq 7.5~M_\odot$ $\rm \left( M_{CO}\geq 0.76~M_\odot \right)$}\label{cburning}

In stars with the initial mass $\rm M\geq 7.5~M_\odot$ (CO core masses at core He depletion $\rm M_{CO}\geq 0.76~M_\odot$), the maximum temperature within the CO core reaches the threshold value for the C-burning ignition. Depending on the initial mass, C-ignition may occur in the center or off-center in a (partial) degenerate environment. 

\begin{figure*}[ht!]
\epsscale{0.8}
\plotone{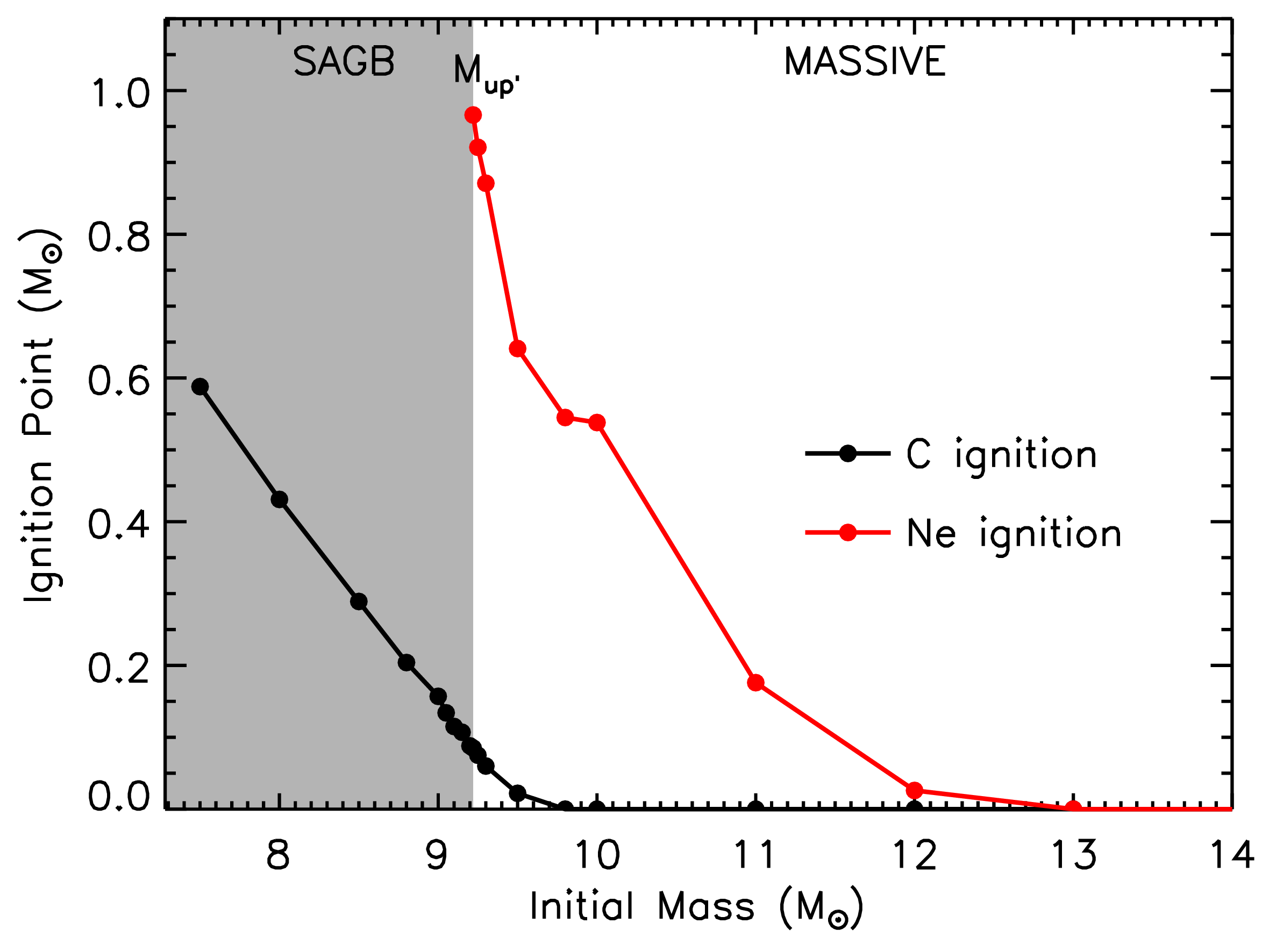}
\caption{Ignition mass coordinate of C- (black line and filled dots) and Ne-burning (red line and filled dots) as a function of the initial mass (see legend)\label{fig:mc_ign}}
\end{figure*}

\begin{figure*}[ht!]
\epsscale{0.8}
\plotone{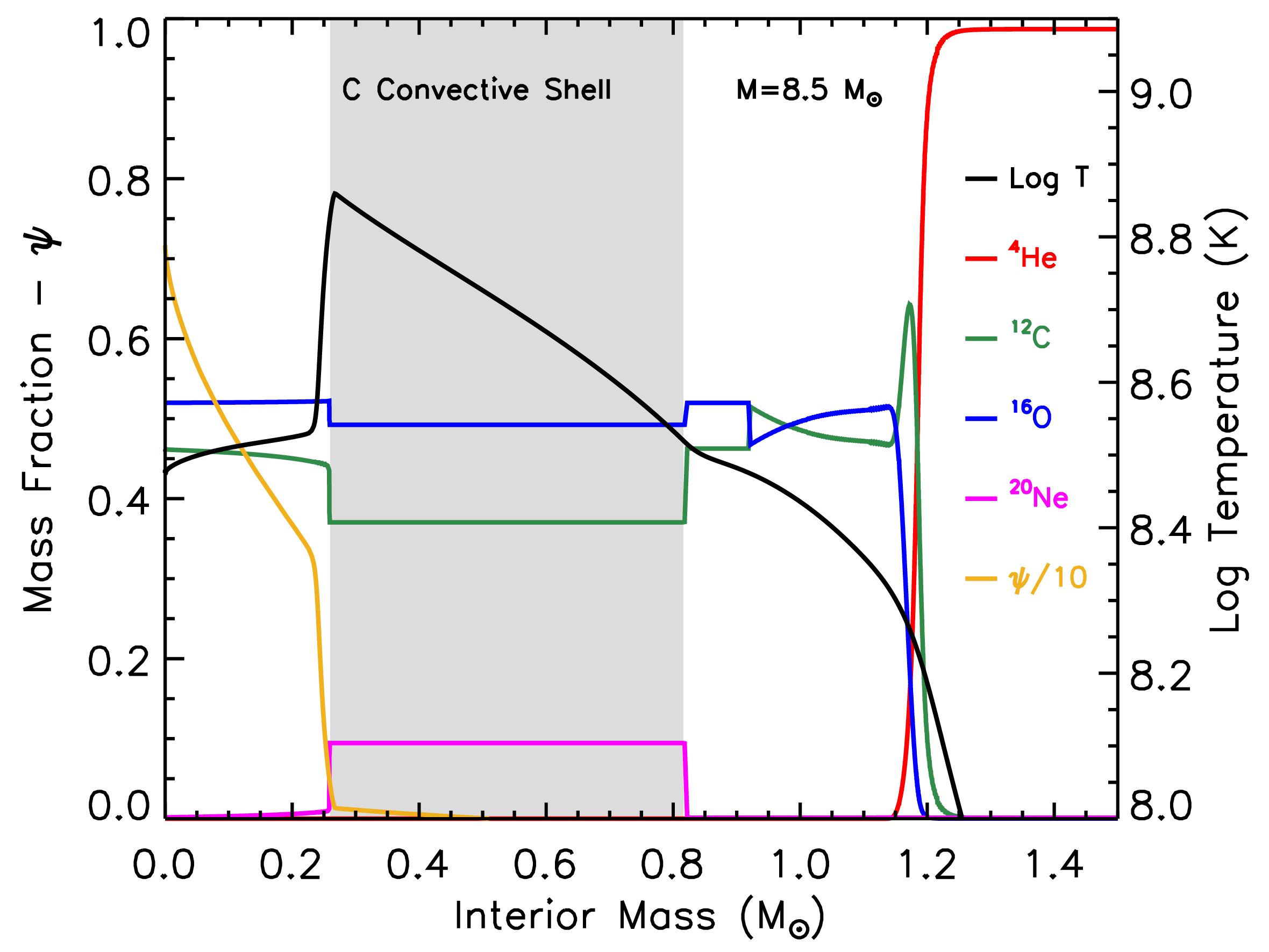}
\caption{Selected interior properties of a $\rm 8.5~M_\odot$ model during the off-center C-burning (see the legend). The chiemical composition and the degeneracy parameter are reported on the right y-axis, while the temperature is reported on the right y-axis. The degeneracy parameter is divided by 10 in order to improve the readability.\label{fig:offcburn-0010600}}
\end{figure*}

\begin{figure*}[ht!]
\epsscale{0.8}
\plotone{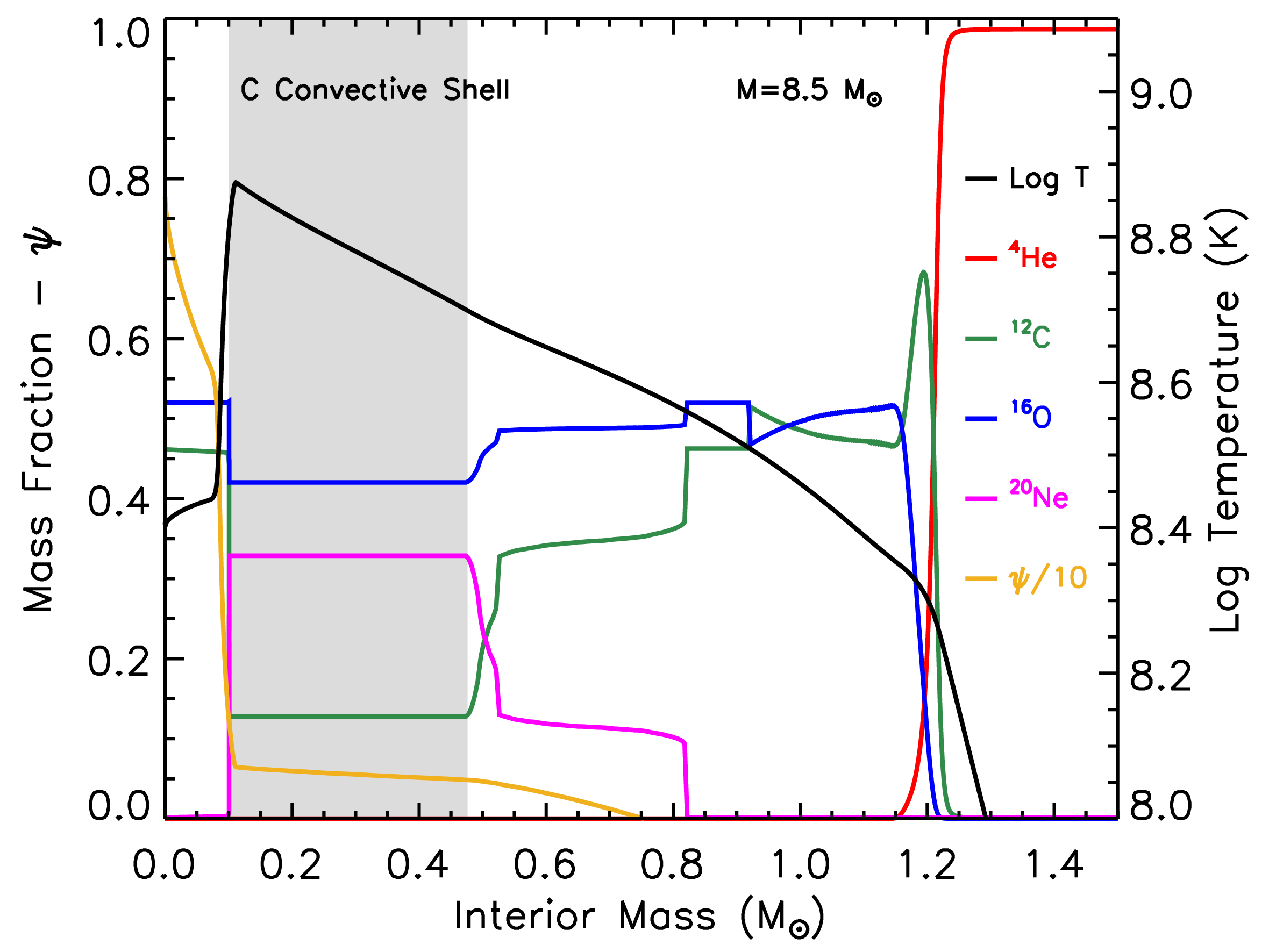}
\caption{Selected interior properties of a $\rm 8.5~M_\odot$ model during the off-center C-burning (see the legend). The chiemical composition and the degeneracy parameter are reported on the right y-axis, while the temperature is reported on the right y-axis. The degeneracy parameter is divided by 10 in order to improve the readability.\label{fig:offcburn-0028000}}
\end{figure*}

\begin{figure*}[ht!]
\epsscale{0.9}
\plotone{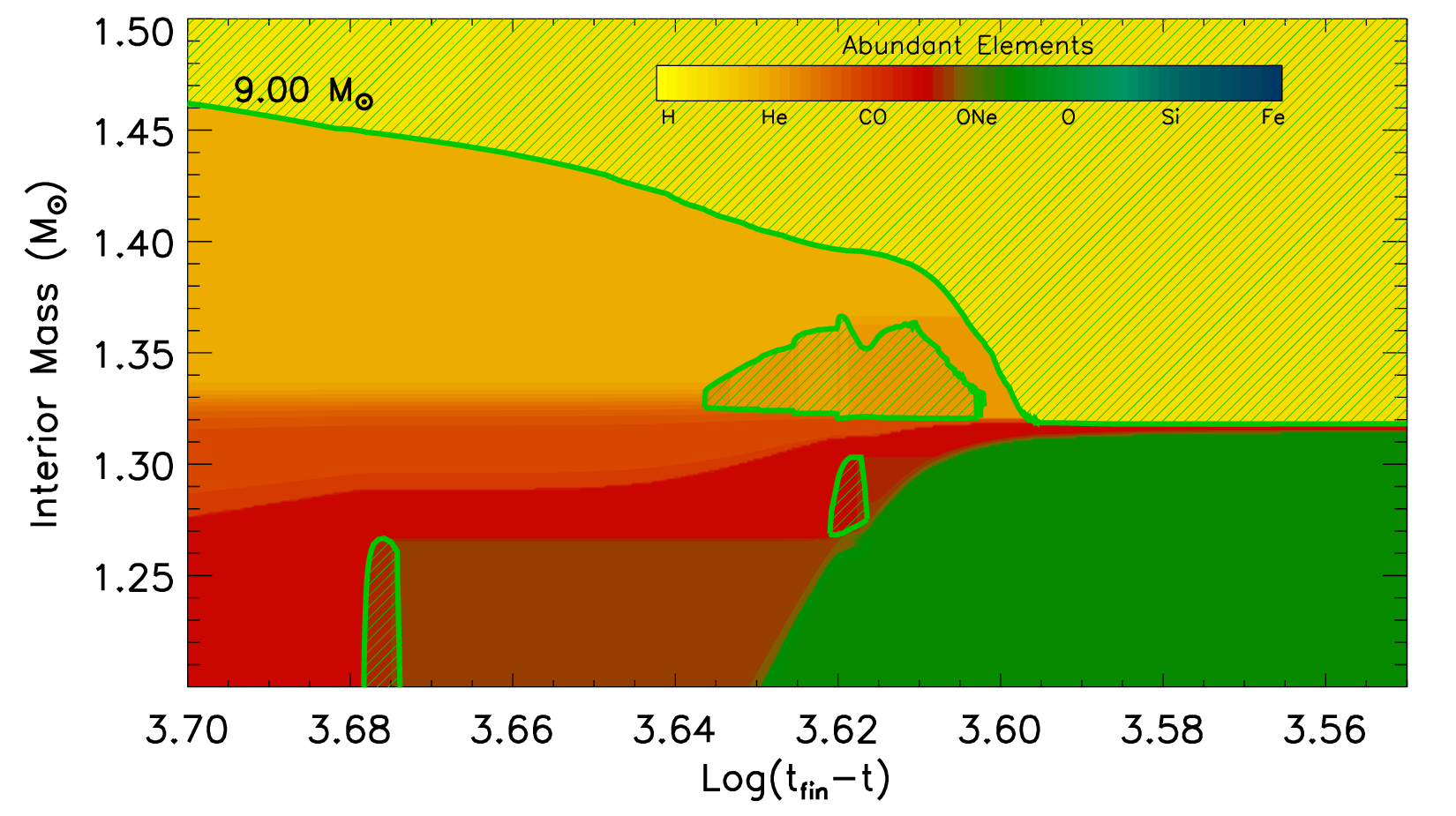}
\caption{Convective (green shaded areas) and chemical (color codes reported in the color bar) internal history of the $\rm 9~M_\odot$ model during the late phase of the second dredge-up. In the x-axis is reported the logarithm of the time till the end of the evolution ($\rm t_{\rm fin}-t$) in units of yr.\label{kip2dupTPzoom}}
\end{figure*}

%{\it Stars with initial mass $\rm 7.5\leq M/M_\odot\leq 9.5$}. 
In stars with the initial mass $\rm 7.5\leq M/M_\odot\leq 9.5$ ($\rm 0.76\leq M_{\rm CO}/M_\odot\leq 1.16$) the interplay between processes (2) and (3) (see above) induces the formation of an off-center peak temperature that progressively increases locally and moves outward in mass while the CO core becomes progressively more degenerate. Therefore, for these stars the C-ignition occurs off-center and, in general, as the initial mass increases the mass coordinate corresponding to the ignition point becomes closer to the center, ranging from $\rm 0.59~M_\odot$ to $\rm 0.02~M_\odot$ for the $\rm 7.5~M_\odot$ and $\rm 9.5~M_\odot$ models, respectively (black line in Figure \ref{fig:mc_ign}). The off-center ignition is generally accompanied by the formation of a convective zone driven by the high flux produced by the C-burning reactions in a  (partial) degenerate environment ($\rm \psi_{ign.} \sim 2.5$, see Table \ref{tab_main_prop}). When the energy flux produced by the nuclear burning reduces, the convective zone vanishes. The disappearance of this convective shell is then followed by a number of other convective zones associated to the C burning front that progressively moves toward the center and locally removes the degeneracy. In general, the number of convective zones that follow the first one, decreases as the initial mass of the star increases (Figure \ref{kiptot}). 
%The speed at which the C burning front moves toward the center is driven by the transfer of heat from the burning zone at high temperature to the more internal zones at much lower temperature, that in turn is controlled by the CBF treatment (see section \ref{code}). 
Typical internal structures corresponding to two different stages during this phase are shown in Figures \ref{fig:offcburn-0010600} and \ref{fig:offcburn-0028000} for the $\rm 8.5~M_\odot$ model. Note that $\rm ^{12}C$ is not completely exhausted in the zones where the C burning front has passed. The C burning front, marked by the maximum temperature, reaches the center in stars with the initial mass $\rm M\geq 9.2~M_\odot$ and then it begins moving outward in mass where $\rm ^{12}C$ is still abundant, inducing again the formation of a number of successive convective shells (Figure \ref{kiptot}). In the lower mass models, on the contrary, the maximum temperature never reaches the center. However, the following shell C burning phase develops in these stars similarly to the more massive ones.
%At variance with one could expect, before the C burning front reaches the center, it begins to move outward in mass, where $\rm ^{12}C$ is still abundant, inducing again the formation of a number of successive convective shells (Figure \ref{kiptot}). 
In all these models, after the last C convective episode, C burning proceeds in a radiative shell, reducing progressively the CO rich zone confined between the ONeMg core, resulting from the shell C burning, and the He rich zone (the red zone in Figure \ref{kip2dupTPzoom}). It is worth noting that in all these models some $\rm ^{12}C$ remains unburnt in the central zone. The mass fraction of the unburnt $\rm ^{12}C$, as well as the mass of the core where this quantity is larger than 0.01
%The mass of the core with $\rm ^{12}C$ left unburnt 
decreases with increasing the initial mass. In particular, Figure \ref{fig:c12} (the blue line) shows that the mass of the central zone where the $\rm ^{12}C$ mass fraction is larger than 0.01 decreases from $\rm \sim 0.33~M_\odot$ in the $\rm 7.5~M_\odot$ model to $\rm \sim 0.01~M_\odot$ in the $\rm 8.8~M_\odot$ and disappears for larger initial masses. 
%This peculiar configuration of the core, made of a CO-rich central zone surrounded by a ONe rich zone that lies below another CO rich zone, will not change till the end of the evolution of the star. 
Figure \ref{kiptot} shows how the configuration of the ONeMg (green zone)/CO (red zone) cores change as a function of the initial mass.

%{\it Stars with initial mass $\rm M\geq 9.8~M_\odot$}. 
In stars with the initial mass $\rm M\geq 9.8~M_\odot$ (CO core masses at core He depletion $\rm M_{CO}\geq 1.2~M_\odot$), C burning is ignited in the center and develops in a convective core. Once the $\rm ^{12}C$ is depleted in the center, C burning shifts in shell and, as in the case of the lower mass models, induces the formation of a number of convective shells (Figure \ref{kiptot}). These stars behave like the classical massive stars \citep{cl20,lc18}.

\subsection{Evolution of the convective envelope. Stars with initial mass $\rm M\leq 10.0~M_\odot$ $\rm \left( M_{CO}\leq 1.20~M_\odot \right)$}
As already mentioned in section \ref{hedepletion}, after core He depletion the H burning shell is switched off by the advancing He burning shell and this may induce the convective envelope to penetrate into the He rich layer (second dredge up). %The occurrence as well as the development of this phenomenon depends on the initial mass. 
This phenomenon occurs only in stars with the initial mass $\rm M\leq 10~M_\odot$ although in different evolutionary stages \citep[see also section 4.2 and Fig. 4.3 of][for a discussion on the mechanism of the convective penetration]{SugiNom1980}.
More specifically, the second dredge up 
%(1) begins after core He depletion and goes to completion before the onset of the thermal pulses in stars with initial mass $\rm M\leq 7.0~M_\odot$ (the second dredge-up for these stars has been already discussed in section \ref{sec:agb} and will not be repeated here); 
(1) begins before carbon ignition and goes to completion after carbon burning in stars with the initial mass $\rm 7.5\leq M/M_\odot\leq 8.0$ (compare the black line with the cyan and the blue line in Figure \ref{fig:mcore_he}); (2) begins after carbon ignition and goes to completion after carbon burning in stars with the initial mass $\rm 8.5\leq M/M_\odot\leq 10.0$ (compare the black line with the cyan, blue and magenta lines in Figure \ref{fig:mcore_he}). The second dredge-up for stars with the initial mass $\rm M\leq 7.0~M_\odot$ has been already discussed in section \ref{sec:agb} and will not be repeated here.

In general, during the second dredge up the convective envelope penetrates into the He core and therefore brings to the surface material that has been processed by H burning through the CNO cycle. This means that during this phase the surface $\rm ^{12}C$ mass fraction decreases, while the surface $\rm ^{14}N$ mass fraction increases (Figure \ref{c12supall}).

In stars with the initial mass $\rm 8.50\leq M/M_\odot\leq 9.20$ the C-burning shell is efficient enough to progressively swith off the He burning shell. As a consequence, in these stars, the convective envelope penetrates deep enough to dredge up the products of the He burning. Once this deep penetration occurs the surface abundance of $\rm ^{12}C$ suddenly increases while the one of $\rm ^{14}N$ progressively decreases (see Figure \ref{c12supall}). In general, the larger the mass the larger the increase of the surface $\rm ^{12}C$ abundance.

It is worth noting that 
%The behavior of this phenomenon, however, is different depending on the initial mass. While in the $\rm 8.5~M_\odot$ ($\rm M_{CO}=0.92~M_\odot$) model the convective envelope penetrates into a zone that was partially processed by a radiative He burning shell, 
in stars with $\rm 8.8\leq M/M_\odot\leq 9.20$ ($\rm 0.98\leq M_{\rm CO}/M_\odot\leq 1.08$) 
a He convective shell forms during the second dredge-up. Such an occurrence slows down slightly the penetration of the convective envelope into the He core. Once the He convective shell disappears, the penetration resumes (Figure \ref{kip2dupTPzoom}) and the products of the He burning, mainly (primary) $\rm ^{12}C$ (Figure \ref{c12supall}), are mixed up to the surface. Note that, in this case, the increase of the surface $\rm ^{12}C$ abundance at the end of the second dredge-up is mainly due to the penetration of the convective envelope into the extinguished He convective shell and to a lesser extent to the penetration into the CO core. The dredge-up of the He burning products has been named "corrosive second dredge-up" by \cite{gilpons+13} and \cite{doherty+14} but, rather, we refer to it as an early third dredge-up (E3DU) because this phenomenon is identical to the one occurring during the inter-pulse phase of TP-AGB stars (see above), although in this case it occurs well before the beginning of the onset of the thermal pulses (see below). It is worth noting that in none of these models the He convective shell and the convective envelope interact with each other, i.e., we do not find the so called "dredge-out" episode, phenomenon found for the first time by \cite{ritossa+99} and later also in other more recent studies \citep[see, e.g.,][]{siess07,gilpons+13,doherty+15,jones+16}. This is probably due to the fact that while in the previous mentioned studies a convective overshooting is considered for any convective zone (cores and shells), in these calculations we assume some amount of extramixing only at the edge of the H convective core, at the base of the convective envelope and at the base of the convective shells during the off-center C- and Ne-ignition (see section \ref{code}).
As a final comment we point out that in all the models mentioned above the increase of the surface carbon abundance due to the E3DU is in any case very mild. In the $\rm 9.2~M_\odot$ model, where the E3DU
produces the largest effects, the enhancement of the surface carbon abundance is a factor of $\sim 2$ compared to the value at core He depletion. Since during the first dredge up the surface carbon abundance decreases by a factor of $\sim 2$ compared to the initial one, the result is that the surface carbon mass fraction after the E3DU is roughly equal to the initial one. For this reason, also in this case we decided to not take into account carbon enhanced opacity tables.

In all the stars with the initial mass $\rm 7.5\leq M/M_\odot\leq 9.2$, during the second dredge-up, the envelope expands and cools down in order to reabsorb the energy produced by the more internal nuclear burning shells. However, the rate at which the convective envelope penetrates in mass is higher than the rate at which it cools down, therefore the temperature at the base of the convective envelope increases progressively until the H burning shell is re-ignited. After the convective envelope reaches its maximum depth, the H burning shell begins to advance in mass. In this phase the He burning shell is progressively reignited in those stars where it was switched off. From this stage onward, for all the stars the evolution is characterized by a double shell burning, where the two H- and He-burning shells advance in mass with a similar rate. In fact, in this phase the H-burning luminosity is a factor of $\sim 7$ higher than that of the He-burning (see the leftmost values in Figures \ref{l3aall}), that corresponds roughly to the ratio between the energy provided by the CNO cycle and the energy provided by the $3\alpha$ reactions. This means that the two shells burn the same amount matter per second. This stage ends with the onset of the thermal pulses \citep[see next section and also][for discussion on the evolution of this phase]{SugiNom1975}.

In stars with the initial mass $\rm 9.22\leq M/M_\odot\leq 10.00$ neon burning is ignited off-center during the second dredge-up and before the bottom of the convective envelope enters into the region enriched by the He burning products \citep[see also][]{Nomoto84}. Once neon is ignited the evolution of the star becomes fast enough that the zones above the He cores remain essentially freezed. %As an example, the evolution of the $\rm 9.20~M_\odot$ after the onset of the thermal pulses lasts at least $\rm \sim 10^{3}~yr$, while the evolution of the $\rm 9.22~M_\odot$ after the off-center Ne ignition lasts $\rm \sim 6~yr$.
For this reason in these stars the products of He burning are never brought to the surface and therefore no increase of the surface $\rm ^{12}C$ is found.

\subsection{Evolution during the TP-SAGB phase: stars with mass $\rm 7.5\leq M/M_\odot\leq 9.20$}

After the H-burning shell has been re-ignited, stars in the mass range $\rm 7.5\leq M/M_\odot\leq 9.20$ enter a classical thermally pulsing phase where the two H- and He-shells alternatively activate above a degenerate ONeMg core which is surrounded by a thin zone enriched in CO, the latter left by the He burning shell. The general properties of these stars during this phase, named TP-SAGB, has been reviewed and described in detail in literature (see section \ref{sec:intro}), therefore we will focus here mainly on how these properties change as a function of the initial mass. The main evolutionary properties during the TP-SAGB phase of stars in this mass interval ($\rm 7.5\leq M/M_\odot\leq 9.20$) are reported in Tables \ref{tpprop750}, \ref{tpprop800}, \ref{tpprop850}, \ref{tpprop880}, \ref{tpprop900}, \ref{tpprop905}, \ref{tpprop910}, \ref{tpprop915}, \ref{tpprop920}.

In general each thermal pulse is characterized by the following phenomena: (1) a strong activation of the He burning reactions followed by the formation of a He convective zone and by a peak in the He luminosity ($\rm L_{He}$); (2) the disappearance of the He convective zone and the steady He-shell burning phase that accretes the CO core; (3) the switch off of the H-burning shell; (4) the penetration of the convective envelope that may erode in some cases the He core (third dredge up, 3DU); (5) the reactivation of the H shell and the switch off of the He burning shell; (6) the steady H-shell burning phase where the He core increases and the convective envelope recedes in mass (interpulse phase), until the next pulse is ignited. A schematic view of this phase can be found, e.g., in Doherty et al. 2017 (Figure 5).

Figure \ref{l3aall} shows the luminosity of the H- and He-burning shells as a function of time for the AGB and selected SAGB models. Moving from AGB stars ($\rm M=7.0~M_\odot$) to SAGB stars ($\rm 7.5\leq M/M_\odot\leq 9.20$) the maximum luminosity of the He burning shell reached during each thermal pulse decreases while the frequency of the TPs increases. This is due to the fact that the core mass becomes more massive and hotter as the initial mass of the star increases (see Doherty et al. 2017 and references therein). The increase of both the $\rm ^{4}He$ and $\rm ^{12}C$ abundance after the second dredge-up contribute to increase the frequency of the thermal pulses in stars with the initial mass $\rm M\geq 8.5~M_\odot$ because it makes the H-shell more efficient.

Figure \ref{tpzoomalllin} shows a zoom of selected models during the last few computed thermal pulses. Moving from the $\rm 7~M_\odot$ to the $\rm 9.2~M_\odot$ the following things are worth to be noted: (1) the reduction of the size of the He convective shell from $\rm \sim 10^{-4}~M_\odot$ to $\rm \sim 10^{-5}~M_\odot$; (2) the strong reduction of the interpulse time from $\rm \sim 10^{3}~yr$ to $\rm \sim 10~yr$; (3) the progressive reduction of the 3DU that disappears in stars with mass $\rm M\geq 9.0~M_\odot$. 
%A zoom of the last thermal pulse is shown in Figures \ref{tpzoomalllin2}, \ref{tpzoom915lin2} (lower right panel) and \ref{tpzoom920lin2} (lower right panel), where the shape and time duration of the He convective shell and the efficiency of the 3DU can be appreciated. 
It is also worth to be mentioned that, in general, the higher the mass the higher the number of thermal pulses occurring before the beginning of the formation of a He convective shell associated to each thermal pulse (see Tables \ref{tpprop700}-\ref{tpprop920}).

Figure \ref{tceall} shows that the maximum temperature reached at the base of the convective envelope ($\rm T_{BCE}$) is in the range $80-110~\rm MK$ and scales roughly with the initial mass, i.e., the larger the mass the larger $\rm T_{BCE}$. In general this quantity increases slightly during the TP phase but it may also show a non monotonic behavior as a function of time if some other energy sources are activated inside the CO core, as in the case of the more massive models ($\rm M\geq 9.05~M_\odot$) where the URCA processes become efficient (see below). 

The mass loss, during this phase plays a key role because it competes with the increase of the CO core in reducing the H-rich envelope and therefore in determining the duration of the TP phase. The typical mass loss rate averaged over the last few thermal pulses is in the range $\rm 1-3 \sim 10^{-5}~M_\odot/yr$, the higher values reached by the more massive models.

The computations are stopped after a sufficient number of thermal pulses have been computed to safely extrapolate the evolution of these stars during the TP phase (see section \ref{finalfate}).
%arbitrarily stopped after a substantial number of thermal pulses have been computed for each model (see Tables \ref{tpprop700}-\ref{tpprop920}). The reason is that, giving the typical rates of CO core growth and mass loss, reaching the end of the evolution, marked by the removal of the entire H-rich envelope or by the activation of the electron captures on $\rm ^{24}Mg$ (see below), would require the computation of thousands of thermal pulses. Assuming, in the most favorable case, the computation of 1-2 TPs per days it would require at least 2 years (or even more) of continuous computer time for each evolution, that it is not feasible at the moment.

\begin{figure*}[ht!]
\epsscale{0.8}
\plotone{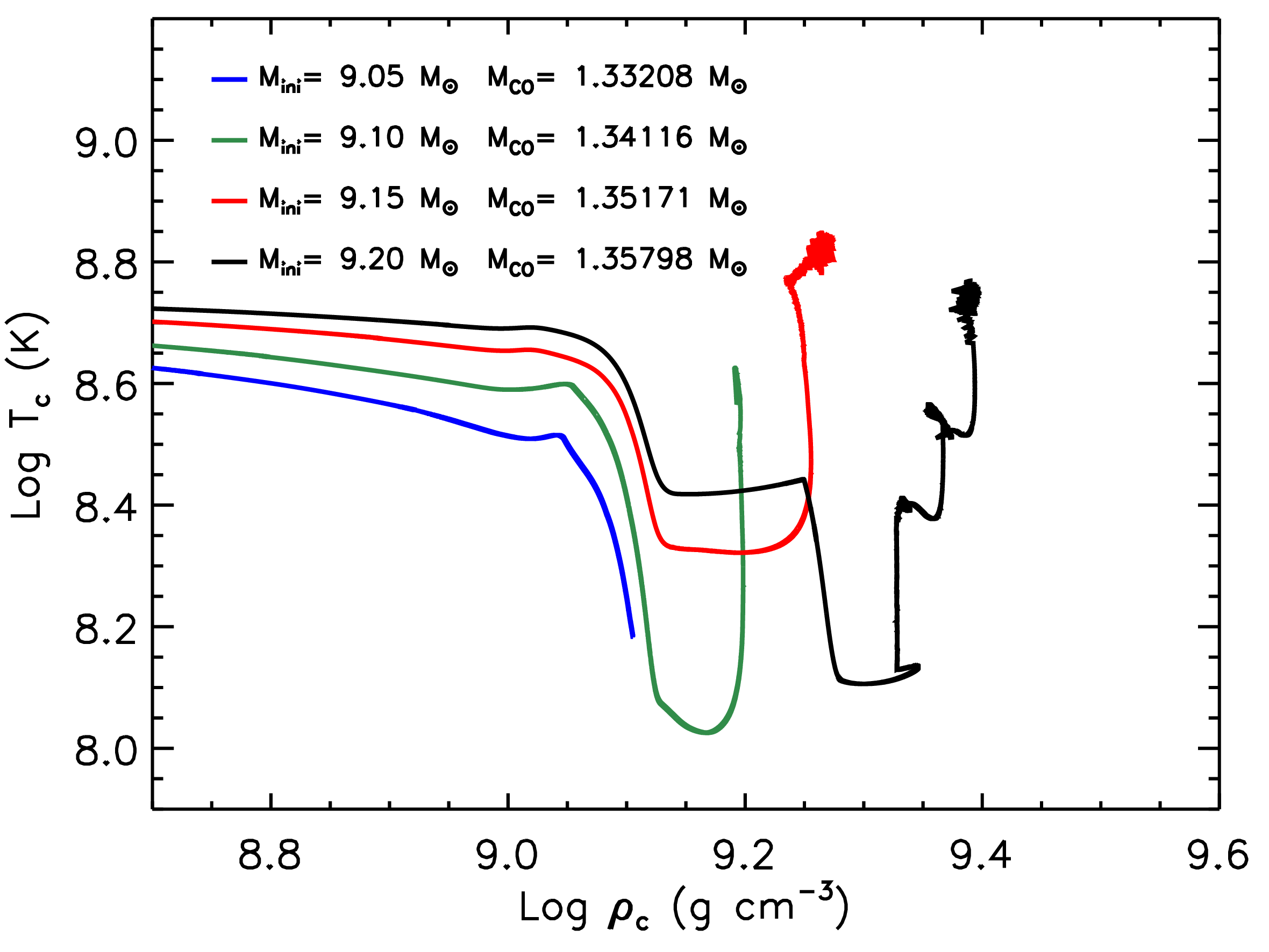}
\caption{Central temperature as a function of the central density during the late stages of models in which the URCA processes are activated.\label{tcrocallzoom}}
\end{figure*}

\begin{figure*}[ht!]
\epsscale{1.0}
\plotone{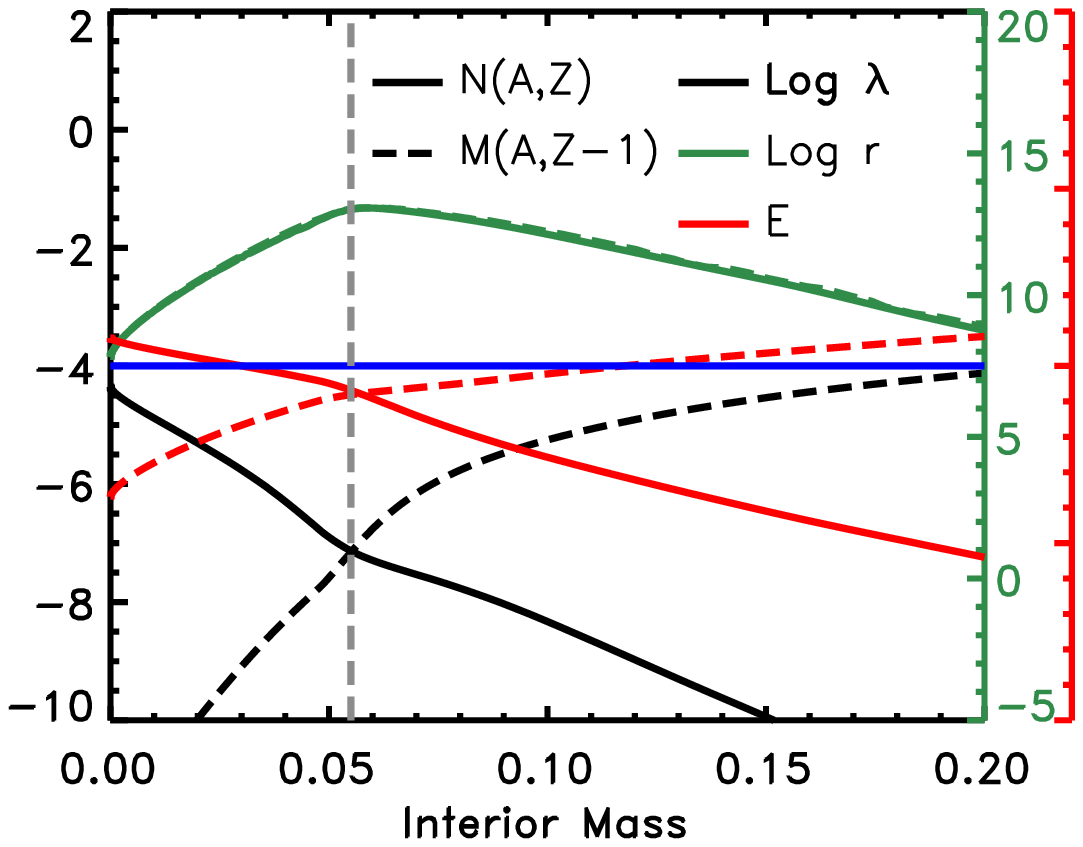}
\caption{Selected properties of a model in which a generic URCA pair $^{A}(N,M)$ is efficient. Upper letf panel: abundance in mass fraction as a function of the interior mass of the two interacting nuclei $N$ and $M$. Upper right panel: density as a function of the interior mass. Lower left panel: number of captures/decays per unit time in s ($\lambda$, black lines, right y-axis), number of reactions occurring per unit mass and unit time in $\rm g~s^{-1}$($r$, green lines, right y-axis) and nuclear energy of the URCA pair in MeV ($E$, red lines, right y-axis); the solid line refers to the electron capture while the dashed lines to the beta decay. Lower right panel: nuclear energy of the URCA pair in erg per unit mass and unit time ($\varepsilon=-r E_{\rm \nu}$); the black line refers to the electron capture, the red line to the beta decay and the orange thick line to the total of the URCA pair. In all the panels the grey vertical dashed line marks the URCA shell. \label{sketchurca}}
\end{figure*}

\subsection{Stars with mass $\rm 9.05\leq M/M_\odot\leq 9.20$: URCA processes}
In stars with the more massive ONeMg degenerate cores ($\rm 9.05 \leq M/M_\odot \leq 9.20$) the density increases enough (Figure \ref{tcrocallzoom}) that the Fermi energy becomes close to the threshold value for the electron captures on a number of nuclear species that are quickly followed by the beta decays. 
In this situation, given two generic nuclear species $N(A,Z)$ and $M(A,Z-1)$, the two reactions $N(A,Z)+e^{-} \rightarrow M(A,Z-1)+\nu$ and $M(A,Z-1) \rightarrow N(A,Z)+e^{-}+\bar{\nu}$, written in a compact form as $^{A}(N,M)$, are in equilibrium. The reaction pair $^{A}(N,M)$ is called URCA process. 
%and may heat or cool the matter depending on the physical conditions. 

The effect of the activation of an URCA process can be explained with the aid of Figure \ref{sketchurca}, that shows the properties of a model in which a generic URCA pair $^{A}(N,M)$ is efficient. If we define $\lambda$ the number of captures/decays per unit time, the lower left panel shows that as the density decreases (i.e. the interior mass increases) the electron capture rate ($\rm \lambda_{\rm ec}$, black solid line) decreases while the beta decay rate increases ($\lambda_{\beta}$, black dashed line). The density at which $\lambda_{\rm ec}=\lambda_{\beta}$ is called URCA shell ($\rho_{\rm crit}$) and is marked in all the panels of Figure \ref{sketchurca} by a vertical dashed line. Inside the mass coordinate corresponding to the URCA shell, $\rho > \rho_{\rm crit}$ and $\lambda_{\rm ec} \gg \lambda_{\beta}$. Outside the URCA shell, on the contrary, $\rho < \rho_{\rm crit}$ and $\lambda_{\rm ec} \ll \lambda_{\beta}$. As mentioned above, in this situation the two reactions are in equilibrium,
% perche' i tempi di raggiungimento dell'equilibrio sono piu' brevi dei tempi scala evolutivi
that means that the number of reactions occurring per unit mass and unit time $r$ of the two processes coincide ($r_{\rm ec}=r_{\beta}$) and show a maximum corresponding to the URCA shell. Since $r=\lambda Y$ (where $Y=X/A$ is the abundance by number, $X$ the abundance in mass fraction and $A$ the atomic weight), this also implies that the equilibrium abundances of the two nuclei satisfy the relation $Y(N)/Y(M)=\lambda_{\beta}/\lambda_{\rm ec}$. As a consequence $X(M) \gg X(N)$ inside the URCA shell while $X(N) \gg X(M)$ outside the URCA shell (upper left panel in Figure \ref{sketchurca}).

The energy released by the electron capture $E_{\rm ec}$ and by the beta decay $E_{\rm \beta}$ are given by (see Miyaji et al. 1980 and Suzuki et al. 2016):

\begin{eqnarray}
E_{\rm ec}=Q_{\rm nuc}-E_{\nu}+\mu_{\rm e} \nonumber \\
\nonumber \\
E_{\rm \beta}=Q_{\rm nuc}-E_{\nu}-\mu_{\rm e} \nonumber
\end{eqnarray}

\noindent
where $Q_{\rm nuc}$ is the mass defect between reactants and products, $E_{\nu}$ is the neutrino energy loss (in absolute value) and $\mu_{\rm e}$ is the chemical potential of the electrons. 
%The energy released by each process per unit mass and unit time is therefore $\varepsilon=r \cdot E$. 
When an URCA pair is in equilibrium $r_{\rm ec}=r_{\beta}$, therefore, the total net energy released per unit mass and unit time in this case will be simply $\varepsilon=-r\left[E_{\rm \nu,ec}+E_{\nu,\beta}\right]$, i.e., it will be always negative and will show a deep minimum corresponding roughly to the URCA shell (lower right panel of Figure \ref{sketchurca}). Thus, in general, we can identify inside a model various cooling zones associated to the URCA shells of the various URCA pairs. It goes without saying, however, that only URCA pairs involving nuclear species with sizable abundances will have some effect on the evolution of the model. In addition to that, as the core of the star contracts, the density increases, the URCA shell of any given URCA pair shifts outward in mass and constitutes an outward moving "cooling wave".

\begin{figure*}[ht!]
\epsscale{0.8}
\plotone{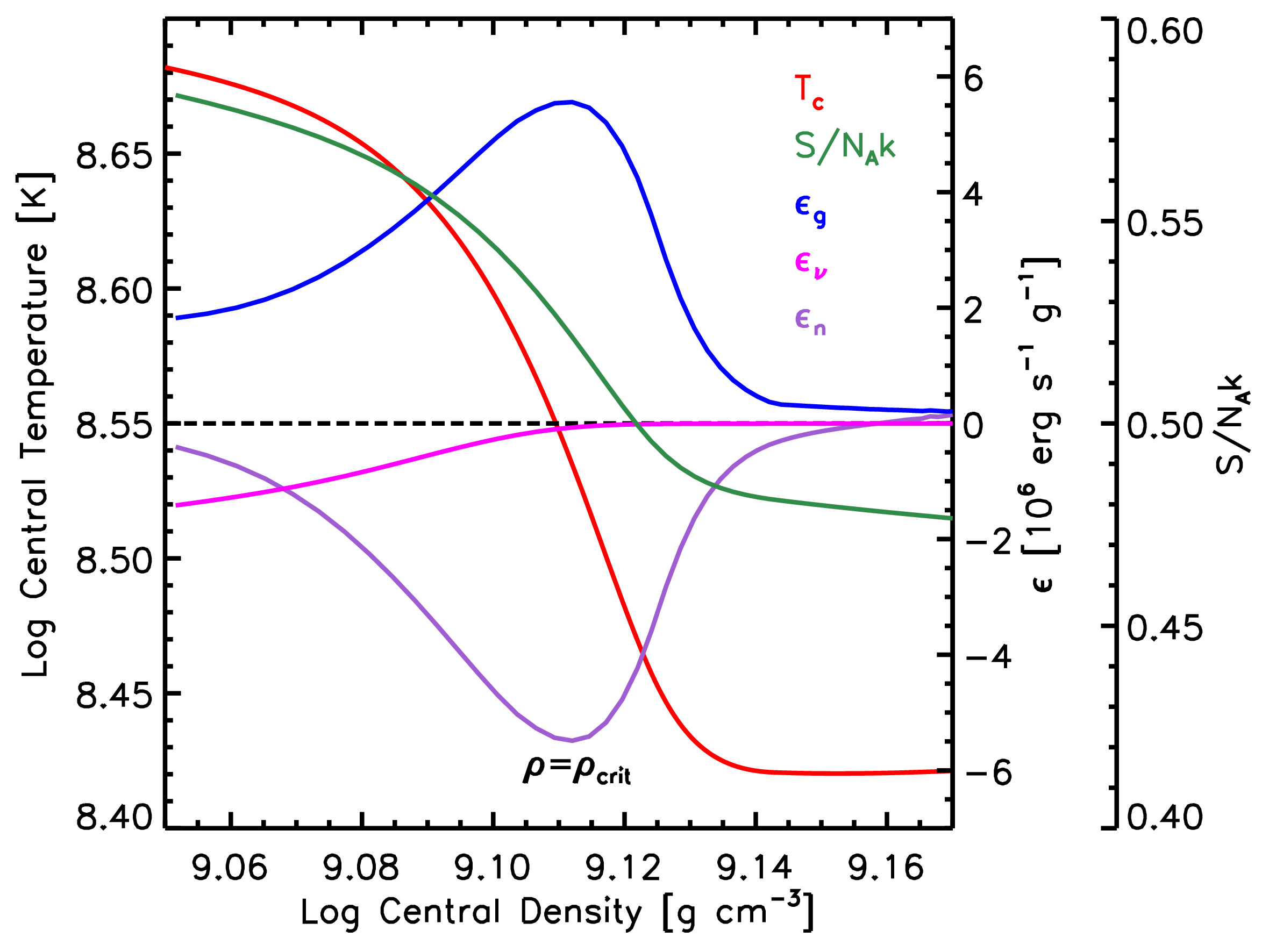}
\caption{Selected central quantities as a function of the central density of the $\rm 9.2~M_\odot$ model during the activation of the $\rm ^{25}(Mg,Na)$ pair: the temperature (red line, left y-axis); the adminesional entropy per baryon (green line right y-axis); the nuclear (purple), gravitational (blue) and neutrino (magenta) energies (right y-axis).\label{tcrocmovurca}}
\end{figure*}

Having said this, in the following we will describe first the evolution of the $\rm 9.2~M_\odot$.
%, that we consider as representative of the group of models for which the URCA processes become efficient ($\rm 9.05 \leq M/M_\odot \leq 9.20$).
As the central density increases above $\rm Log \left[\rho_c~(g~cm^{-3})\right] \sim 9.0$ two URCA pairs activate and produce some effect on the interior of the star. The first one is the $\rm ^{25}(Mg,Na)$, at the density $\rm Log \left[\rho_c~(g~cm^{-3})\right]\sim 9.11$, while the second one is the $\rm ^{23}(Na,Ne)$, at density $\rm Log \left[\rho_c~(g~cm^{-3})\right] \sim 9.25$. The effect of each URCA pair episode at the center is that of a cooling phase, followed by a roughly isothermal evolution (black line in Figure \ref{tcrocallzoom}). The cooling phase corresponds to the stage when the central density is close to the URCA shell ($\rho_{\rm c}\sim\rho_{\rm crit}$) while the isothermal evolution corresponds to the stage when the URCA shell leaves the center and shifts outward in mass. In order to describe in more detail these phases we show in Figure \ref{tcrocmovurca} some properties of the center of the model during the activation of the $\rm ^{25}(Mg,Na)$ pair, i.e. the first important URCA pair.

\begin{figure*}[ht!]
\epsscale{0.8}
\plotone{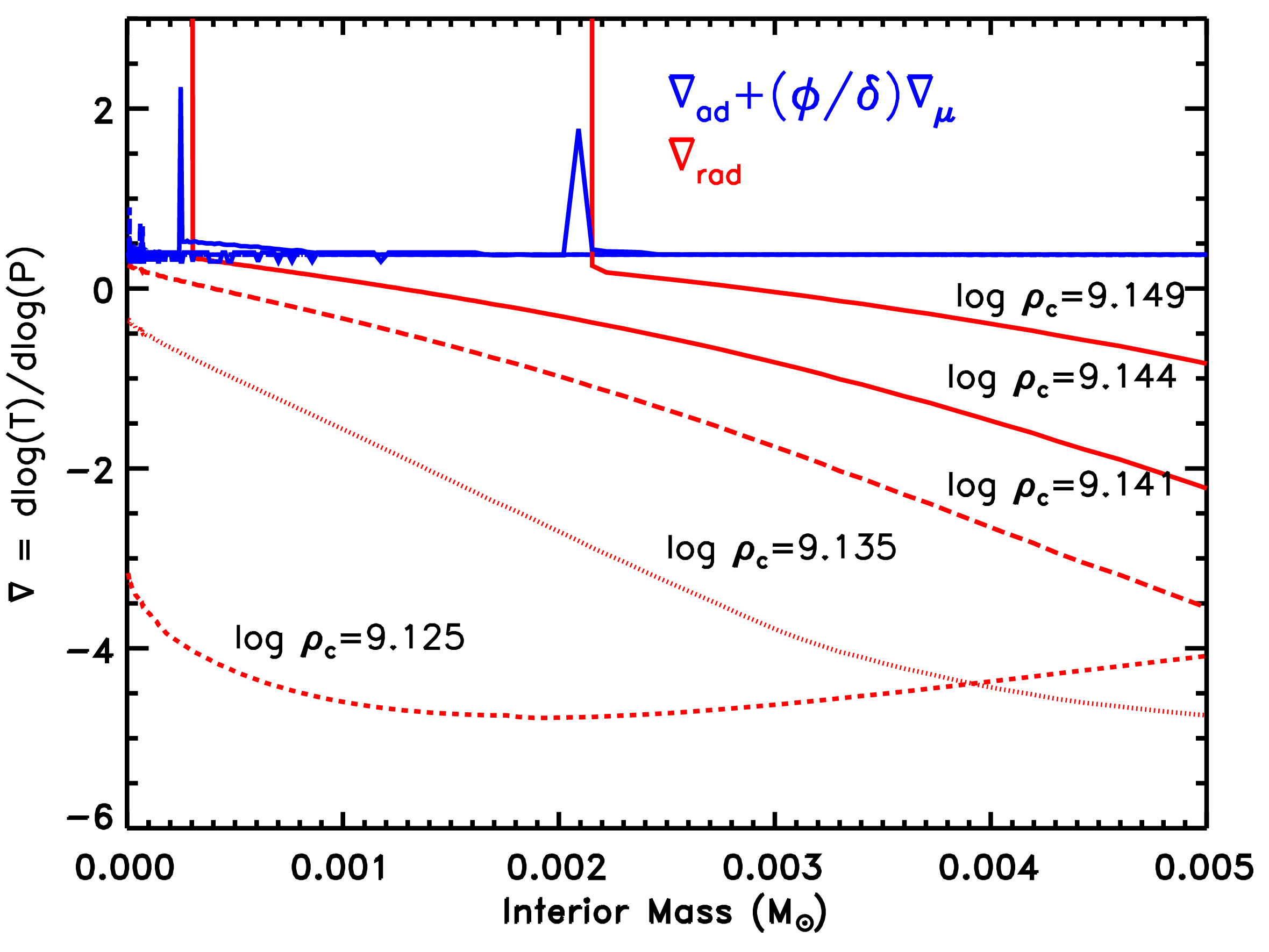}
\caption{Temperature gradients (see the legend) as a function of the interior mass of the $\rm 9.2~M_\odot$ star at selected times, during the formation of the convective core associated with the $\rm ^{25}(Mg,Na)$ URCA pair, marked by the values of the central densities in units of $\rm g~cm^{-3}$. According to the adopted stability criterion (see text), convection sets in when the radiative gradient becomes larger than the adiabatic one. Let us remind that the actual temperature gradient used is the adiabatic one in the convective zones and the radiative one in the radiative layers.\label{gradmovurca}}
\end{figure*}

As the central density approaches the URCA shell, the nuclear energy ($\varepsilon_{\rm n}$) that, as already mentioned above, is dominated by the neutrino emission due to the URCA pair ($-E_{\rm \nu}$), decreases dramatically. During this phase, the gravitational energy ($\varepsilon_{\rm g}$) increases while the (thermo)neutrino losses ($\varepsilon_{\rm \nu}$) progressively decrease due to the lowering of the central temperature and the net result is that the total energy is negative. This implies a substantial reduction of the central temperature ($T_{\rm c}$). As the central density continues to increase ($\rho_{\rm c}>\rho_{\rm crit}$), the URCA shell leaves the center and moves outward in mass (driving a outward moving of a cooling wave), therefore in the center the nuclear energy begins to increase toward values it had before the activation of the URCA pair (i.e. it tends toward $\sim 0$). During this phase the gravitational energy decreases progressively reabsorbing partially the increase of the nuclear energy, while the neutrino energy losses become negligible compared to the nuclear and the gravitational energies, because of the low temperature. The net effect is that in this phase the total energy progressively increases toward less negative values. The energy imbalance between the center and the location of the URCA shell produces also an increase in the radiative gradient in the core (see dashed, dotted and long-dashed lines in Figure \ref{gradmovurca}). When the central density becomes higher than $\rm Log \left[\rho_c~(g~cm^{-3})\right] \sim 9.141$ the total net energy becomes positive, the radiative gradient overcomes the adiabatic one and the center of the star becomes convective. Note that the gradient of chemical composition around the center is not high enough to stabilize the zone against the onset of convection (Figure \ref{gradmovurca}). This is also confirmed by a test evolution in which we adopted the Ledoux criterion during this phase. \cite{jones+13,Taka+13,Zha+19} did not find the formation of the convective core in their models. Such a difference might be due to the difference in the treatment of convection and the zoning, which could affect the gradient of the chemical composition, in the evolutionary codes.

\begin{figure*}[ht!]
\epsscale{1.0}
\plotone{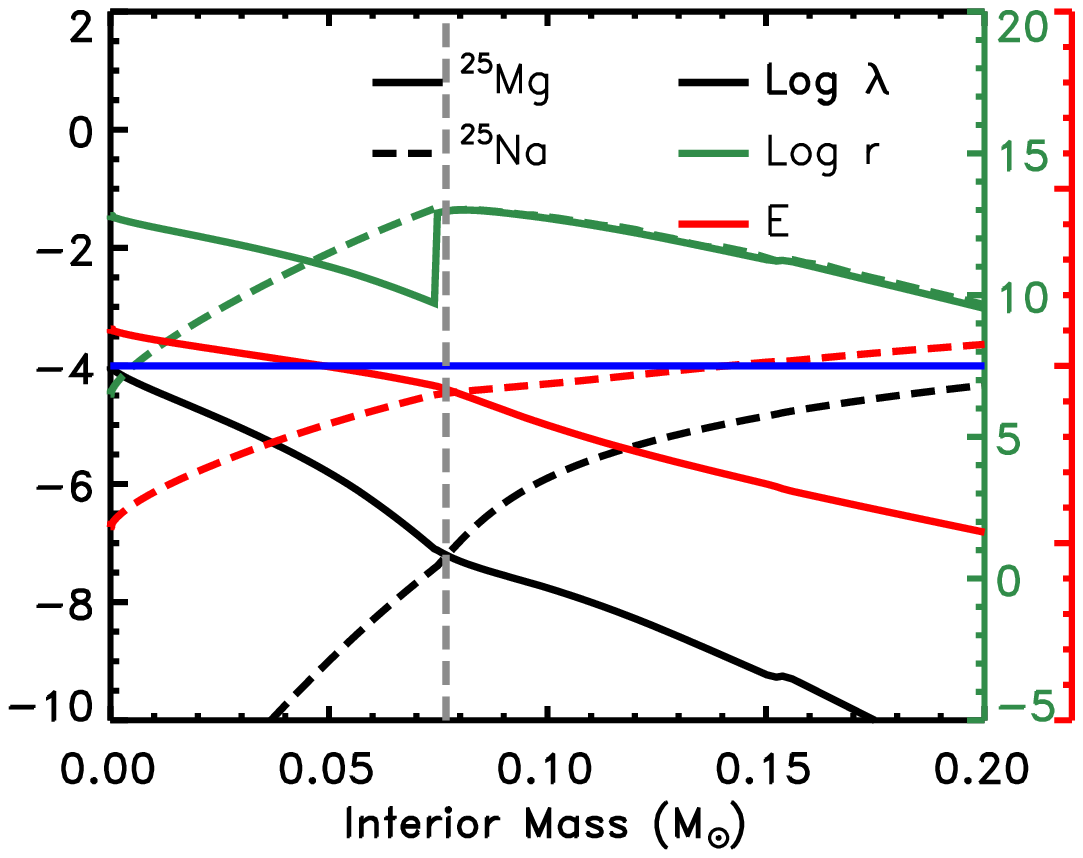}
\caption{As Figure \ref{sketchurca} in the case of a $\rm 9.20~M_\odot$ model during the phase in which a convective core induced by the $\rm ^{25}(Mg,Na)$ URCA pair is formed. The vertical grey dashed line marks the $\rm ^{25}(Mg,Na)$ URCA shell.\label{snap_urca_detailepsi_25_100}}
\end{figure*}

When the convective core sets in, it has a strong effect on the equilibrium of the URCA pair reaction. The reason is the following.
In a radiative environment the $\rm ^{25}Mg$ abundance in the central zones is the result of the equilibrium between electron capture and beta decay and has an increasing profile from the center toward the URCA shell (like the one shown in the upper left panel of Figure \ref{sketchurca}). Once convection sets in, it forces the $\rm ^{25}Mg$ abundance to increase in the inner zones and to decrease in the outer ones compared to the radiative case (compare the upper left panels of Figures \ref{sketchurca} and \ref{snap_urca_detailepsi_25_100}). As a consequence the two reactions of the $\rm ^{25}(Mg,Na)$ pair are not anymore in equilibrium but, on the contrary, $r_{\rm ec}>r_{\beta}$ roughly in the inner half of the convective core while $r_{\rm ec}<r_{\beta}$ in the outer half (solid and dashed green lines in the). In this case the total net energy released per unit mass and unit time is given by $r_{\rm ec}E_{\rm ec}+r_{\beta}E_{\rm \beta}$. Since $E_{\rm ec}$ is positive in roughly the inner half of the convective core and negative outward in mass while $E_{\rm \beta}$ is always negative (see the solid and dashed red lines in the lower left panel of Figure \ref{snap_urca_detailepsi_25_100}), the total energy released by the $\rm ^{25}(Mg,Na)$ pair is positive in roughly the inner half of the convective core and negative in the remaining half. (see the orange line in the lower right panel in the figure), the zero value corresponding roughly to the mass coordinate where $r_{\rm ec}=r_{\beta}$. The continuous ingestion of a higher $\rm ^{25}Mg$ abundance from the outer radiative layers, produces an increase of the nuclear energy close to the center that induces the convective zone to extend even more driving in this way a progressive increase of the convective core (Figure \ref{gradmovurca}).  
During the phase characterized by the increase of the convective core the central density increases progressively at almost constant temperature (Figure \ref{snap_urca_detailepsi_25_100}).

\begin{figure*}[ht!]
\epsscale{1.0}
\plotone{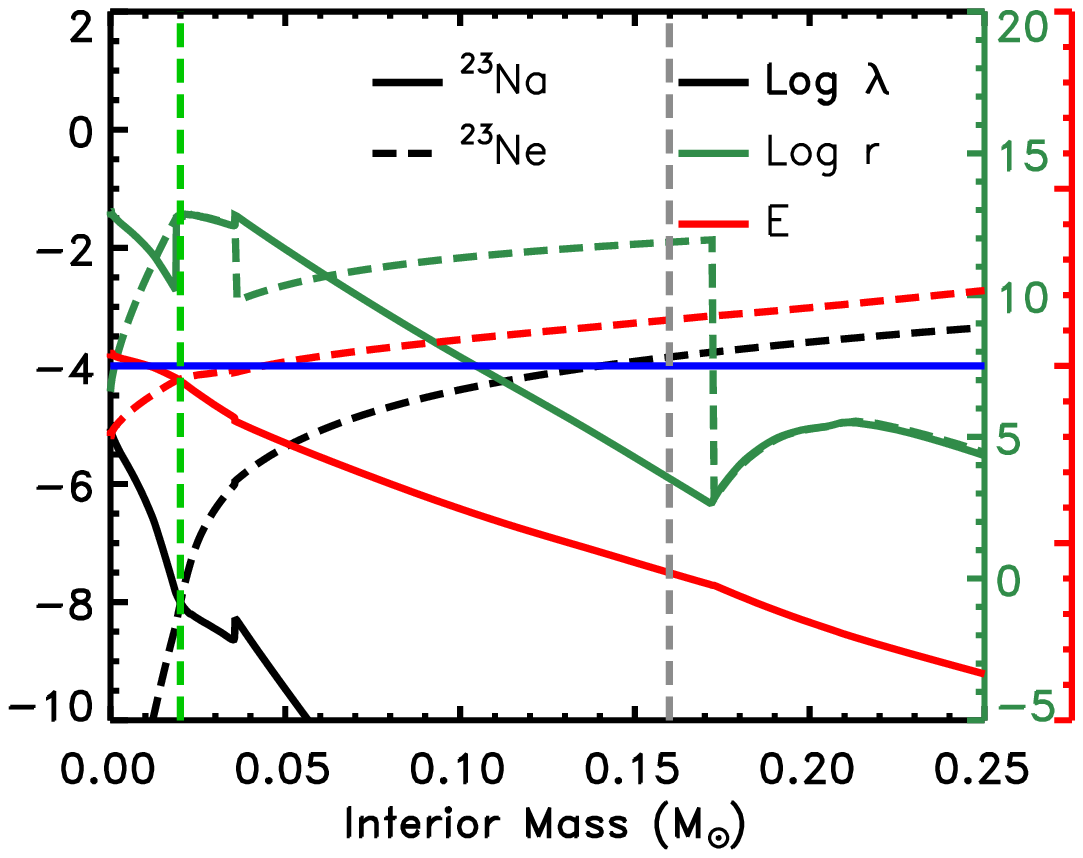}
\caption{Same as Figure \ref{snap_urca_detailepsi_25_100} but during a phase in which a convective core, driven by the $\rm ^{23}(Na,Ne)$ URCA pair, and a convective shell, induced by the $\rm ^{25}(Mg,Na)$ URCA pair, are formed. The grey and green vertical dashed line mark the $\rm ^{25}(Mg,Na)$ and the $\rm ^{23}(Na,Ne)$ URCA shells, respectively.\label{snap_urca_detailepsi_23_208}}
\end{figure*}

\begin{figure*}[ht!]
\epsscale{1.0}
\plotone{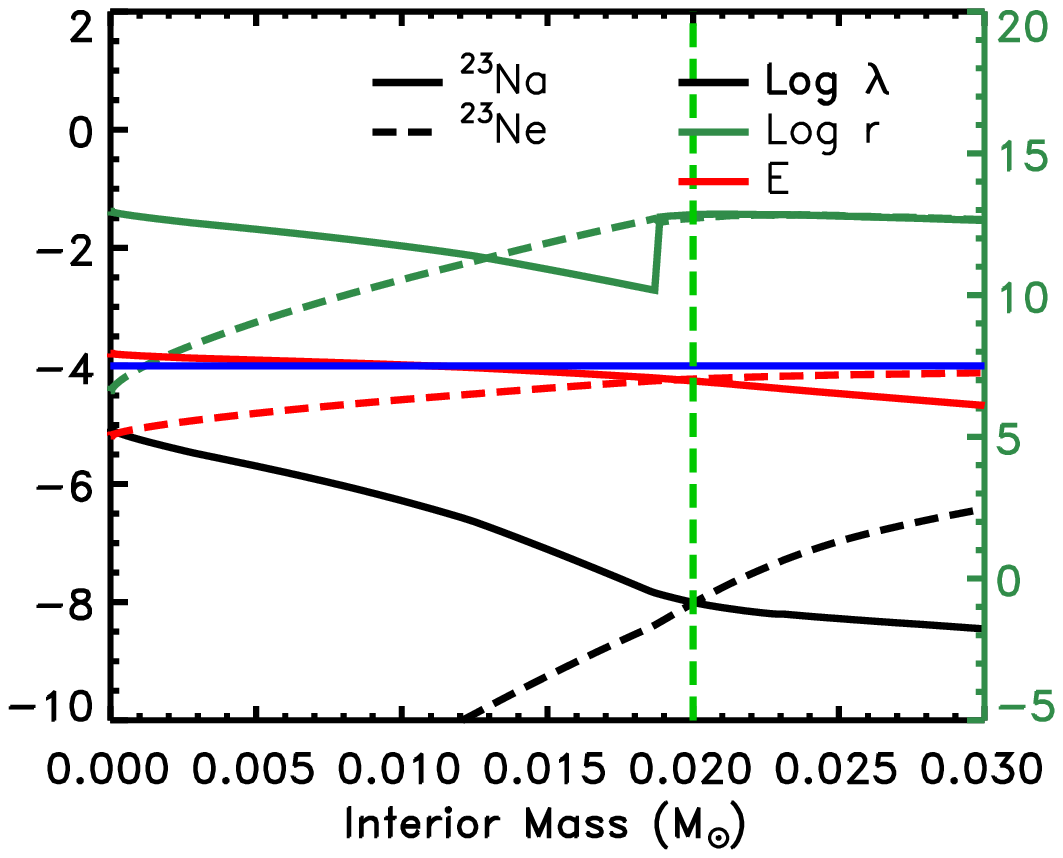}
\caption{Same as Figure \ref{snap_urca_detailepsi_23_208} but zoomed in the inner $\rm 0.03~M_\odot$.\label{snap_urca_detailepsi_23zoom}}
\end{figure*}

When the central density approaches $\rm log \left[\rho_c~(g~cm^{-3})\right] \sim 9.248$ the URCA pair $\rm ^{23}(Na,Ne)$ starts activating and producing some effects on the structure of the star. The evolution of the center during this phase is similar to that already discussed for the $\rm ^{25}(Mg,Na)$ pair. The initial phase is characterized by a cooling, due to the electron capture on $\rm ^{23}Na$, that makes the center of the star radiative and forces the the convective core driven by the $\rm ^{25}(Mg,Na)$ pair to become a convective shell that shifts progressively outward in mass. Then, after the URCA shell of the pair $\rm ^{23}(Na,Ne)$ leaves the center and moves outward in mass a convective core forms that increases progressively in mass while the center contracts at almost constant temperature (Figure \ref{tcrocallzoom}). 
A typical model during this phase is shown in Figures \ref{snap_urca_detailepsi_23_208} and \ref{snap_urca_detailepsi_23zoom}.
The inner $\rm 0.02~M_\odot$ zones are convective and show the typical behavior already discussed in the case of the $\rm ^{25}(Mg,Na)$ pair. In particular the rate of the electron capture on $\rm ^{23}Na$ dominates over the decay of the $\rm ^{23}Ne$ in approximately the inner half of the convective core while the $\rm ^{23}Ne$ decay prevails on the electron capture on $\rm ^{23}Na$ in the remaining half. As a consequence, the nuclear energy is positive in the zones where the electron capture dominates and negative where the beta decay prevails. Inside the convective core, $\rm ^{23}Ne$ is much more abundant than $\rm ^{23}Na$, that, on the contrary, dominates in the outer radiative layers. During this phase the convective shell driven by the $\rm ^{25}(Mg,Na)$ URCA pair is roughly confined between the $\rm ^{23}(Na,Ne)$ URCA shell, from the bottom, and the $\rm ^{25}(Mg,Na)$ URCA shells from the top. Note also the radiative zone that separates the convective core and the convective shell; in this zone the $r_{\rm ec}=r_{\beta}$. It is worth mentioning at this point that a similar evolution has been already found by \cite{ritossa+99}. In particular, in Figure 25 they show the properties of their last computed model characterized by a convective core driven by the $\rm ^{23}(Na,Ne)$ pair and by a convective shell driven by the $\rm ^{25}(Mg,Na)$ pair. The chemical composition as well as the various contributions to the total energy generation are extremely similar to what we find. In particular, they also find that in each convective region the URCA pair releases a positive energy in the inner zone and a negative one in the outer layers (see Panel c in their Figure 25). At variance with what we and \cite{ritossa+99} find, the formation of a convective core and of a convective shell during this phase is not addressed by \cite{jones+13,Taka+13,Zha+19}. 
%{\bf The origin of this difference is difficult to understand because of the many differences among the various stellar evolution codes. Note, in addition, that in the case of \cite{Taka+13} and \cite{Zha+19} the thermal pulses are not followed in a self consistent way but, on the contrary, they are simulated by means of a continous accretion of CO matter onto the degenerate CO core after it is completely peeled off once the ONe core is formed.}
We do not have a clear explanation for that, hence what we can say is that the origin of such a difference could be due to the difference in the numerical treatment of convection in the stellar evolution code.

\begin{figure*}[ht!]
\epsscale{1.0}
\plotone{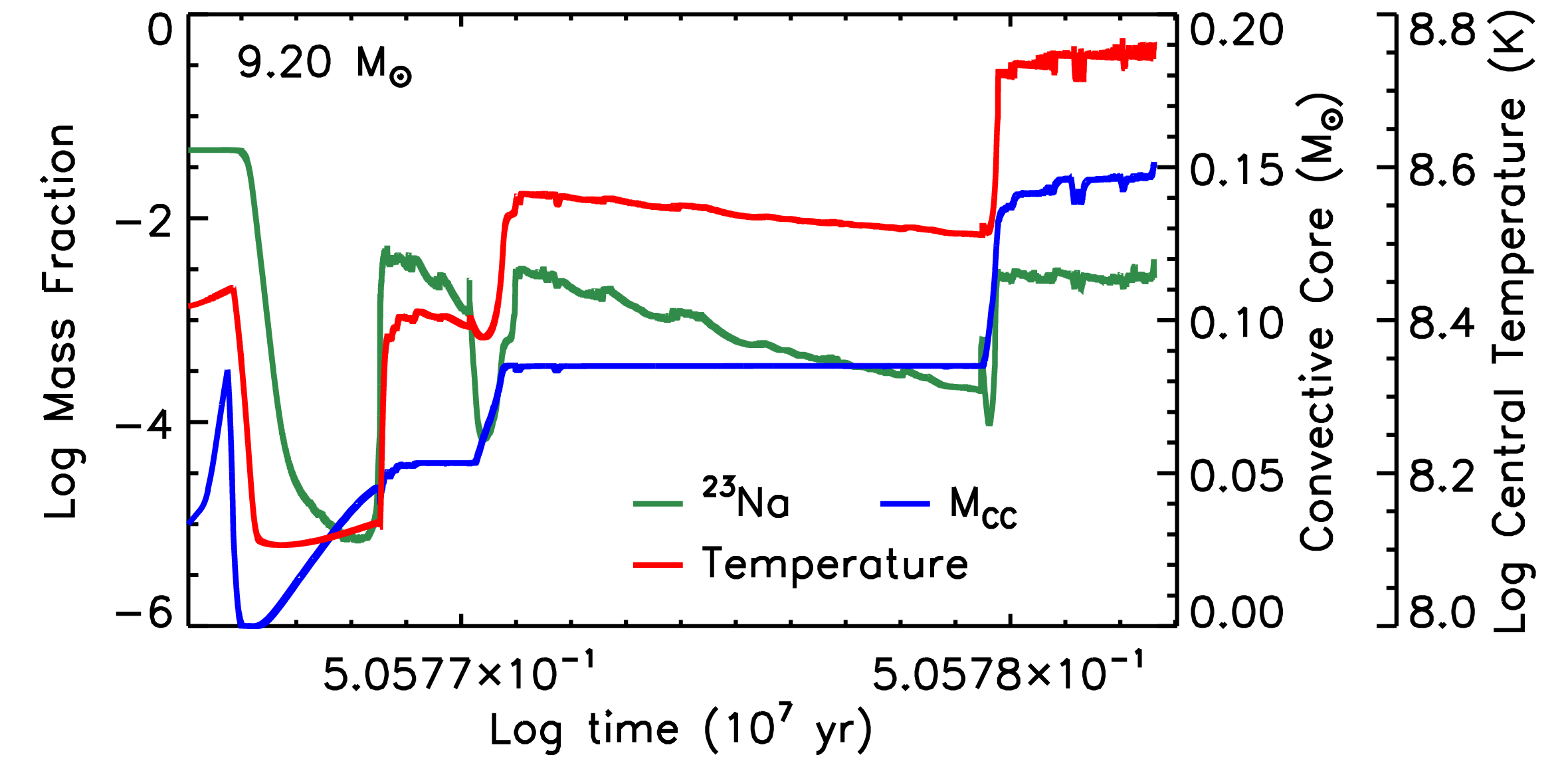}
\caption{Selected quantities of the $\rm 9.20~M_\odot$ model during the breathing pulses induced by the $\rm ^{23}(Na,Ne)$ URCA pair: the central $\rm ^{23}Na$ mass fraction (green line, left y-axis); the mass of the convective core (blue line, right y-axis); the central temperature (red line, right y-axis).\label{na23cent}}
\end{figure*}

\begin{figure*}[ht!]
\begin{center}
\includegraphics[width=.45\linewidth]{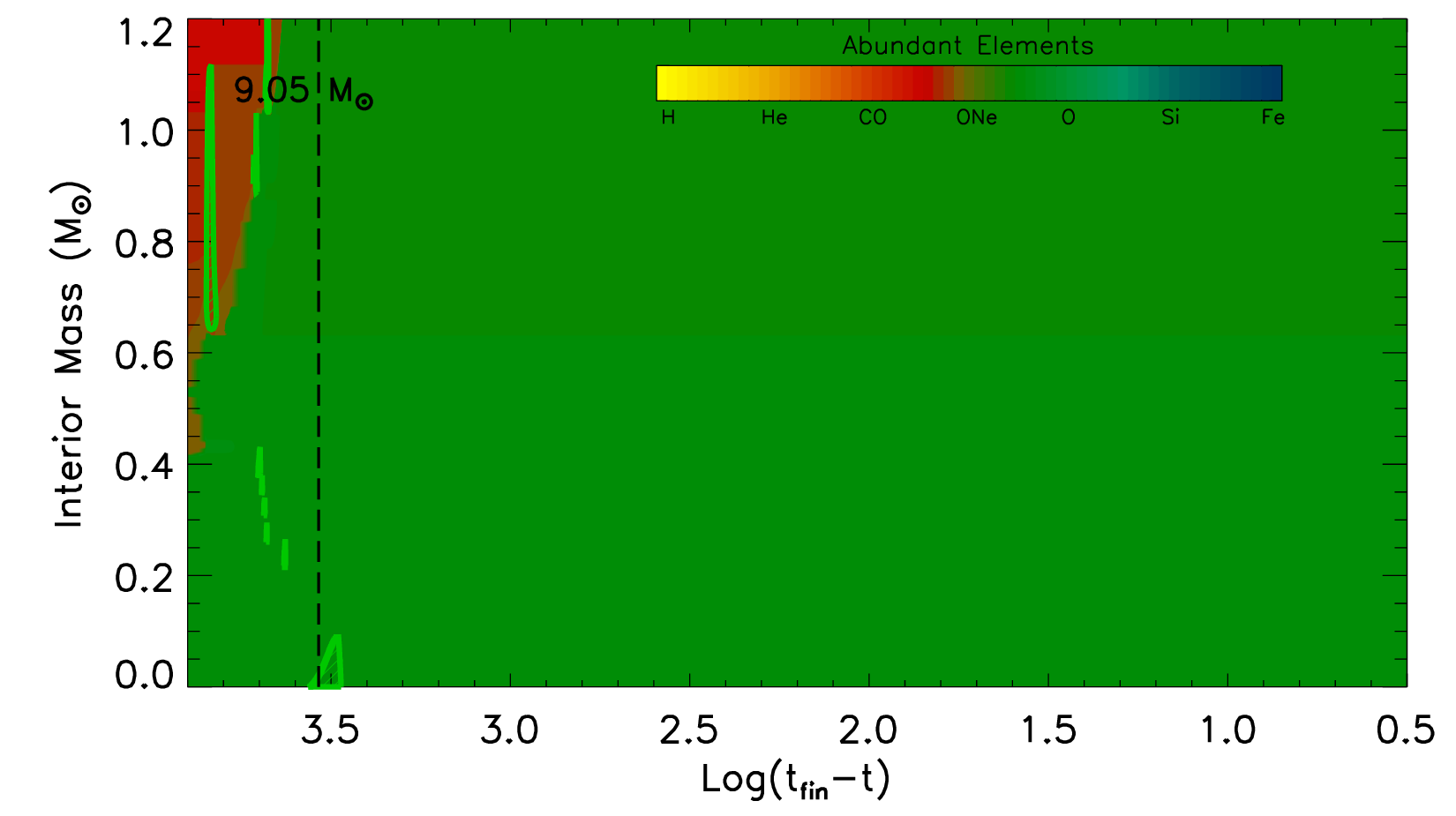}\quad\includegraphics[width=.45\linewidth]{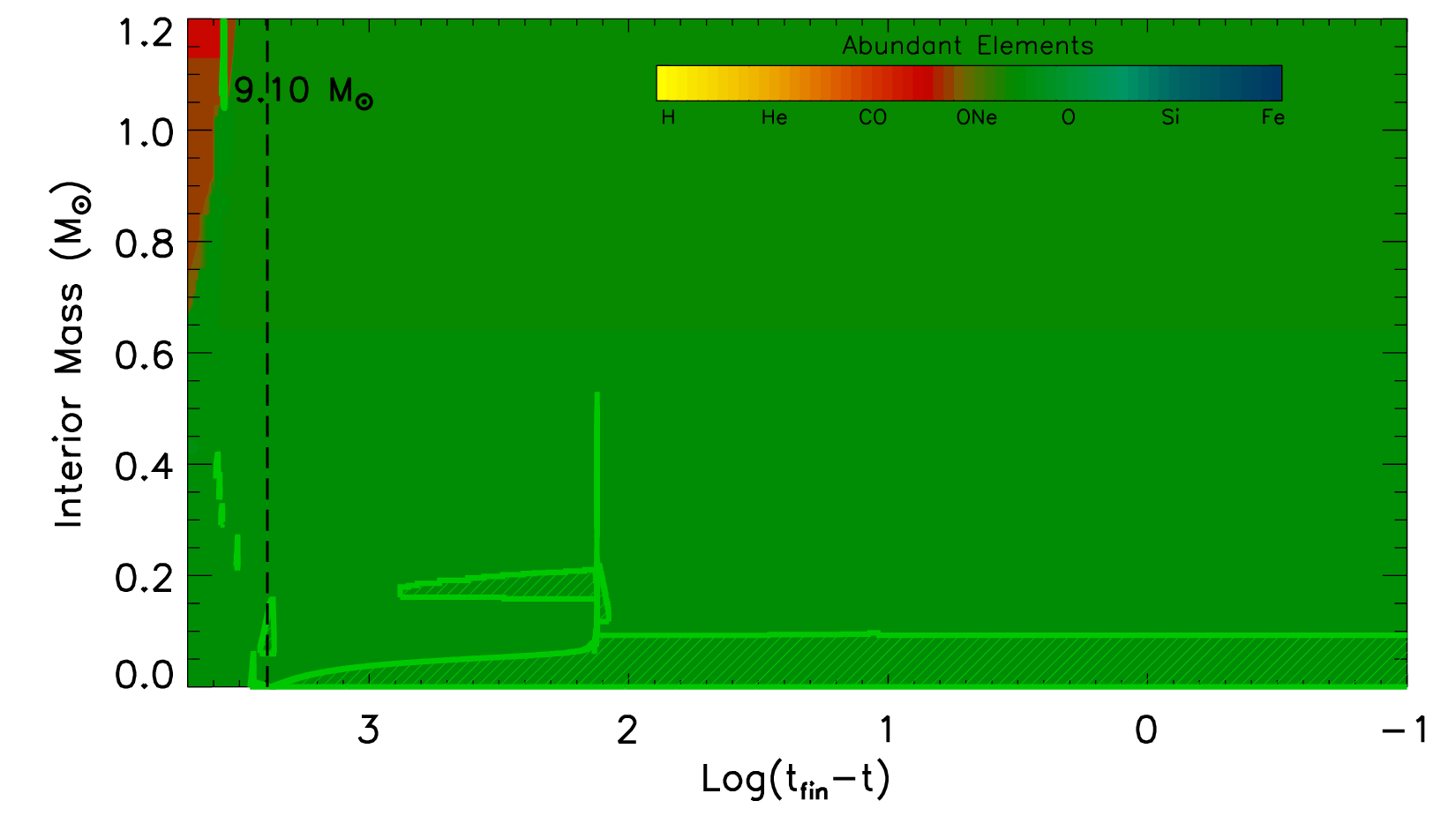}
\\[\baselineskip]% adds vertical line spacing
\includegraphics[width=.45\linewidth]{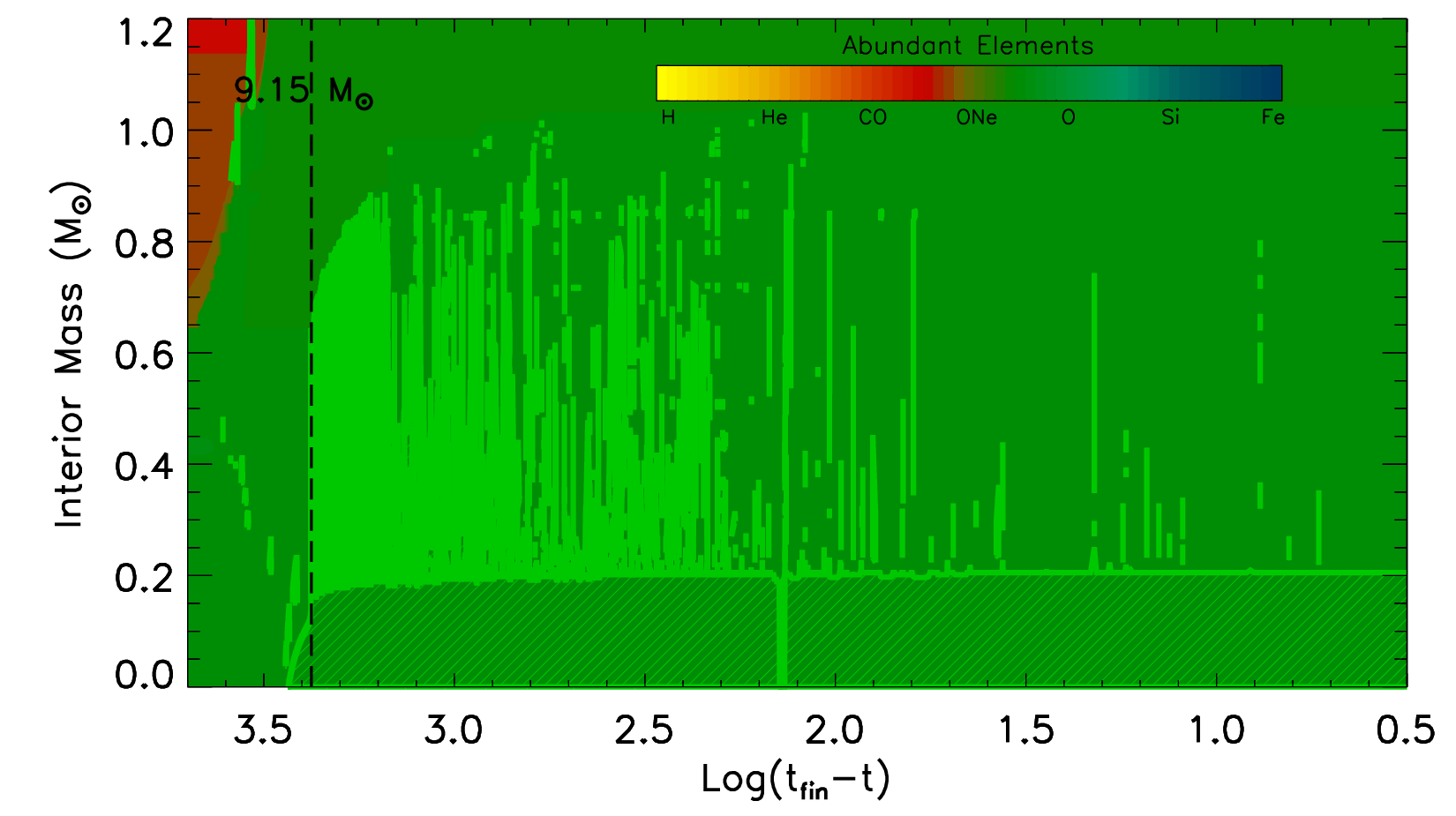}\quad\includegraphics[width=.45\linewidth]{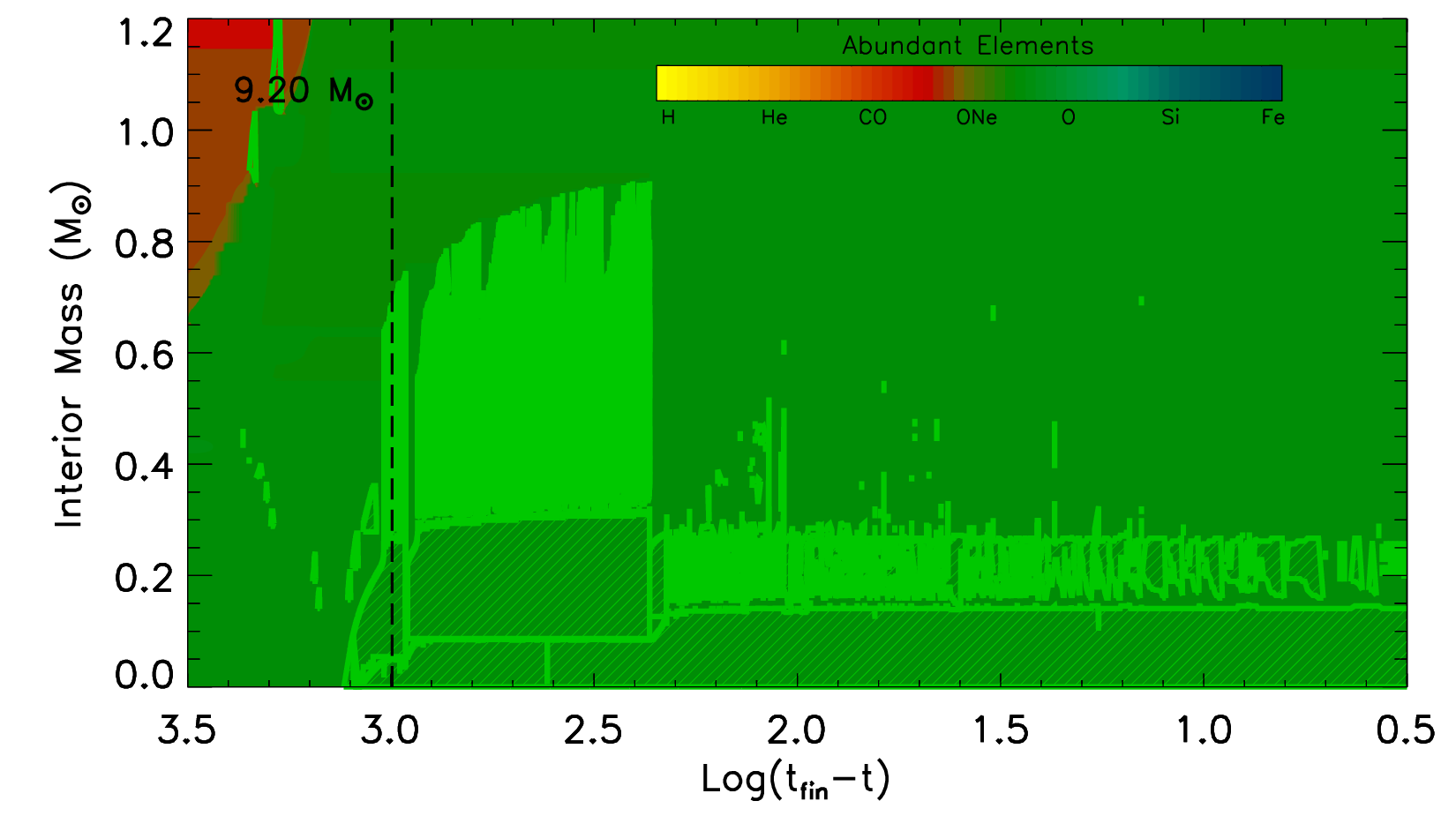}
\end{center}
\caption{Convective (green shaded areas) and chemical (color codes reported in the color bar) internal history during the phase when the URCA processes are active. In the x-axis is reported the logarithm of the time till the end of the evolution ($\rm t_{\rm fin}-t$) in units of yr.\label{kipurca}}
\end{figure*}

The evolution of the center, following the formation of the convective core driven by the $\rm ^{23}(Na,Ne)$ pair, is characterized by an increase of the central temperature at almost constant density interspersed with phases where the density increases at almost constant temperature (black line in Figure \ref{tcrocallzoom}). The reason for such a behavior is due to the fact that the convective core, after it is formed, begins to progressively increase in mass because of the increase of the nuclear energy produced close to the center due to the ingestion of fresh $\rm ^{23}Na$ present in the outer radiative zones. When the $\rm ^{23}Na$ abundance mixed into the convective core is comparable or even larger than the one initially present, the increase of the nuclear energy is not reabsorbed but, on the contrary, it drives a farther increase of the convective core on a very short timescales, compared to the previous evolution. This is a runaway that looks like the breathing pulse phenomenon occurring during the core He burning (see above). During this phase the central $\rm ^{23}Na$ mass fraction increases to values as high as $\sim 3\cdot 10^{-3}$, i.e., more than two order of magnitudes compared to the abundance present before the beginning of this process (see the first increase of the central $\rm ^{23}Na$ abundance, green line, in Figure \ref{na23cent}). Since the matter is highly degenerate, the increase of the nuclear energy due to this process induces an increase of the central temperature at constant density (Figure \ref{na23cent}). We call this phenomenon TIR, i.e., Temperature Increase due to a Runaway.
%Let us remind that when the matter is highly degenerate, the pressure depends only on the density and not on the temperature, therefore a local heating due to a release of nuclear energy cannot be reabsorbed by an expansion (reduction of the density) but, on the contrary, it drives a farther heating and therefore an increase of the temperature (see the first rise in the temperature, red line, in Figure \ref{na23cent}). 
The increase of the convective core eventually ceases when the $\rm ^{23}Na$ ingested from the radiative zones is such that it does not alter significantly the nuclear energy. This happens when the mass of the convective core is $\rm \sim 0.06~M_\odot$. During the following evolution the excess of the nuclear energy is progressively reabsorbed, the mass of the convective core and the central temperature remains essentially constant while the central $\rm ^{23}Na$ abundance progressively decreases towards values similar to those corresponding to the beginning of this process. This stage coincides also with the onset of the thermal pulses. 
%The increase of the mass of the convective core also induces a slight contraction of the core. 
The following evolution of the star is characterized by two other similar processes (see the last two sharp increase of the central temperature in Figure \ref{na23cent}) that rise the central temperature to values as high as ${\rm Log} \left[T_{\rm c}~{(\rm K)}\right] \sim 8.8$. 
During the phase characterized by the thermal pulses the convective shell driven by the $\rm ^{25}(Mg,Na)$ increases in mass up to $\rm \sim 0.9~M_\odot$ but this has little effect on the interior of the star (lower right panel in Figure \ref{kipurca}).
The calculation of the evolution of this star is then stopped after 102 thermal pulses. The final fate of this star is discussed in the following, however we anticipate here that, on the basis of the results obtained, it is difficult to envisage wether the center of the star will reach the threshold temperature for the activation of the $\rm ^{20}Ne$ photodisintegration or the increase of the central temperature will stop and the core will restart contracting until the density thresholds for the activation of the electron captures on $\rm ^{24}Mg$ first and on $\rm ^{20}Ne$ later are reached. Moreover, an interaction between the convective core and the convective shell cannot be excluded with consequences on the evolution of the star that are difficult to predict.

\begin{figure*}[ht!]
\epsscale{0.8}
\plotone{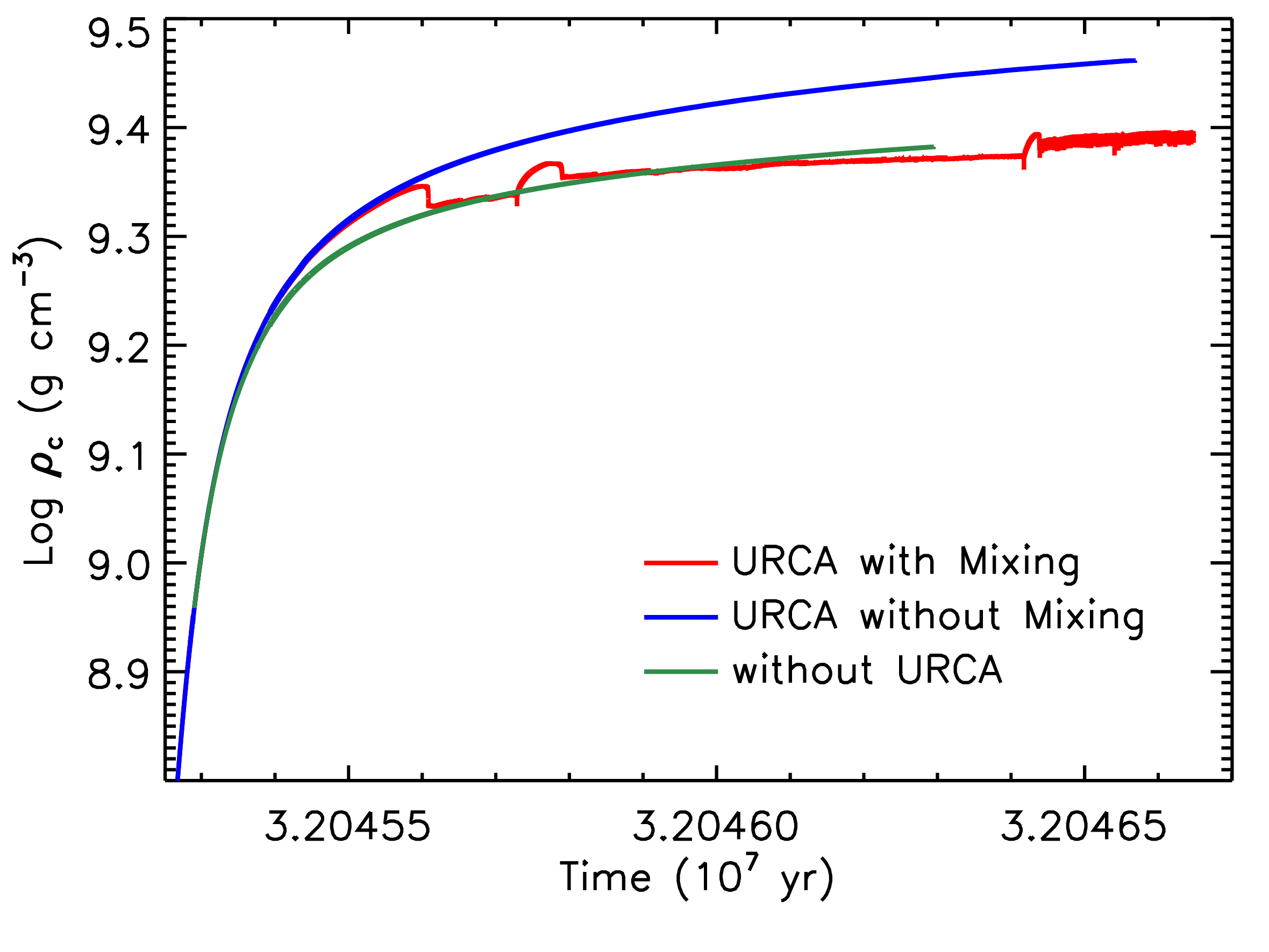}
\caption{Evolution of the central density as a function of time for three models of the initial mass $\rm 9.20~M_\odot$ prior the onset of the TIR (see text), computed with the following assumptions: URCA processes and convective mixing taking account (the reference model, red line); URCA processes taken into account and convective mixing artificially suppressed (blue) line); URCA processes neglected (green line).\label{roctime920}}
\end{figure*}

\begin{figure*}[ht!]
\epsscale{0.8}
\plotone{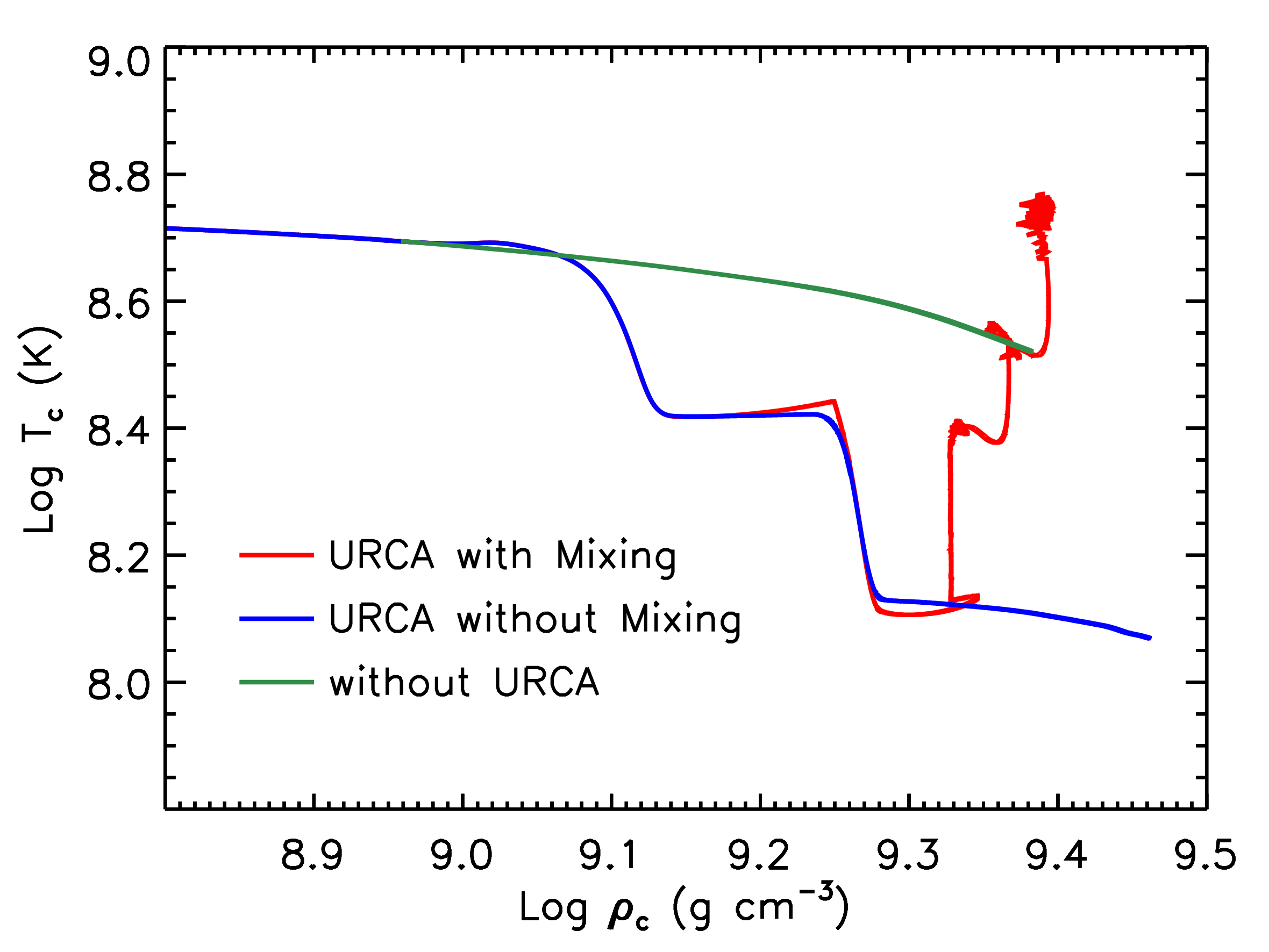}
\caption{Evolution of the central temperature as a function of the central density for three models of the initial mass $\rm 9.20~M_\odot$ prior the onset of the TIR (see text), computed with the following assumptions: URCA processes and convective mixing taking account (the reference model, red line); URCA processes taken into account and convective mixing artificially suppressed (blue) line); URCA processes neglected (green line).\label{tcrocvaritest}}
\end{figure*}

It is interesting to note that the evolution of the center prior the onset of the TIR discussed above, i.e. until the central density approaches the value $\rm Log \left[\rho_c~(g~cm^{-3})\right] \sim 9.35$, is not affected by the efficiency of mixing in the convective zones. Figures \ref{roctime920} and \ref{tcrocvaritest}, in fact, show that, as long as $\rm Log \left[\rho_c~(g~cm^{-3})\right] \leq 9.35$, the evolution of both the central density and the central temperature of the standard model (red lines) is almost identical to the one obtained in a test model in which the mixing is artificially suppressed (blue lines). The differences between the two models appear only when the TIR begins in the standard model, i.e., when the red and blue lines begin to differ each other. The occurrence of this phenomenon has two main effects, i.e., it slows down the contraction of the core (Figure \ref{roctime920}) and induces an increase of the central temperature (Figure \ref{tcrocvaritest}) compared to the case in which the chemical mixing is suppressed. For sake of completeness we report also the results obtained for a test model in which the URCA processes are not included (green lines in Figures \ref{roctime920} and \ref{tcrocvaritest}). In this case the contraction of the core is slower than in the case of the model where the URCA processes are taken into account and mixing is suppressed and similar to the reference case. Moreover, as it is expected the activation of the two URCA pairs $\rm ^{25}(Mg,Na)$ and $\rm ^{23}(Na,Ne)$ reduces the central temperature by a factor of $\sim 3$ (at $\rm Log \left[\rho_c~(g~cm^{-3})\right] =9.35$) compared to the model in which the URCA processes are not included. Let us eventually note at this point that, as is it mentioned above and shown in Figure \ref{tcrocvaritest}, the TIR is associated to the presence of a convective core and occurs in an advanced phase after its formation. For this reason this phenomenon is not found neither by \cite{ritossa+99}, because they stop the calculation too early, nor by \cite{jones+13,Taka+13,Zha+19}, because they do not find the formation of the convective core in their models. 

\begin{figure*}[ht!]
\epsscale{1.0}
\plotone{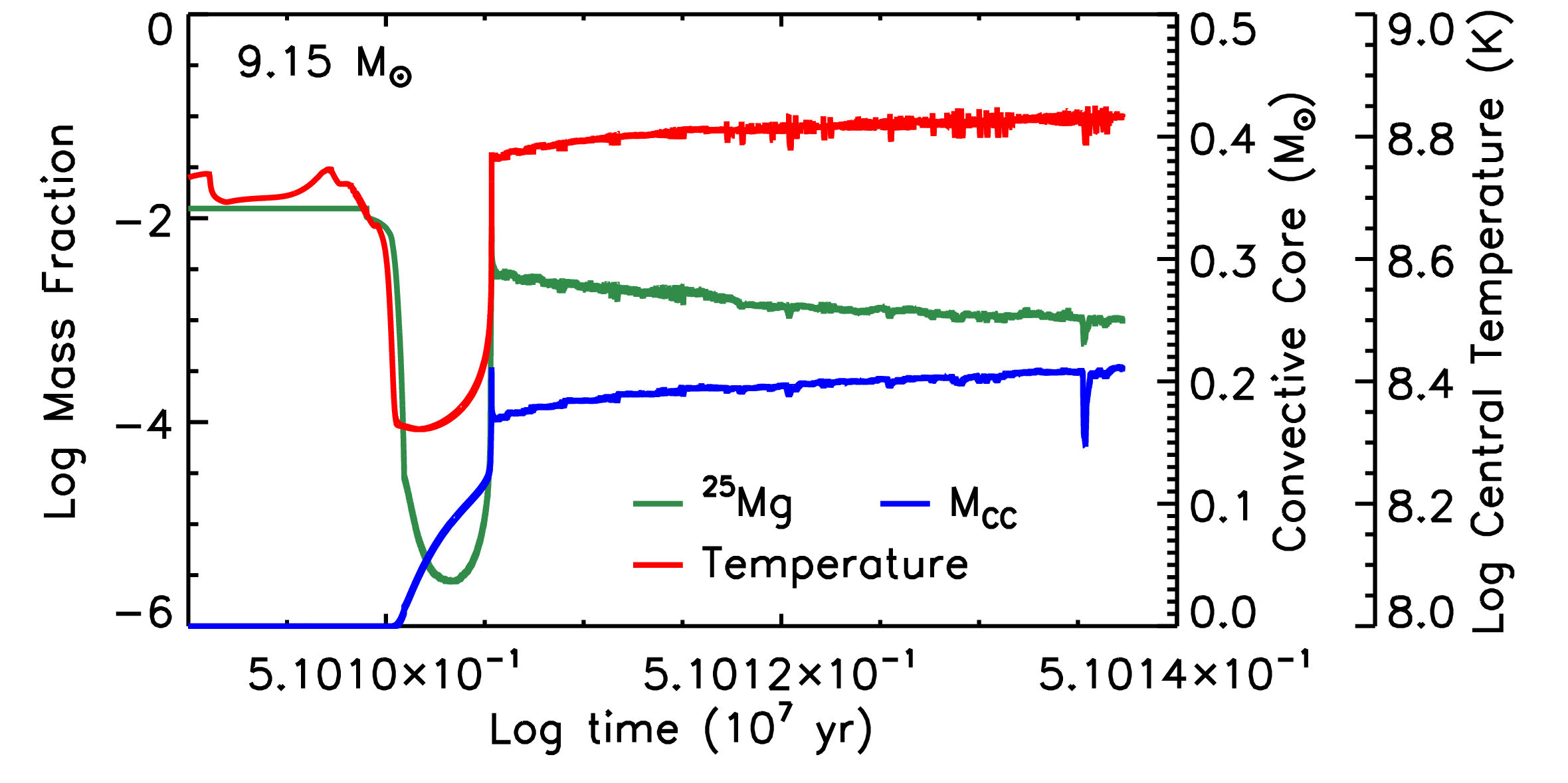}
\caption{Same as Figure \ref{na23cent} but for the $\rm 9.15~M_\odot$ model and during the TIR (see text) induced by the $\rm ^{25}(Mg,Na)$ URCA pair.\label{mg25cent}}
\end{figure*}

The evolution of the center of the $\rm 9.15~M_\odot$ is shown in Figure \ref{tcrocallzoom} (red line). As in the case of the $\rm 9.20~M_\odot$, the first cooling phase is due to the activation of the $\rm ^{25}(Mg,Na)$ URCA pair. The cooling phase ends when the URCA shell shifts from the center outward in mass. Such an occurrence drives the formation of a convective core that progressively increases in mass. At variance with the $\rm 9.20~M_\odot$, in this case the $\rm ^{25}Mg$ ingested from the radiative zones above the convective core is high enough to induce a TIR before the threshold density for the activation of the $\rm ^{23}(Na,Ne)$ URCA pair is reached. During the TIR the central $\rm ^{25}Mg$ mass fraction increases by $\sim 3$ orders of magnitude, i.e., from ${\rm log}~X_{\rm c}\simeq -5.5$ to ${\rm log}~X_{\rm c}\simeq -2.5$ while the central temperature increases from ${\rm log} \left[T_{\rm c}~{(\rm K)}\right] \simeq 8.3$ to ${\rm log} \left[T_{\rm c}~{(\rm K)}\right] \simeq8.75$ (Figure \ref{mg25cent}). As for the $\rm 9.20~M_\odot$ also in this case the TIR is followed by a phase in which the central abundance of the leading isotope (in particular the $\rm ^{23}Na$ for the $\rm 9.20~M_\odot$ and the $\rm ^{25}Mg$ for the $\rm 9.15~M_\odot$ cases, respectively) decreases progressively and the extra energy provided by the TIR is progressively reabsorbed. This stage coincides also with the onset of the thermal pulses. We stopped the calculation during this phase after the completion of 193 thermal pulses (see Table \ref{tpprop915}).

\begin{figure*}[ht!]
\epsscale{1.0}
\plotone{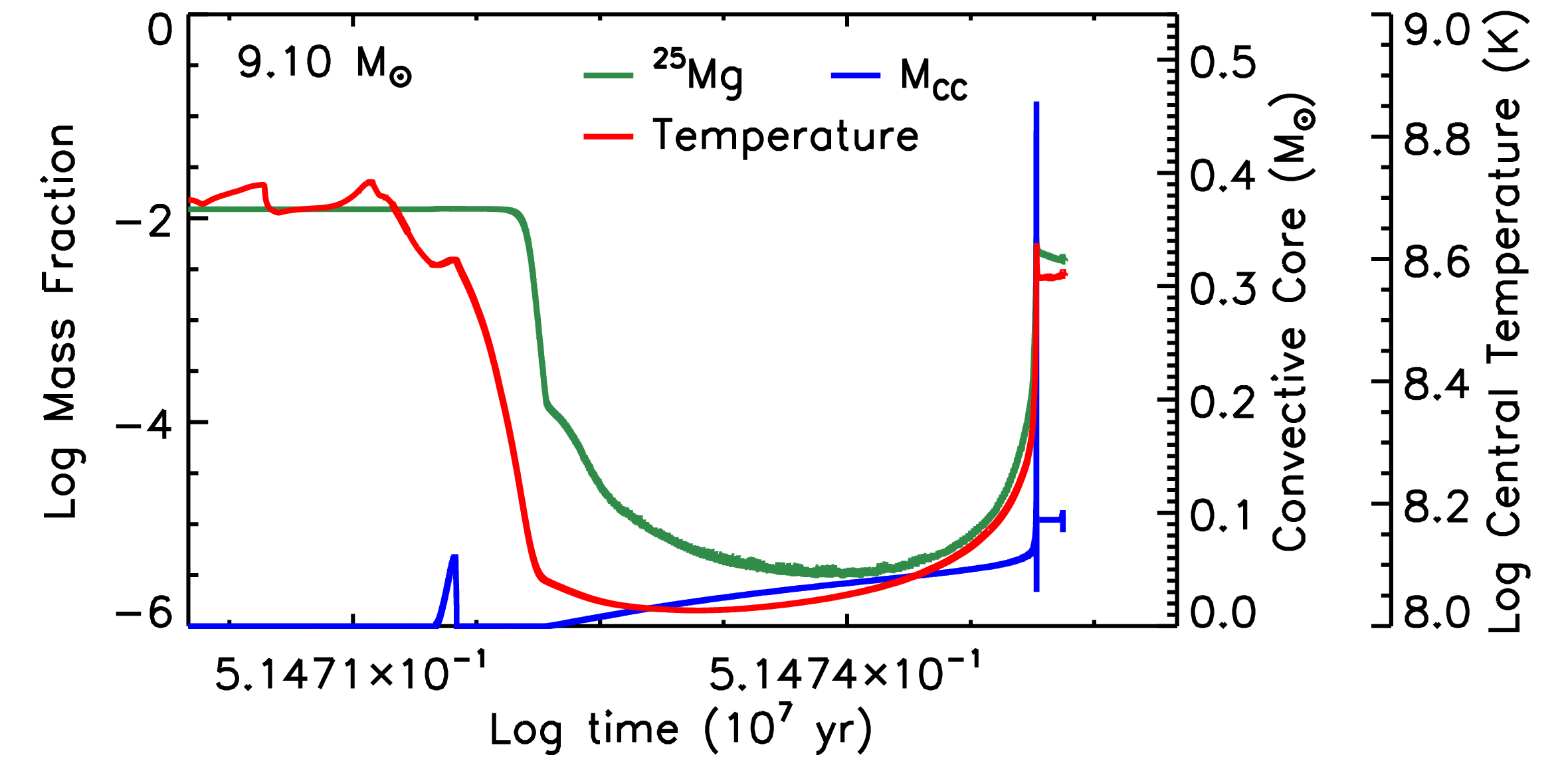}
\caption{Same as Figure \ref{mg25cent} but for the $\rm 9.10~M_\odot$ model.\label{mg25cent910}}
\end{figure*}

The evolution of the $\rm 9.10~M_\odot$ star is similar to the one of the $\rm 9.15~M_\odot$ (Figure \ref{mg25cent910}) and it is followed for 159 thermal pulses.

\subsection{Final fate of stars with initial mass $\rm 7.50\leq
M/M_\odot\leq 9.20$}\label{finalfate}

During the thermal pulses phase the CO core is continuously increased
by the alternate advancing of the He- and H-burning shells. Such an
occurrence induces an increase of the central density.
%that roughly follows the central density vs total mass relation for a
%completely degenerate structure, the total mass being in this case
%the CO core mass.

During the same stage, however, the star looses mass due to stellar
wind and this induces a progressive reduction of the H-rich
envelope. If the CO core mass reaches the value ($\rm M_{CO-ec}$)
corresponding to a central density close to the threshold value for
the activation of the $\rm ^{24}Mg(e^{-},\nu)^{24}Na$ before the
H-rich envelope is completely lost, the core contracts rapidly until
the density approaches the threshold value for the activation of the
$\rm ^{20}Ne(e^{-},\nu)^{20}Fe$ and then the star can potentially
explode as an electron capture supernova \citep{Miyaji+80,Nomoto87,Zha+19}.
%Miyaji et al.(1980), Nomoto (1987), and Zha et al. (2019) found
% it takes only less than $\sim 3 \times 10^3$ yr from the activation
% of $\rm ^{24}Mg(e^{-},\nu)^{24}Na$ to that of $\rm
% ^{20}Ne(e^{-},\nu)^{20}Fe$.
If, on the contrary, the H-rich envelope is completely
lost before the activation of the electron captures on $\rm ^{24}Mg$,
then the final fate of the star will be a ONeMg white dwarf
\citep{Nomoto84}. A self-consistent determination of the competition
between the increase of the CO core mass and the reduction of the
H-rich envelope due to the mass loss would require the calculation of
several thousands of thermal pulses that, at present, is not
feasible. Therefore an estimate of the final fate of these stars must
necessary relies on an "extrapolated" evolution.

\begin{figure*}[ht!]
\begin{center}
\includegraphics[width=.90\linewidth]{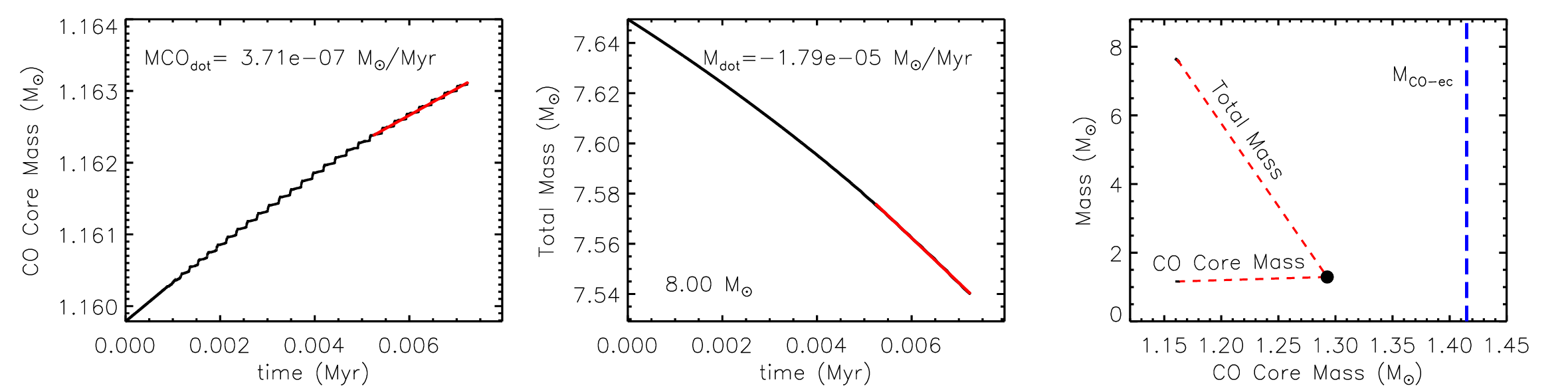}
\\[\baselineskip]% adds vertical line spacing
\includegraphics[width=.90\linewidth]{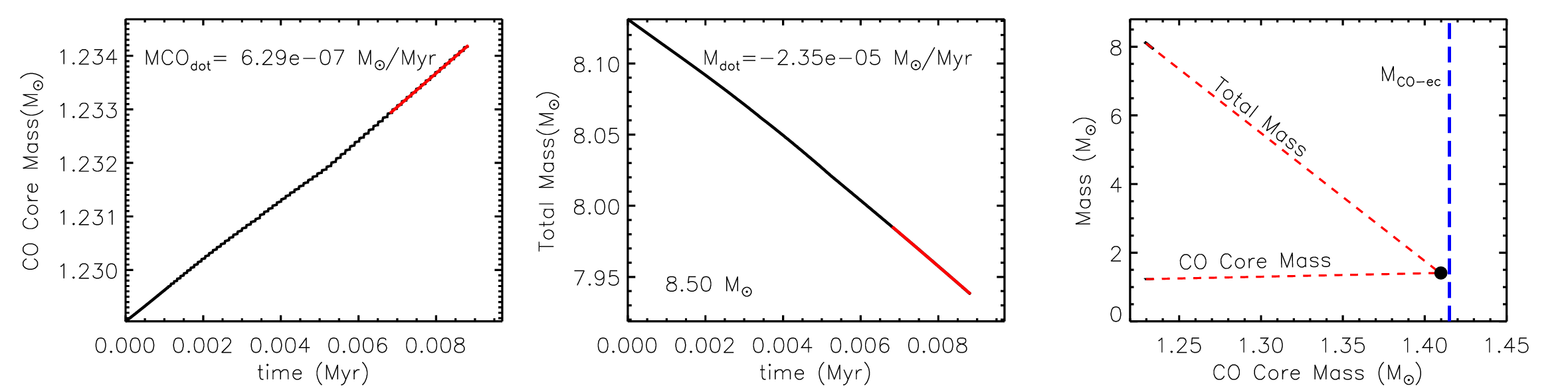}
\\[\baselineskip]% adds vertical line spacing
\includegraphics[width=.90\linewidth]{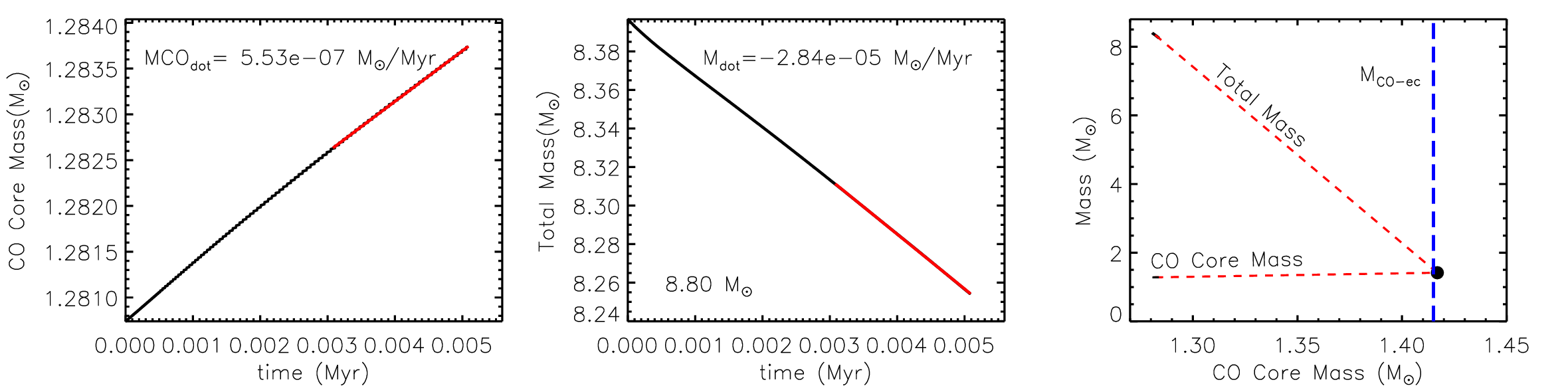}
\\[\baselineskip]% adds vertical line spacing
\includegraphics[width=.90\linewidth]{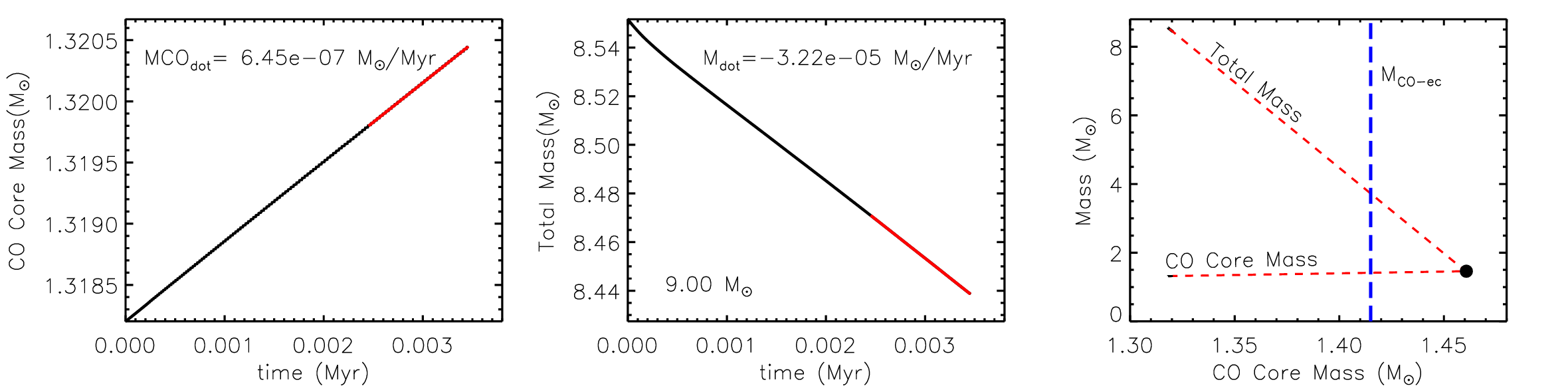}
\\[\baselineskip]% adds vertical line spacing
\end{center}

\caption{CO core mass (left panels) as a function of time, total mass
 (middle panels) as a function of time and total mass as a function
 of the CO core mass (right panels) for the $\rm
 8~,8.50,~8.80,~and~9.00~M_\odot$ models (from the upper to the lower
 line). The time has been reset at the beginning of the thermal
 pulses. The solid red line in the left and middle panels refers to
 the linear regression of the black line over the last few thermal
 pulses, superimposed to the black line itself. The values reported
 in the plots ($\rm MCO_{\rm dot}$ and $\rm M_{\rm dot}$) refer to
 the rate of growth of the CO core and the rate of mass loss obtained
 with the linear regression. The dashed lines in the right panels
 refers to the extrapolation at late times of the various quantities
 shown in the figure, obtained with the linear regression mentioned
 before. The vertical blue dashed line marks the CO core mass
 corresponding to the central density threshold for the activation of
 the electron capture on $\rm ^{24}Mg$ derived as discussed in the
 text \citep{Zha+19}.
%The vertical green dashed line refers to the mass of the ONe core at
%the onset of the electron capture on $\rm ^{20}Ne$ found by
\label{finalfateall}}
\end{figure*}

\begin{figure*}[ht!]
\epsscale{0.8}
\plotone{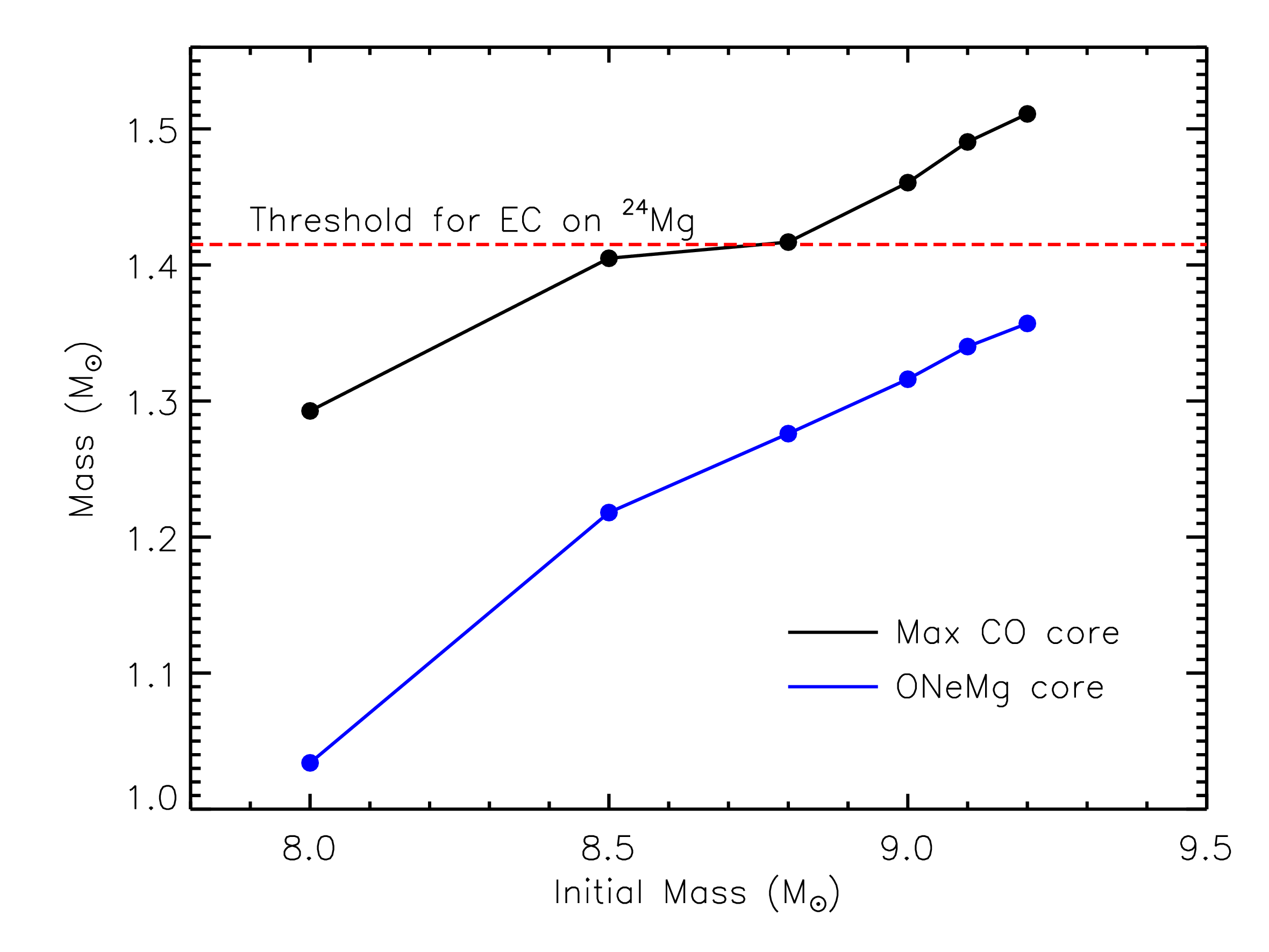}
\caption{Final CO and ONeMg core masses obtained with the "extrapolated
evolution" based on a linear regression (see text). Also shown is the
CO core mass corresponding to the threshold central density for the
activation of the electron capture on $\rm ^{24}Mg$.\label{fig:mcomax}}
\end{figure*}

Figure \ref{finalfateall} shows the time evolution of the CO core mass
(left panels) and of the total mass (central panels) for some selected
models, i.e., the $\rm 8.00~M_\odot$, $\rm 8.50~M_\odot$, $\rm
8.80~M_\odot$ and $\rm 9.00~M_\odot$, starting from the beginning of
the thermal pulses phase. These two quantities show an almost linear
behavior in the last part of the evolution that can be very well
approximated by a linear regression (red lines in the left and middle
panels of the figure). In the above mentioned panels we also show the
average values of the CO core mass growth rate and of the mass loss
rate obtained by such a linear regression. Under the assumption that
the evolution following the last computed model will remain
self-similar, we can easily extrapolate these quantities at late times
(dashed lines in the right panels in the same figure). We are aware
that when the envelope becomes sufficiently small the strength of the
pulses may change and therefore also the behavior of the CO core mass
and of the total mass may change accordingly. However, the importance
of these effects, if they really exist, is difficult to predict a
priori, therefore, as a working hypothesis, we assume a self similar
behavior of the relevant quantities up to the end of the
evolution. The intersection of the two (extrapolated) lines,
corresponding to the total mass and the CO core mass, is the maximum
CO core mass ($\rm M_{CO-max}$, marked in the right panels of the
figure with a black dot) that can be potentially formed before the
envelope of the star is completely removed by the stellar mass
loss. This quantity should be compared with $\rm M_{CO-ec}$, as
defined above. An estimate of this last quantity can be obtained by
solving the stellar structure equations for a completely degenerate
star with a given mass $\rm M$ and a chemical composition typical of
the zones interior to the CO core. In particular, we have taken the
internal composition of the $\rm 9.00~M_\odot$ star model as a
representative one, being that slight variations of the chemical
composition does not affect significantly the total mass-central
density relation obtained in this way. By adopting the public code
provided by F. Timmes \footnote{available at the web site $\rm
https://cococubed.com/code\_pages/coldwd.shtml$}, we find that the
mass corresponding to the threshold density ($\rm log
\left[\rho_c~(g~cm^{-3})\right] =9.6$) for the activation of the 
$\rm ^{24}Mg(e^{-},\nu)^{24}Na$ is $\rm M_{CO-ec}=1.415~M_\odot$ while the
Chandrasekhar mass is $\rm M_{Ch}\simeq1.45~M_\odot$.  $\rm M_{CO-ec}$
is shown in the right panels of Figure \ref{finalfateall} by a
vertical blue dashed line. The right panels of the figure show that
the minimum mass that can potentially explodes as an ECSN is $\rm \sim
8.5-8.8~ M_\odot$.
%{\ bf Note that $\rm M_{CO-ec}$ depends on the $Y_e$ distribution in
 the core.  
 \cite{Zha+19} obtained $\rm M_{CO-ec}=1.36~M_\odot$
 in their evolutionary model, which pushes the minimum mass that can
 potentially explode as an ECSN to $\rm \sim 8.3~M_\odot$ in Figure
 \ref{fig:mcomax}.
It is worth noting that for any given mass, the density obtained
assuming that the structure is fully degenerate is the minimum one,
the reason being that a progressive departure from degeneration allows
for a progressively higher contraction and therefore larger central
densities (for the same mass). This implies that the value of $\rm
M_{CO-ec}$ marked by the blue dashed lines in the above mentioned
figures constitutes an upper limit to this quantity.

%Indeed, \cite{Zha+19} computing the evolution of a CO core accreting
% mass with a rate consistent with the values reported here, found a
% value of $\rm 1.36~M_\odot$, i.e., slightly lower than $\rm
% M_{CO-ec}$ that pushes the minimum mass that can potentially explode
% as ECSN to values in the range $\rm \sim 8.0-8.5~ M_\odot$. The
% difference between the two values of the critical ONe core masses
% above mentioned can be considered as a typical uncertainty on this
% critical quantity.

Figure \ref{fig:mcomax} shows the $\rm M_{CO-max}$ as a function of
the initial mass compared to the $\rm M_{CO-ec}$. Also shown in the
figure is the final ONeMg core mass as a function of the initial mass
under the assumption that the ONeMg core does not increase during the
thermal pulses phase because the accretion rate of the CO core is not
high enough to induce further C burning \citep{NomotoIben85}.

Taking into account all the possible uncertainties, we conclude that
stars in the range $\rm 7.50\leq M/M_\odot\leq 8.00$ will loose their
H-rich envelope before the threshold density for the electron capture
on $\rm ^{24}Mg$ is achieved and therefore they will produce a
ONeMg-WD. Stars in the range $\rm 8.50\leq M/M_\odot\leq 9.20$ on the
contrary will reach such a critical density before the H envelope
reduces enough to quench definitely the H burning shell. Once the $\rm
^{24}Mg(e^{-},\nu)^{24}Na$ is activated the final fate of these stars
(explosion or collapse to a neutron star) depends on both the details
of the explosion modeling and of the initial conditions (see section
\ref{sec:intro}) and cannot be predicted with certainty in this
work. As a final comment, we point out that in stars with the initial
mass $\rm 9.10\leq M/M_\odot\leq 9.20$ the central temperature
increases substantially due to the TIR and therefore the ignition of
the $\rm ^{20}Ne$ photodisintegration before the activation of the
electron capture on $\rm ^{24}Mg$ cannot be excluded.  In that case it
is difficult to predict, a priori, the final fate of these stars.

%on the competition between the energy released by the nuclear burning
% front and the loss of pressure due to the deleptonization occurring
% in the central zones. If the energy released by nuclear burning
% prevails, degeneracy is removed and the star is blown up as a
% thermonuclear electron capture supernova (tECSN) (Miyaji et al. 1980,
% Nomoto \& Kondo 1991; Isern et al. 1991; Canal et al. 1992; Jones et
% al. 2016a; Nomoto \& Leung 2017). If, on the contrary, the
% deleptonization dominates, the collapse cannot be reversed by the
% thermonuclear burning and the star collapses into a neutron star
% (Miyaji et al. 1980; Miyaji \& Nomoto 1987; Nomoto1987; Kitaura et
% al. 2006; Fischer et al. 2010; Jones et al. 2016a). In this case we
% have a collapsing electron capture supernova (cECSN). Which outcome
% is realized depends on both the details of the explosion modeling and
% the initial conditions and cannot be predicted in this work.

\subsection{Evolution toward core collapse: stars with $\rm M\geq 9.22~M_\odot$ $\rm \left( M_{CO}\geq 1.08~M_\odot \right)$}
\begin{figure*}[ht!]
\epsscale{0.8}
\plotone{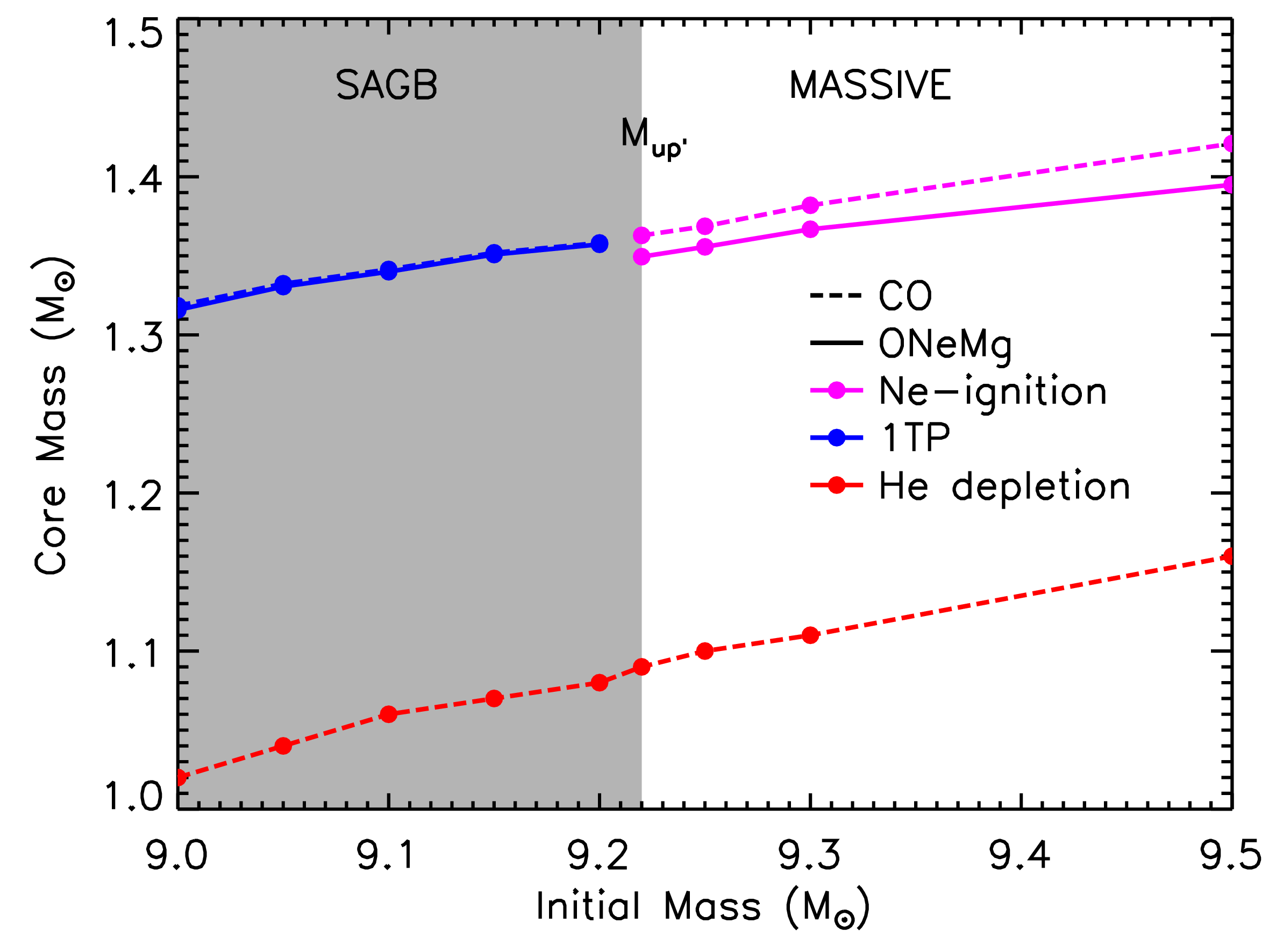}
\caption{CO core mass (dahsed line) and ONeMg core mass (solid line) as a function of the initial mass at various evolutionary stages: core He depletion (red line and filled dots); first thermal pulse (blue line and filled dots); Ne ignition (magenta line and filled dots).\label{fig:mcore_one_zoom}}
\end{figure*}

\begin{figure*}[ht!]
\begin{center}
\includegraphics[width=.45\linewidth]{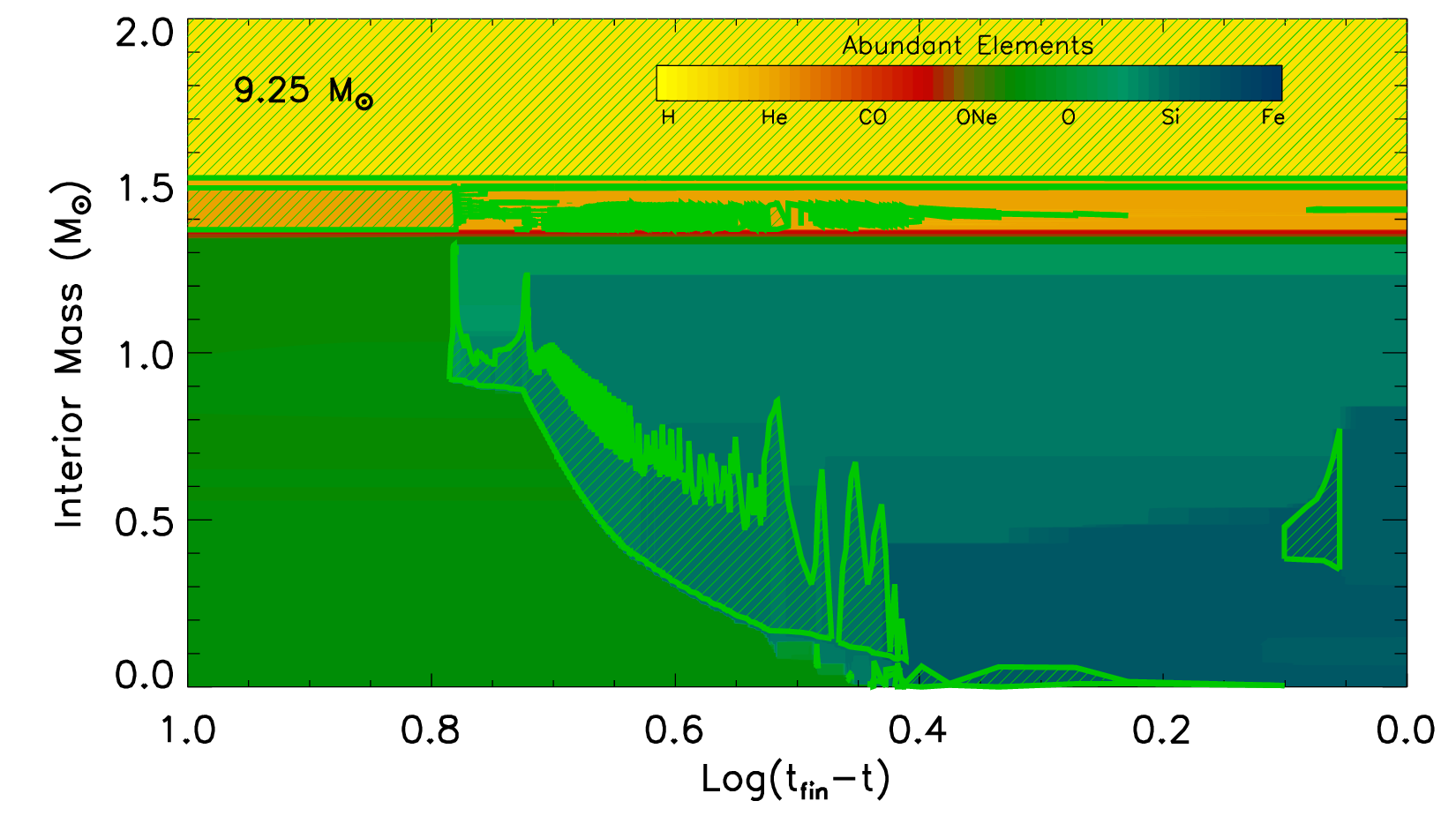}\quad\includegraphics[width=.45\linewidth]{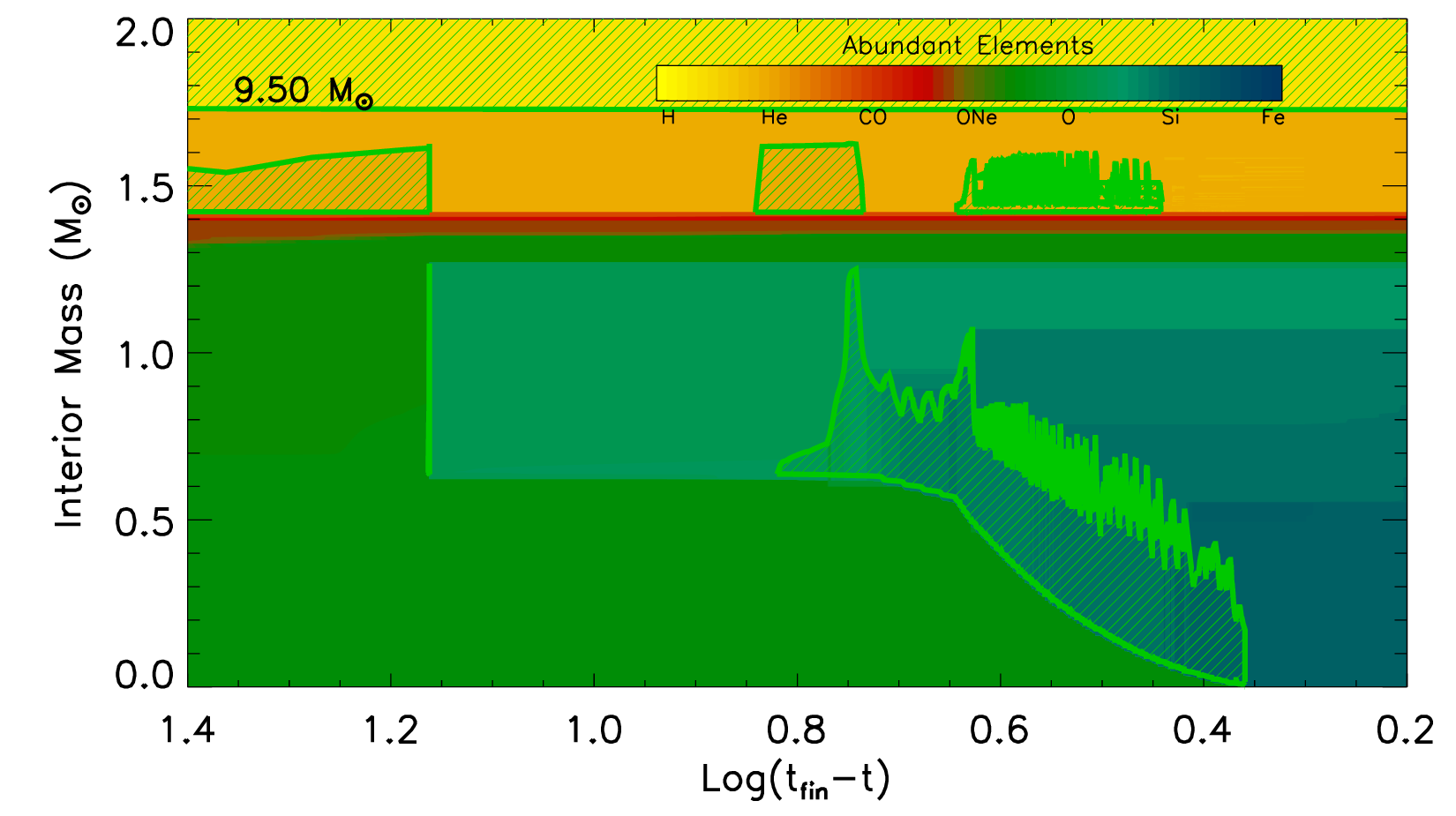}
\end{center}
\caption{Convective (green shaded areas) and chemical (color codes reported in the color bar) internal history during off-center Ne burning for the $\rm 9.25~M_\odot$ (left panel) and the $\rm 9.50~M_\odot$ (right panel) model.In the x-axis is reported the logarithm of the time till the end of the evolution ($\rm t_{\rm fin}-t$) in units of yr.\label{kipneonmassicce}}
\end{figure*}

Stars with the initial mass $\rm M\geq 9.22~M_\odot$ form an ONeMg core in which the maximum temperature reaches the threshold value for the Ne ignition. The thermal behavior of the ONeMg core depends on both the behavior of the C burning shell and the convective history of the CO core that, in turn, depend in general on the CO core mass at core He depletion. In the present set of models we find that the minimum CO core mass at core He depletion for the activation of Ne burning is $\rm M_{CO}= 1.08~M_\odot$, that corresponds to a CO core and an ONeMg core masses at Ne ignition of $\rm M_{CO}=1.363~M_\odot$ and $\rm M_{ONeMg}=1.349~M_\odot$, respectively (Figure \ref{fig:mcore_one_zoom} and Table \ref{tab_main_prop2}). It is interesting to note that Ne ignition is activated before the ONeMg core (which coincides with the C burning shell by definition) approaches the CO core (i.e. the He burning shell), as it happens in all the models that do not ignite Ne and evolve through the SAGB phase. This is the reason why the ONeMg core mass versus the initial mass relation shows a small bending in the transition between SAGB stars and stars that do ignite Ne burning (Figure \ref{fig:mcore_one_zoom}). 
As in the case of C ignition, the mass coordinate corresponding to the Ne ignition decreases progressively as the initial mass increases, ranging from $\rm 0.966~M_\odot$ for the $\rm 9.22~M_\odot$ to 0 for the $\rm 13~M_\odot$, which is the lowest mass that ignites Ne at the center (Figure \ref{fig:mc_ign} and Table \ref{tab_main_prop2}). Off-center Ne burning is ignited under conditions of sizable degeneracy ($\psi \sim 7-5$, in the mass range $\rm 9.22-12.00~M_\odot$), therefore the local nuclear energy release drives a progressive increase of both the temperature and the luminosity and as a consequence the formation of a convective zone. Such a convective zone reaches a maximum extension and then tends to recede in mass as the Ne is progressively depleted. In the lower mass models ($\rm 9.22-9.30~M_\odot$), during this phase, the temperature approaches values as high as ${\rm Log}T~{\rm (K)} \sim 9.3$ and therefore O burning is also ignited before convection quenches. This drives the convective zone to increase again up to a maximum extension. After the O ignition, the Ne/O burning proceeds simultaneously in a convective shell that progressively moves toward the center as the fuel is locally exhausted, the temperature increased and the degeneracy significantly removed (left panel of Figure \ref{kipneonmassicce}). In the more massive models ($\rm 9.50-12.0~M_\odot$), on the contrary, the local temperature does not reach the threshold values for O ignition and therefore the first convective zone quenches and disappears as the Ne is depleted locally. After this first convective episode, contraction resumes and another convective zone forms. 
%During this phase the temperature reaches the threshold value for the ignition of O-burning and therefore. 
From this time onward the evolution of the Ne/O burning front in these more massive models is similar to the one described above for the lower mass stars (see right panel of Figure \ref{kipneonmassicce}).
%During this phase in the lower mass models ($\rm 9.22-9.30~M_\odot$) the temperature approaches values as high as ${\rm Log}T\sim 9.3$ and therefore O burning is also ignited. In the more massive models ($\rm 9.50-12.0~M_\odot$), on the contrary, the local temperature does not reach the threshold values for O ignition. Once the main burning fuel is depleted in the shell the convection is quenched, the degeneracy is locally removed and the burning front moves inward toward zones where the fuel is still abundant and the degeneracy degree is still high, driving the formation of another convective zone. From this time onward, in all the models Ne and O burning proceed simultaneously within such a convective zone that progressively moves toward the center (Figure \ref{kipneonmassicce}) as the fuel (Ne/O) is locally exhausted, the temperature increased and the degeneracy removed.
%\begin{figure*}[ht!]
%\begin{center}
%\includegraphics[width=.45\linewidth]{kip922_neon.eps}\quad\includegraphics[width=.45\linewidth]{kip950_neon.eps}
%\\[\baselineskip]% adds vertical line spacing
%\includegraphics[width=.45\linewidth]{kip1100_neon.eps}\quad\includegraphics[width=.45\linewidth]{kip1500_neon.eps}
%\end{center}
%\caption{.....\label{kipneonmassicce}}
%\end{figure*}
%As in the case of the off-center C ignition, also in this case the propagation of the ONe burning front toward the center is mainly driven by the heat transfer from the burning zone at high temperature, located at the base of the convective shell, and the cooler radiative zones that lie below. This process is again mainly controlled by both the efficiency of the CBF at the base of the convective shell (see above). 
\begin{figure*}[ht!]
\epsscale{0.8}
\plotone{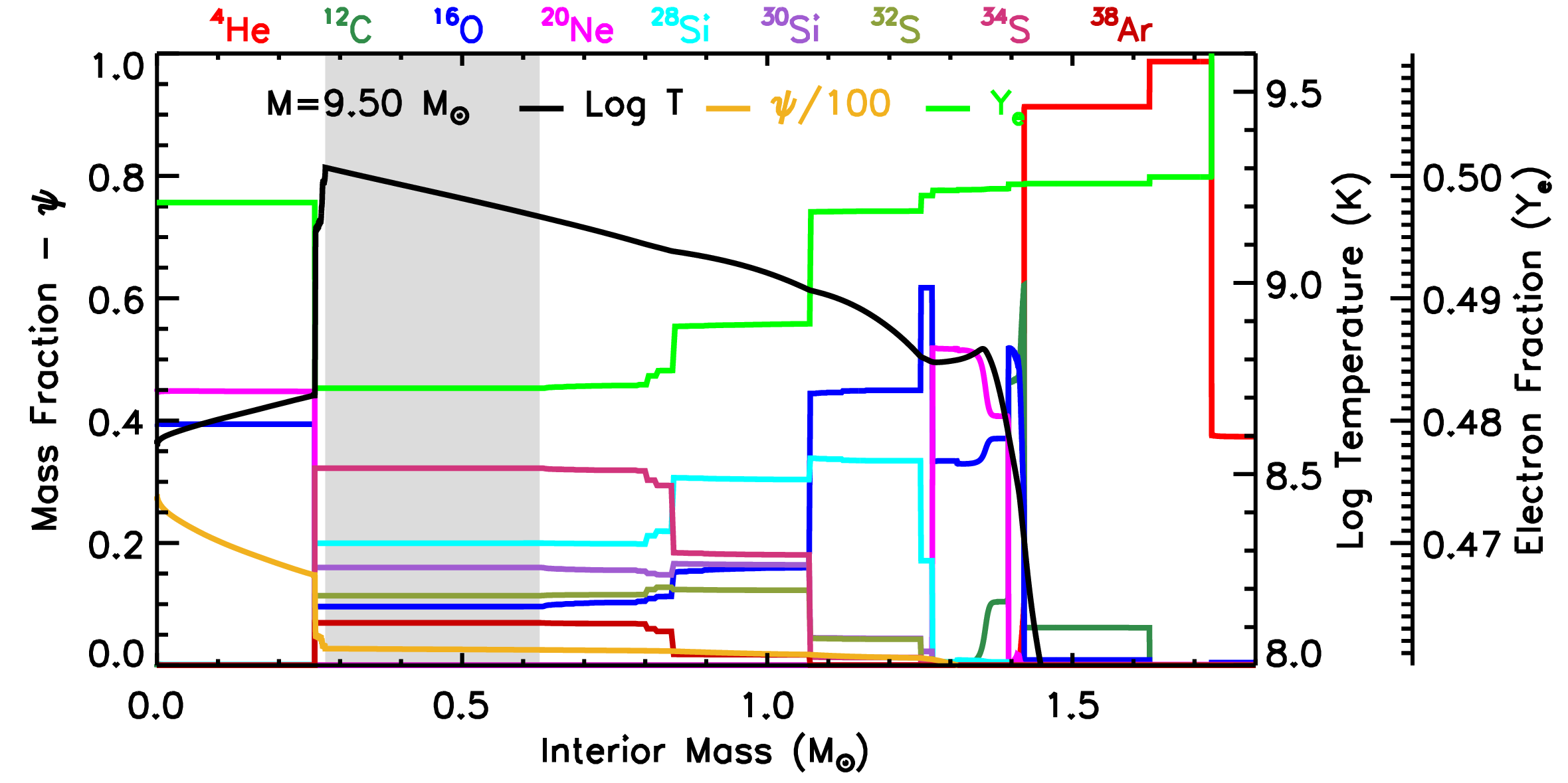}
\caption{Selected chemical and physical properties (see the legend) of the $\rm 9.50~M_\odot$ model during the off-center Neon burning.\label{offneburn-950-0100000}}
\end{figure*}

\begin{figure*}[ht!]
\begin{center}
\includegraphics[width=.45\linewidth]{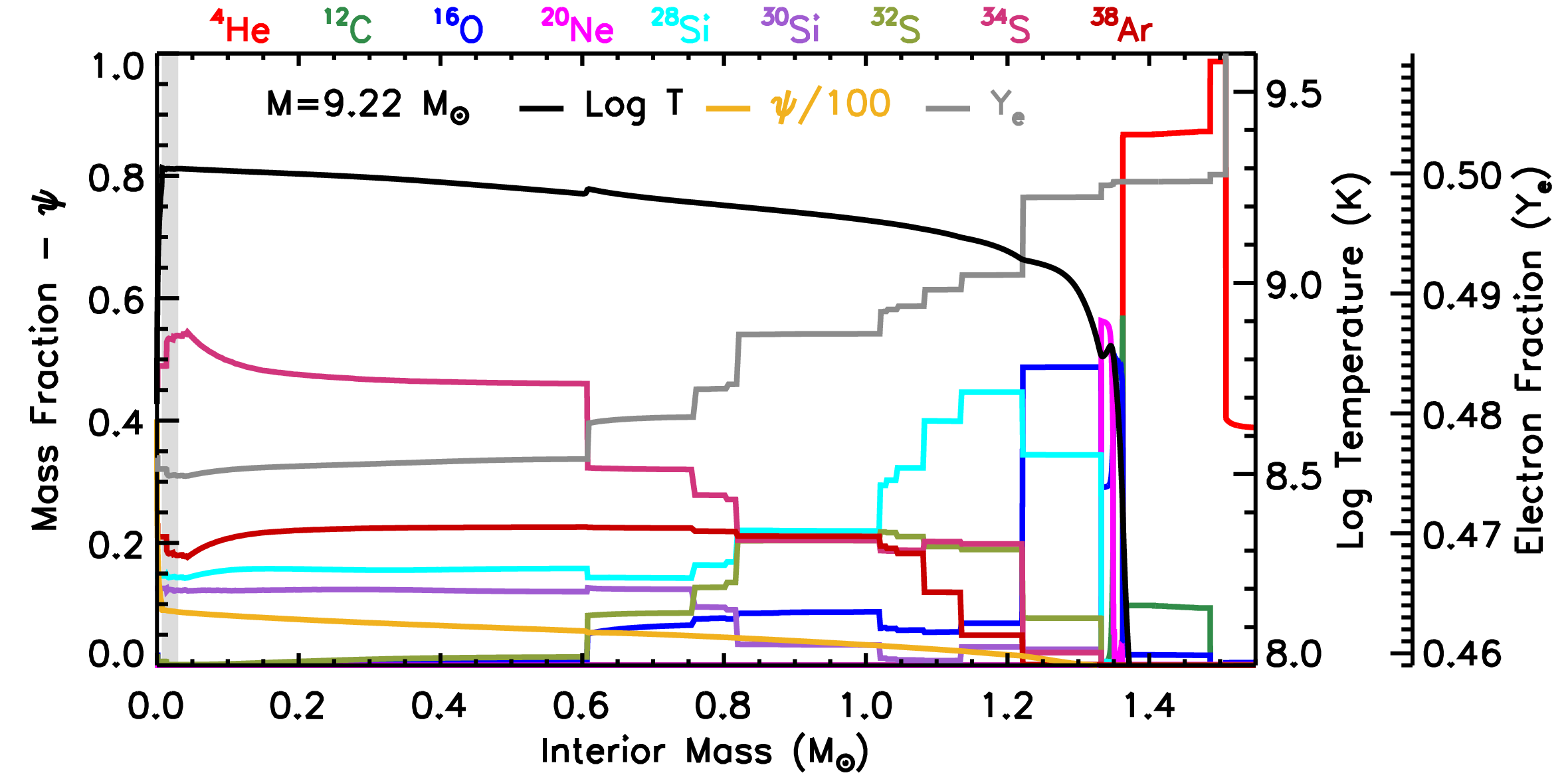}\quad\includegraphics[width=.45\linewidth]{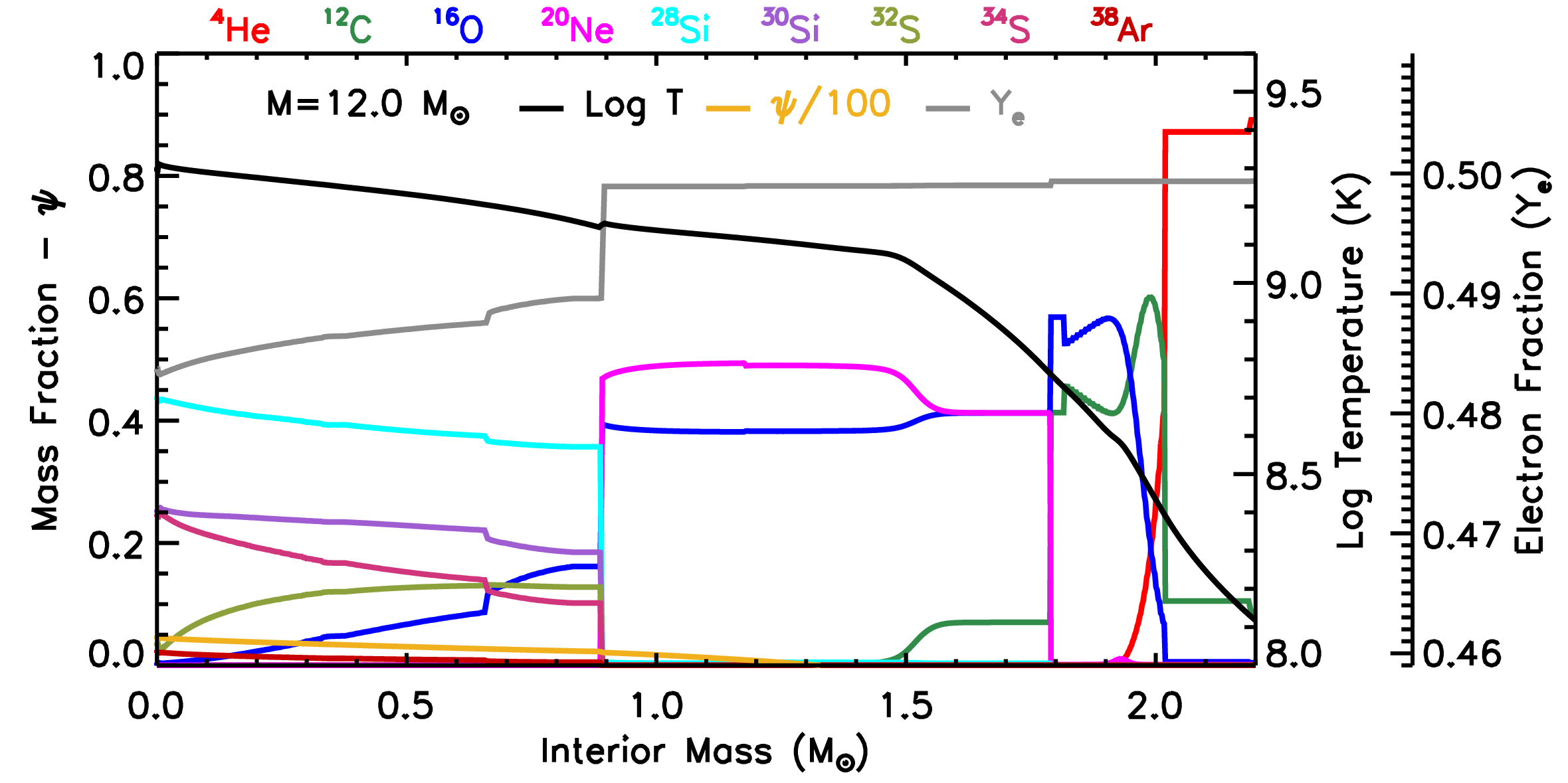}
\end{center}
\caption{Selected chemical and physical properties (see the legend) of the $\rm 9.22~M_\odot$ (left panel) and the $\rm 12.0~M_\odot$ models when the ONe burning front has reached the center.\label{endneoburning}}
\end{figure*}

Figures \ref{offneburn-950-0100000} shows the main properties of a typical model during the propagation of the Ne/O burning front toward the center. The burning is occurring at the base of the convective shell, marked by the grey area, which is at high temperature compared to the inner, much cooler, radiative zones. Because of the efficient electron captures, the main products of the Ne/O burning within the convective shell are $\rm ^{34}S$, $\rm ^{28}Si$, $\rm ^{30}Si$ and $\rm ^{32}S$.
%, to be compared with the products of O burning in a typical massive star ($\rm M\sim20~M_\odot$) that are dominated by $\rm ^{28}Si$ and $\rm ^{32}S$ because of the much lower neutronization of the matter. 
The efficiency of the electron captures, however, decreases as the initial mass of the star increases, therefore the chemical composition left by the Ne/O burning tends to be dominated by less neutron rich isotopes as the initial mass of the star increases. Figure \ref{endneoburning} shows the chemical composition of selected models once the Ne/O burning front has reached the center.
%\begin{figure*}[ht!]
%\begin{center}
%\includegraphics[width=.45\linewidth]{offneburn-922-0182800.eps}\quad\includegraphics[width=.45\linewidth]{offneburn-950-0182800.eps}
%\\[\baselineskip]% adds vertical line spacing
%\includegraphics[width=.45\linewidth]{offneburn-1100-0021400.eps}\quad\includegraphics[width=.45\linewidth]{offneburn-1300-0019600.eps}
%\end{center}
%\caption{.....\label{endneoburning}}
%\end{figure*}

In the $\rm 13.0~M_\odot$ model, Ne burning is ignited at the center and develops in a convective core. Once Ne is depleted in the center the burning shifts outward in mass, in the region a variable composition left by the receding convective core, and drives the formation a convective shell at a mass coordinate of $\rm \sim 0.18~M_\odot$. During this phase the temperature in the shell increases enough that O burning is ignited. Ne and O burning then proceed simultaneously in such a shell that increases progressively in mass up to a maximum extension of $\rm 0.17-1.00~M_\odot$. Once O is exhausted in the shell, the burning shifts inward and drives the formation of a convective core that reaches a maximum extension of $\rm \sim 0.07~M_\odot$ before disappearing at O depletion. Ne and O burning develop in the $\rm 15.0~M_\odot$ model as in a typical massive star.

It is interesting to note at this point that, at variance with off-center C burning, no hybrid CO/ONeMg core is formed as a result of the off-center Ne ignition. All the stars that ignite off-center Ne burning form an O depleted core, i.e., in all these models the ONe burning front reaches the center. This result is consistent with what has been found by \cite{WH2015} and can be understood because we are using a similar approach to treat the CBF. On the contrary, \cite{jones+13} found a case in which the ONe burning front does not propagate toward the center, leading the star to reach central densities high enough for the activation of the electron capture on $\rm ^{20}Ne$ and then to explode as ECSN. This different behavior can be due to the fact that \cite{jones+13} do not include in the code any specific treatment for the CBF, therefore their models cannot be directly compared to ours.

\begin{figure*}[ht!]
\epsscale{0.8}
\plotone{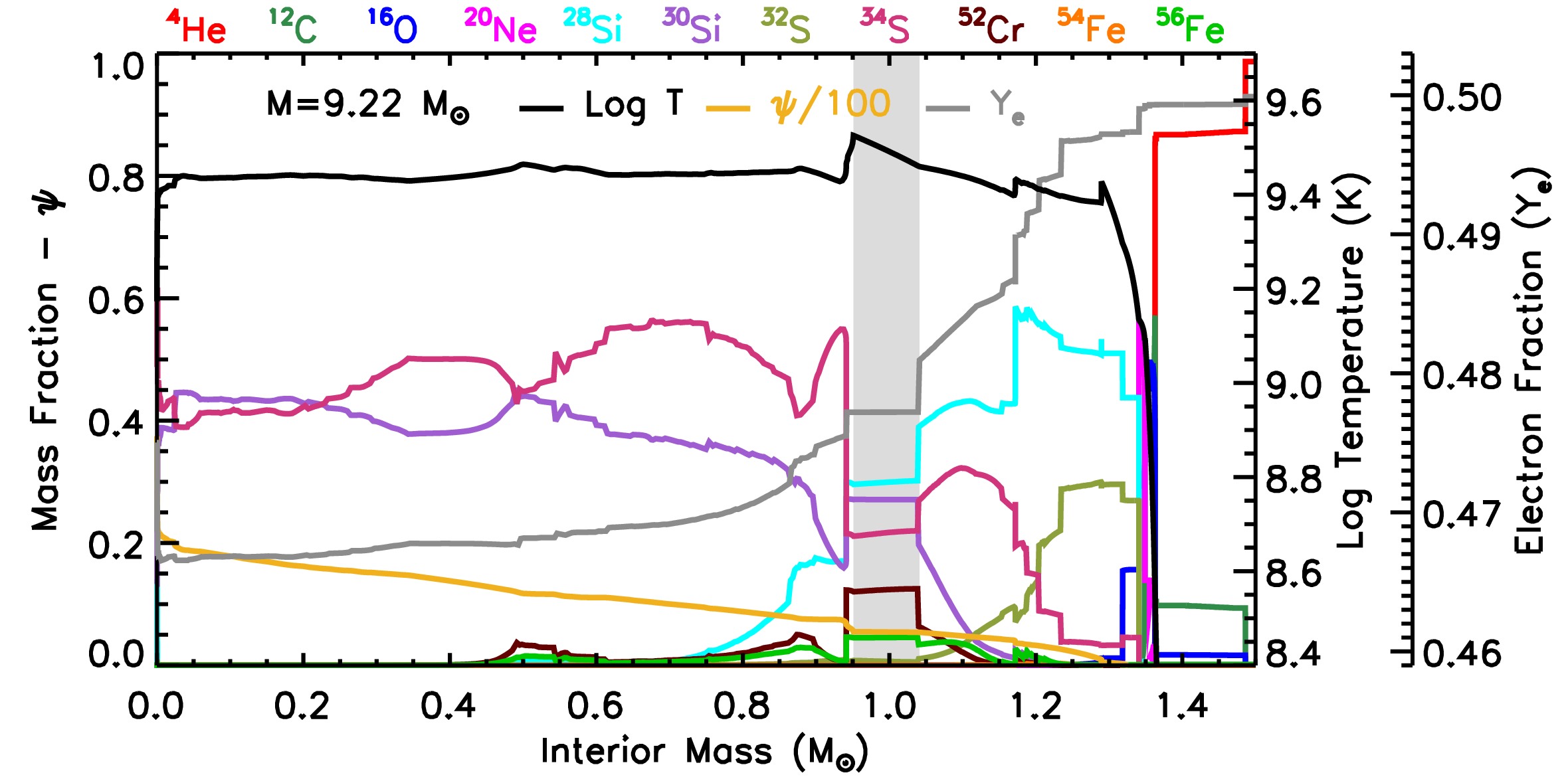}
\caption{Selected chemical and physical properties (see the legend) of the $\rm 9.22~M_\odot$  at off-center SiS ignition.\label{offsiburn-922-0187000}}
\end{figure*}

\begin{figure*}[ht!]
\epsscale{0.8}
\plotone{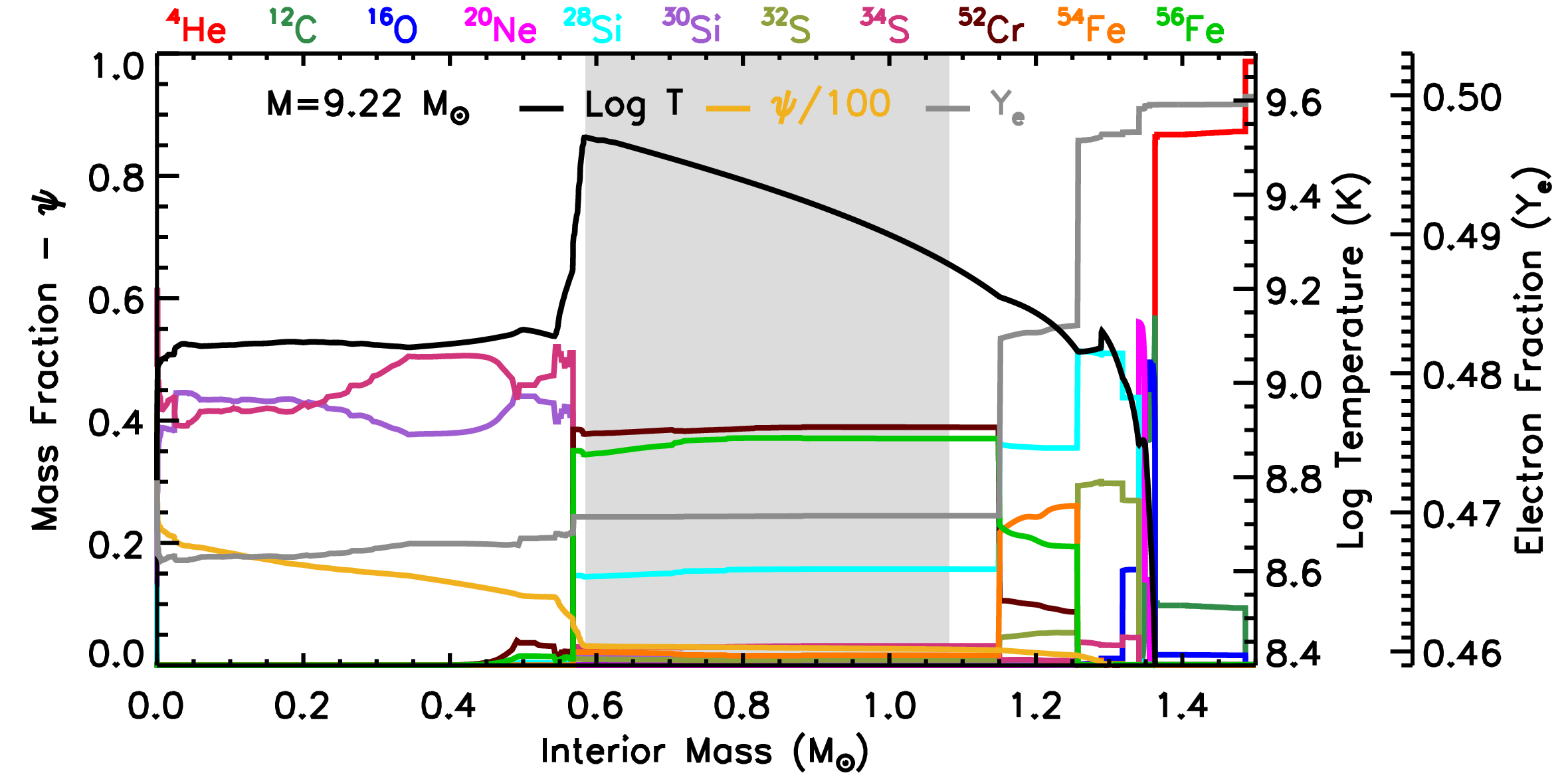}
\caption{Selected chemical and physical properties (see the legend) of the $\rm 9.22~M_\odot$ during the off-center SiS burning.\label{offsiburn-922-0192200}}
\end{figure*}

\begin{figure*}[ht!]
\epsscale{0.8}
\plotone{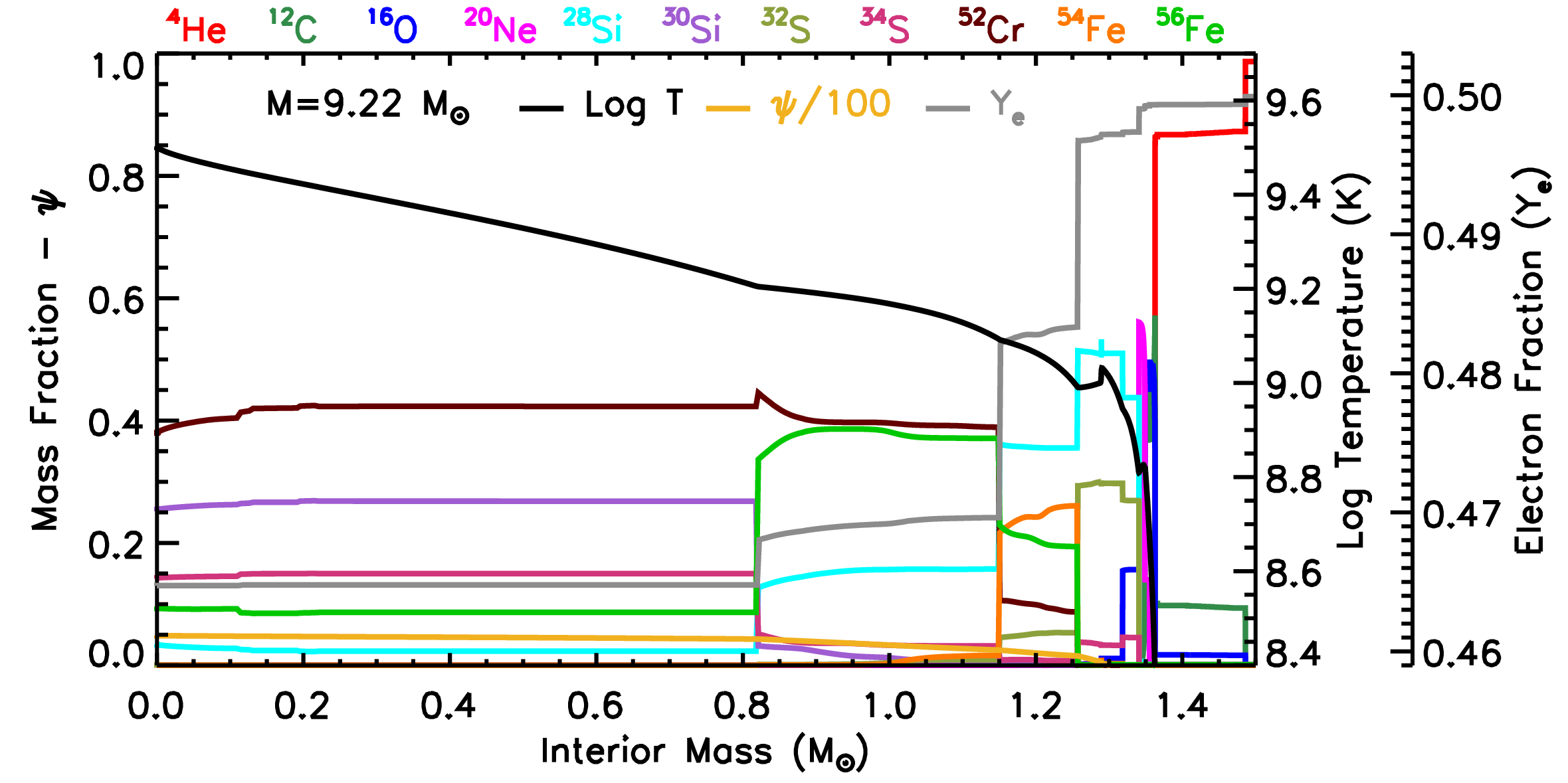}
\caption{Selected chemical and physical properties (see the legend) of the $\rm 9.22~M_\odot$ when the SiS burning front has reached the center.\label{offsiburn-922-0206300}}
\end{figure*}

\begin{figure*}[ht!]
\epsscale{0.8}
\plotone{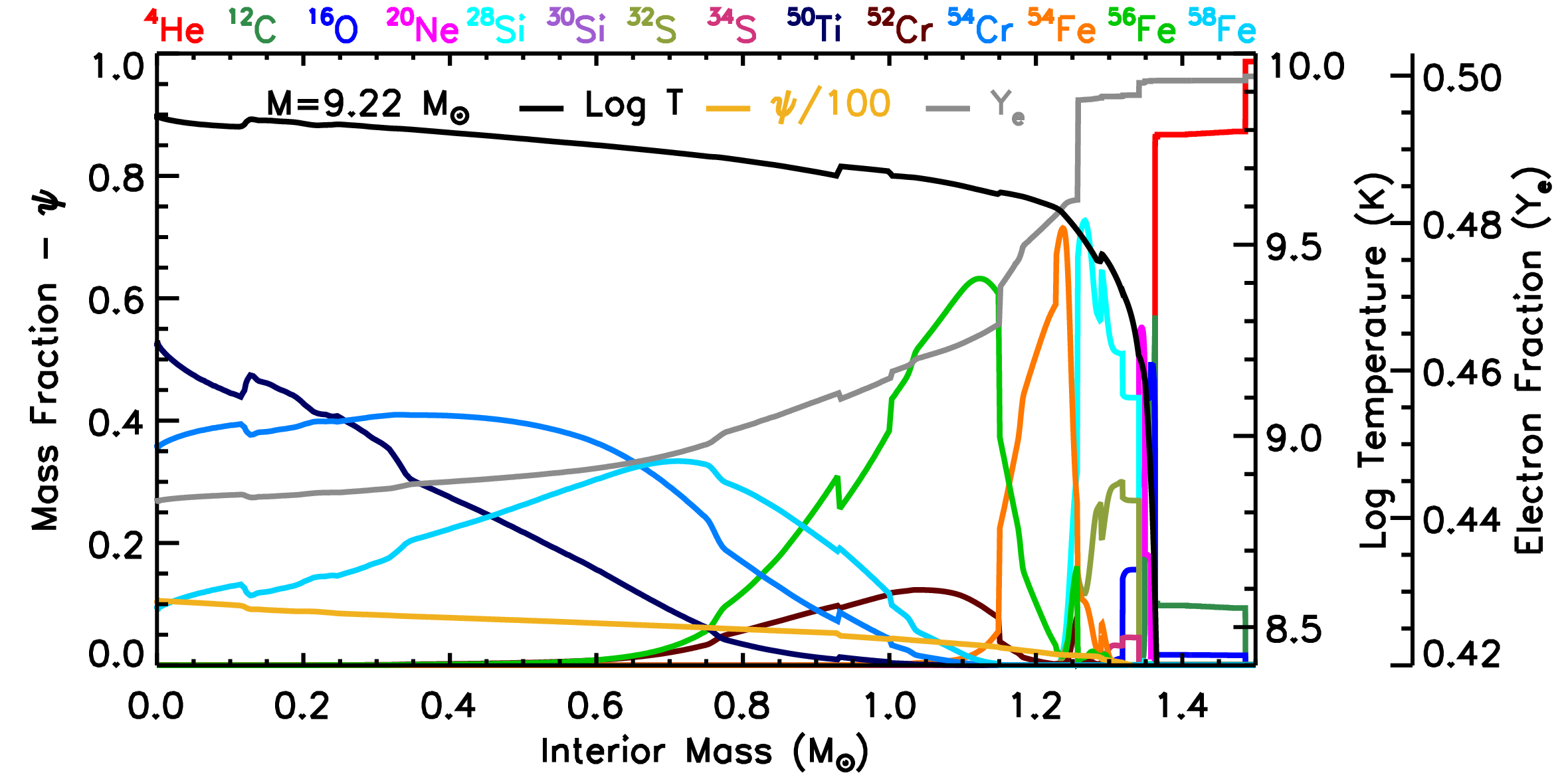}
\caption{Selected chemical and physical properties (see the legend) of the $\rm 9.22~M_\odot$ at the presupernova stage.\label{offsiburn-922-0209800}}
\end{figure*}

In the $\rm 9.22~M_\odot$, after the Ne/O burning front has reached the center the most abundant nuclear species in the O exhausted core are $\rm ^{34}S$ ($\sim0.48$), $\rm ^{38}Ar$ ($\sim0.22$), $\rm ^{28}Si$ ($\sim0.16$) and $\rm ^{30}Si$ ($\sim0.13$) (left panel of Figure \ref{endneoburning}). O burning, that shifts in a shell, settles at a mass coordinate of $\rm \sim 0.6~M_\odot$, where O is still quite abundant. Shell O burning moves outward in mass inducing the formation of three consecutive convective shells. During this phase, in the inner core, $\rm ^{38}Ar$ and $\rm ^{28}Si$ are converted into $\rm ^{34}S$ and $\rm ^{30}Si$ that increase to $\sim0.70$ and $\sim0.28$ in mass fraction, respectively. When the O burning shell has reached $\rm \sim 1.30~M_\odot$, a nuclear burning is ignited at $\rm \sim 0.95~M_\odot$, at a temperature of $\rm \sim 3\cdot 10^{9}~K$ (Figure \ref{offsiburn-922-0187000} shows the physical and chemical structure of the star at this stage). 
During the initial phase of this burning the rearrangement of the matter is such that $\rm ^{34}S$ and $\rm ^{30}Si$ are depleted while $\rm ^{28}Si$, $\rm ^{52}Cr$, $\rm ^{54}Fe$ and $\rm ^{56}Fe$ are produced in a convective shell that increases progressively in size. Once $\rm ^{34}S$ and $\rm ^{30}Si$ are exhausted in the shell, convection quenches and the nuclear burning front shifts inward in mass, where $\rm ^{34}S$ and $\rm ^{30}Si$ are still abundant, and induces the formation of a convective shell that, once again, reaches a maximum extension and then quenches. The burning front, then, propagates in this way progressively toward the center.
A typical model during this phase in shown in Figure \ref{offsiburn-922-0192200}. As the $\rm ^{34}S$-$\rm ^{30}Si$ burning front moves inward in mass the interplay between local burning and convective mixing is such that $\rm ^{52}Cr$ tends to be the dominant nuclear species, followed by $\rm ^{30}Si$ and $\rm ^{34}S$, not completely depleted, and finally by $\rm ^{56}Fe$. 
The physical and chemical structure of the star once the burning front has reached the center is shown in Figure \ref{offsiburn-922-0206300}. The residual $\rm ^{28}Si$ ($\sim 0.02$ in mass fraction) is then eventually burnt in a convective core that increases in size up to $\rm \sim 0.9~M_\odot$ and leaves a chemical composition dominated by $\rm ^{52}Cr$ ($\sim 0.60$) and $\rm ^{56}Fe$ ($\sim 0.28$).
During the following evolution the core contracts and heats up and the matter is converted to iron peak ("Fe") isotopes. The composition of the "Fe" core is dominated by the most abundant isotopes of matter at the nuclear statistical equilibrium corresponding to values of the temperature and the density progressively higher and of the electron fraction progressively lower due to the efficient electron captures. The "Fe" core mass at the presupernova stage is $\rm M_{Fe}=1.257~M_\odot$ and its composition is dominated by $\rm ^{50}Ti$, $\rm ^{54}Cr$ and $\rm ^{58}Fe$ (Figure \ref{offsiburn-922-0209800}). All the other relevant physical quantities of the model at the presupernova stage are reported in Table \ref{tab_main_prop2}.

The evolution of the stars in the range $\rm 9.25-12.0~M_\odot$ after the Ne/O burning front has reached the center and up to the presupernova stage is similar to that of the $\rm 9.22~M_\odot$. The only difference is the mass coordinate corresponding to the Si-S ignition. In particular, in the models with mass in the range $\rm 9.25-9.50~M_\odot$, the Si-S is ignited at a mass coordinate that progressively decreases as the mass increases, i.e., it is $\rm \sim 0.52~M_\odot$, $\rm \sim 0.07~M_\odot$ and $\rm \sim 0.005~M_\odot$ for the $\rm 9.25~M_\odot$, the $\rm 9.30~M_\odot$ and the $\rm 9.50~M_\odot$, respectively (see Table \ref{tab_main_prop2}). 
%Note that in the $\rm 9.30~M_\odot$ and the Si-S ignition is close enough to the center that the burning front reaches the center after the first convective shell episode. Another difference with the $\rm 9.22~M_\odot$ is the maximum extension of the $\rm ^{28}Si$ convective core that is $\rm \sim 1.05~M_\odot$, $\rm \sim 1.14~M_\odot$ and $\rm \sim 1.11~M_\odot$ for the  $\rm 9.25~M_\odot$, the $\rm 9.30~M_\odot$ and the $\rm 9.50~M_\odot$, respectively. The final "Fe core" are similar in these two models ($\rm M_{Fe}=1.263~M_\odot$ and $\rm M_{Fe}=1.260~M_\odot$) and similar to that of the $\rm 9.22~M_\odot$ (see Table \ref{tab_main_prop2}).
\begin{figure*}[ht!]
\epsscale{0.8}
\plotone{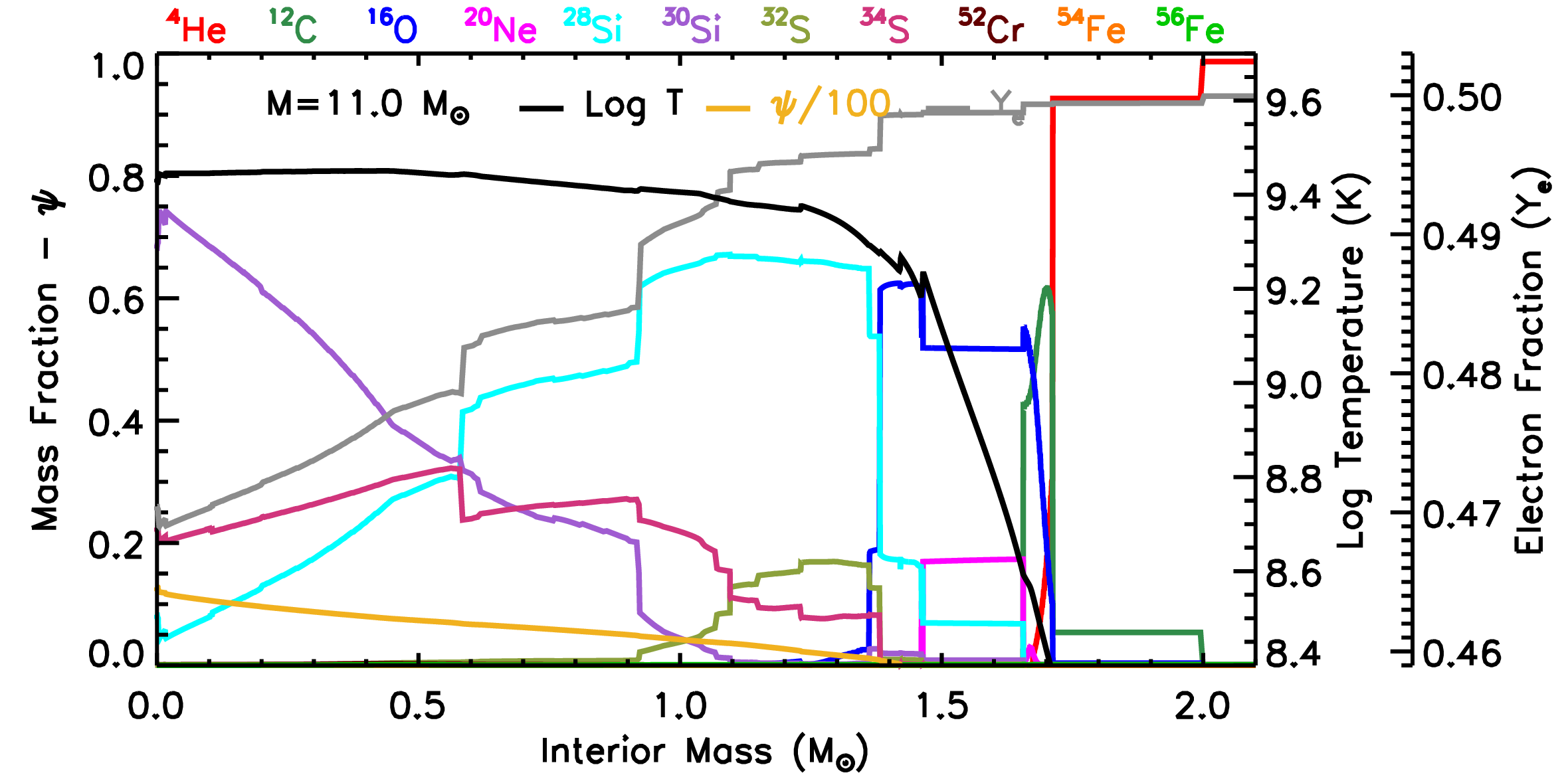}
\caption{Selected chemical and physical properties (see the legend) of the $\rm 11.0~M_\odot$ during the shell O burning phase.\label{offsiburn-1100-0026200}}
\end{figure*}
In the models with mass in the range $\rm 9.80-12.0~M_\odot$, the $\rm ^{28}Si$ is not completely exhausted in the inner core during the shell O burning, as it happens in the lower mass models, and it shows a gradient (see Figure \ref{offsiburn-1100-0026200}). The sizeble abundance of $\rm ^{28}Si$ and its profile induces an off-center nuclear ignition at a mass coordinate that decreases as the initial mass increases, ranging from $\rm \sim 0.387~M_\odot$ for the $\rm 9.80~M_\odot$ to $0$ for the $\rm 13.0~M_\odot$ that is the lowest mass model that ignites Si burning centrally and behaves, during this phase, as a typical massive star. The $\rm 10.0~M_\odot$ model is an outlier in this scheme because for some reasons, difficult to understand, a sizeable abundance of $\rm ^{28}Si$ $\sim 0.19$ (in mass fraction) is left in the center (in the inner $\rm \sim 0.020~M_\odot$) at the end of the shell O burning phase, therefore in this models the Si ignition point is more internal than either the $\rm 9.80~M_\odot$ and the $\rm 11.0~M_\odot$ models (Table \ref{tab_main_prop2}).

Table \ref{tab_main_prop2} reports all the main physical properties of all these models at the presupernova stage.

\section{Summary and Conclusions}
In this paper we computed the evolution of stars with the initial mass in the range $\rm 7.00-15.00~M_\odot$ from the pre-main sequence phase up to the presupernova stage or up to an advanced stage of the thermally pulsing phase, depending on the initial mass. The main goal of these calculations is to study in detail the evolutionary behavior of stars across the transition from AGB and SABG stars, ECSNe and CCSNe progenitors.

\begin{figure*}[ht!]
\begin{center}
\includegraphics[scale=0.25]{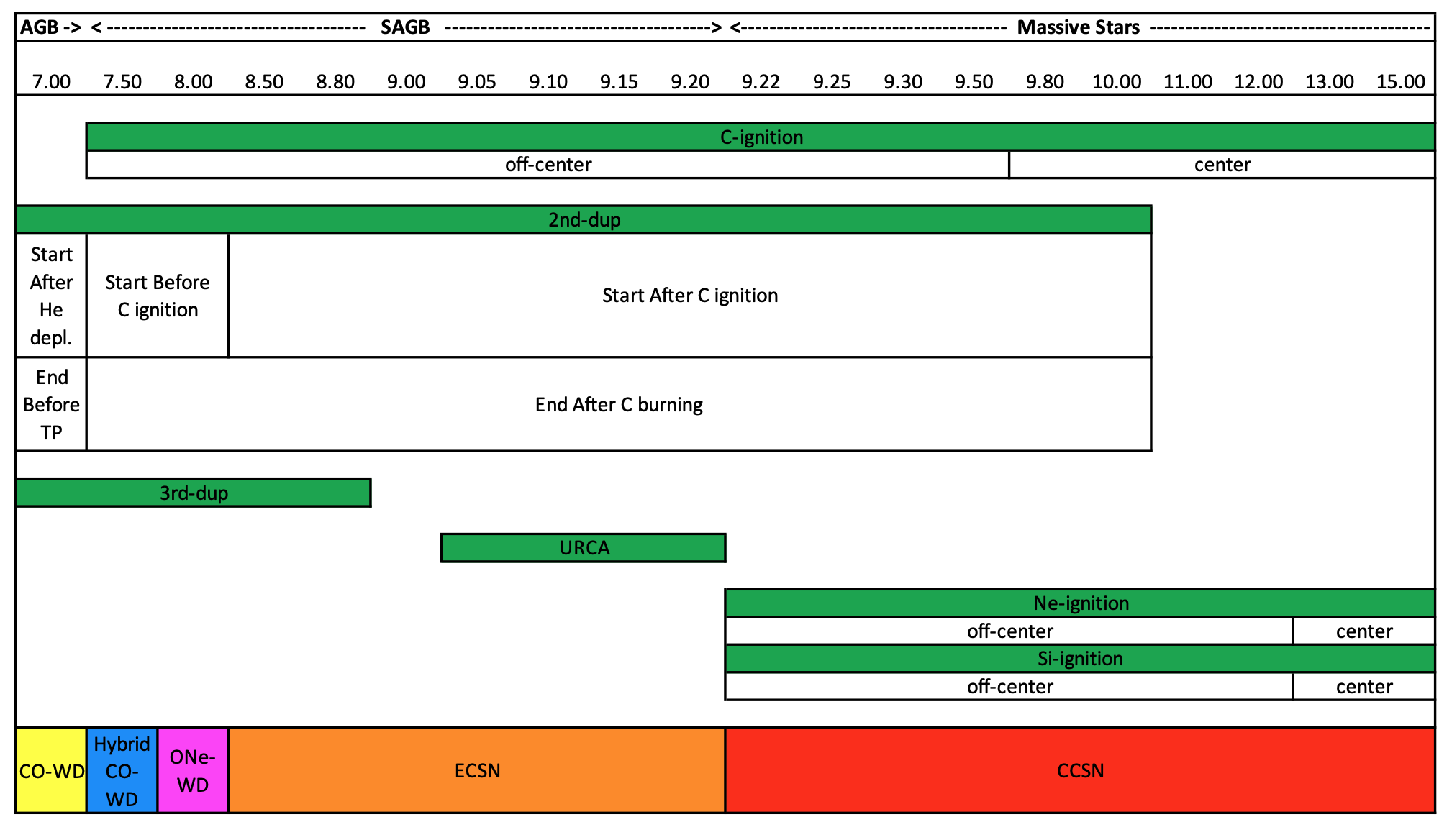}
\caption{Schematic view of some of the evolutionary properties and expected final fate.\label{summary}}
\end{center}
\end{figure*}

A summary of our results are shown in Figure \ref{summary} and will be discussed below.

All the stars in the mass range studied here evolve through the core H and core He burning stages. 

Stars with the initial mass $\rm M<7.50~M_\odot$  develop a degenerate CO core in which the temperature remains below the threshold value for the ignition of the C burning reactions. These stars, then, evolve through the TP-AGB phase and eventually end their evolution forming a CO-WD surrounded by material ejected during the previous evolutionary phases, i.e., a planetary nebula.

In stars with the initial mass $\rm M\ge 7.50~M_\odot$, on the contrary, the temperature in the CO core becomes high enough to allow the ignition of the C burning reactions. In particular, stars with the initial mass in the range $\rm 7.50-9.50~M_\odot$ ignite C off-center, the C ignition point decreasing from $\rm 0.588~M_\odot$ for the $\rm 7.50~M_\odot$ to $\rm 0.022~M_\odot$ for the $\rm 9.50~M_\odot$. Stars with the initial mass $\rm M>9.50~M_\odot$ ignite C centrally. This feature is mainly due to the fact the degree of degeneracy in the CO core decreases progressively as the initial mass increases. In all the stars, the result of the C burning is the production of an ONeMg core with the exception of the $\rm 7.50~M_\odot$ in which the C burning front quenches before reaching the center and therefore a sizable amount of $\rm ^{12}C$ is left unburnt in the inner $\rm \sim 0.3~M_\odot$. In this case a hybrid CO core is formed, i.e., a CO core in which the central part is enriched by a mixture of O and Ne, resulting from the quenching of the off-center C burning.

After core He depletion, in stars with the initial mass $\rm M<11.00~M_\odot$ the convective envelope penetrates into the He layer and the second dredge up (2nd-dup) takes place. The evolutionary stage at which this phenomenon begins and goes to completion (i.e., when the convective envelope reaches the maximum depth) depends on the initial mass. In particular it begins (1) after core He depletion for the $\rm 7.00~M_\odot$ star; (2) before C ignition for the $\rm 7.50-8.00~M_\odot$ stars and (3) after C ignition for the $\rm 8.50-10.00~M_\odot$ stars. The convective envelope reaches its maximum depth during the 2nd-dup (1) before the beginning of the TP phase for the $\rm 7.00~M_\odot$ and (2) after the C burning phase for the $\rm 7.50-10.00~M_\odot$ stars.

In stars with the initial mass in the range $\rm 7.50-9.20~M_\odot$ the maximum temperature in the ONeMg core (the hybrid CO core for the $\rm 7.50~M_\odot$) does not reach the threshold value for the ignition of Ne burning. Therefore, these stars evolve through the TP-SAGB phase. As the initial mass increases, the maximum luminosity of the He burning shell reached during each pulse decreases while the frequency of the thermal pulses increases. This is due to the increasing size of the ONeMg core with the initial mass. This also implies that the third dredge-up (3rd-dup), i.e. the penetration of the convective envelope into the He core, decreases progressively as the initial mass increases, disappearing for stars with $\rm M\ge 9.00~M_\odot$. In stars with the initial mass in the range $\rm 9.05-9.20~M_\odot$ the central density becomes high enough that the URCA pair $\rm ^{25}(Mg-Na)$ is activated. This induces a cooling of the center of the star while the core is still contracting followed by a phase of contraction at constant temperature. In the $\rm 9.20~M_\odot$ the density increases enough to reach the threshold for the activation of the URCA pair $\rm ^{23}(Na-Ne)$. Also in this case the center initially cools down and then evolves at constant temperature. During these phases a convective core forms and increases in mass progressively causing, in stars with the mass $\rm 9.10-9.20~M_\odot$, a phenomenon similar to the breathing pulses in core He burning. This phenomenon happens after the activation of the $\rm ^{25}(Mg-Na)$ URCA pair, in stars with the initial mass $\rm 9.10-9.15~M_\odot$, and after the activation of the $\rm ^{23}(Na-Ne)$ URCA pair in the $\rm 9.20~M_\odot$ model and induces a substantial increase of the central temperature (TIR). The final fate of all these stars that do not ignite Ne burning depends on the competition between the increase of the CO core, that may lead to the potential explosion of the star once the central density reaches the threshold value for the ignition of the electron capture on $\rm ^{24}Mg$, and the reduction of the envelope due to the mass loss. The detailed calculation of such a competition would require the calculation of several thousands of thermal pulses (together to the URCA pairs in the more massive ones) that is not feasible with the network adopted in this work and with the computers presently available. For this reason, the final fate of these stars has been estimated by means of "extrapolated" evolutions. According to these extrapolations, and taking into account all the possible uncertainties, we predict that in stars with the initial mass in the range $\rm 7.50-8.00~M_\odot$ the mass loss is efficient enough to reduce the total mass before the CO mass reaches the critical value for the activation of the electron capture on $\rm ^{24}Mg$. These stars, therefore, will end their life producing an ONeMg-WD (hybrid CO-WD in the $\rm 7.50~M_\odot$ star). Stars with the initial mass in the range $\rm 8.50-9.20~M_\odot$ develop CO cores massive enough to reach the activation of the electron capture on $\rm ^{24}Mg$ before the envelope is completely removed by the mass loss and therefore can explode as electron capture supernovae or collapse to a neutron star, the actual outcome depending on the details of the explosion modeling and the initial conditions and cannot be predicted with certainty in this work. Let us eventually remark that in stars with the initial mass $\rm 9.10-9.20~M_\odot$ the increase of the central temperature due to the TIR up to the threshold value for the ignition of the $\rm ^{20}Ne$ photodisintegration, before the activation of the electron capture on $\rm ^{24}Mg$, cannot be excluded. In such a case the prediction of the final fate of these stars is difficult to predict a priori. 

In stars in the mass range $\rm 9.22-15.00~M_\odot$ the maximum temperature in the ONeMg core reaches the threshold value for the ignition of Ne burning.
In stars with the initial mass in the range $\rm 9.22-12.00~M_\odot$, Ne burning is ignited off-center, the mass coordinate of the ignition point decreasing progressively with increasing the mass. The off-center Ne ignition induces the temperature to increase above the threshold value for the ignition of O burning and therefore in these stars Ne and O burning occurs simultaneously. The Ne/O burning front then shifts progressively toward the center until an O exhausted core is formed. Note that, at variance with the off-center C ignition, no hybrid ONeMg core is formed as a result of the off-center Ne/O burning. In stars with the mass $\rm M\geq 13.00~M_\odot$ the Ne burning is ignited centrally. While in the $\rm 13.00~M_\odot$ star, Ne and O burning occur simultaneously, in the $\rm 15.00~M_\odot$ star they develop in two different separate stages, as it happens in the classical massive stars. Also in these stars the final result of Ne and O burning is the formation of an O exhausted core. 

The evolution after either center and off-center Ne/O burning is characterized by the O shell burning that shifts progressively outward in mass and leads to the Si-S ignition. This burning starts off-center in stars with the initial mass in the range $\rm 9.22-12.00~M_\odot$ and the main fuel is $\rm ^{34}S$ and $\rm ^{30}Si$, in the lower mass models ($\rm 9.22-9.50~M_\odot$), and $\rm ^{28}Si$, in the more massive ones. This is due to the fact the lower mass models evolve at lower entropy and therefore in these stars the electron captures are more efficient in reducing the electron fraction. As in other previous off-center burning, also in this case the Si-S burning front propagates toward the center, followed by a shell Si-S burning phase until an Fe core is formed. Also in this case the Si-S burning front does not quench before reaching the center and therefore no hybrid Si-S core is formed. Si burning is ignited centrally in stars with the initial mass $\rm M\geq 13.00~M_\odot$ and is followed by a shell Si burning phase like in the classical massive stars until an Fe core is formed. The final fate of all the stars in the mass range $\rm 9.22-15.00~M_\odot$ is therefore an explosion as core collapse supernovae. As a final comment, let us note that the luminosity of the lower mass star that explodes as CCSN (see Figure \ref{fig:hrtot}) is compatible with the estimate of the minimum luminosity for the progenitors of SNIIP derived from the analysis of the high resolution images obtained by space and ground-based telescopes \citep{Smartt15}.

\begin{acknowledgments}
This work has been mainly supported by the Theory Grant "Evolution, nucleosynthesis and final fate of stars in the transition between AGB and Massive Stars" (PI M. Limongi) of the INAF Fundamental Astrophysics Funding Program 2022-2023. This work has also been partially supported by the World Premier International Research Center Initiative (WPI), MEXT, Japan. KN has been supported by the Japan Society for the Promotion of Science (JSPS) KAKENHI grants JP20K04024, JP21H04499, and JP23K03452. ML warmly thanks Toshio Suzuki for providing clarifications on the correct use of the weak rates for the electron capture and beta decays of the URCA processes.
\end{acknowledgments}

\newpage

\newpage

\startlongtable
\begin{deluxetable*}{lcccccccccc}\label{tab_main_prop}
\centerwidetable
\tablewidth{0pt}
\tabletypesize{\footnotesize}
%\tabletypesize{\scriptsize}
\movetableright=1cm
\tablecolumns{11}
\tablecaption{Main properties of the computed models}
\tablehead{\colhead{} & \colhead{7.00} &  \colhead{7.50} &  \colhead{8.00} & \colhead{8.50} & \colhead{8.80} & \colhead{9.00} & \colhead{9.05} & \colhead{9.10} & \colhead{9.15} & \colhead{9.20} } 
\startdata    
\cutinhead{H burning}                                       
  $\rm M_{CC}$             &  2.38     & 2.55     & 2.78     & 3.01      & 3.12      &  3.25       & 3.27        &  3.29     &  3.30    & 3.32     \\   
  $\rm t$                  &  5.04(7)  & 4.35(7)  & 3.83(7)  & 3.4(7)    & 3.18(7)   &  3.05(7)    & 3.02(7)     & 2.98(7)   & 2.95(7)  & 2.93(7)  \\   
  $\rm M_{He }$            &  1.13     & 1.21     & 1.31     & 1.45      & 1.54      &  1.63       & 1.63        & 1.64      & 1.65     & 1.66     \\   
  $\rm M_{tot}$            &  6.97     & 7.47     & 7.96     & 8.46      & 8.76      &  8.96       & 9.01        & 9.06      & 9.11     & 9.16     \\   
\cutinhead{ He ignition}                                                               
  $\rm M_{He}$             &  1.15     & 1.28     & 1.41     & 1.55      & 1.61      &  1.67       & 1.68        & 1.69      & 1.71     & 1.73     \\   
  $\rm M_{CE}$             &  1.74     & 1.91     & 2.10     & 2.26      & 2.37      &  2.47       & 2.49        & 2.52      & 2.55     & 2.57     \\   
\cutinhead{ He burning}                                                                 
  $\rm M_{CC}$             &  0.70     & 0.76     & 0.84     & 0.92      & 0.98      &  1.03       & 1.05        & 1.06      & 1.07     & 1.09     \\    
  $\rm t$                  &  4.85(6)  & 4.12(6)  & 3.45(6)  & 2.98(6)   & 2.76(6)   &  2.62(6)    & 2.61(6)     & 2.58(6)   & 2.54(6)  & 2.51(6)  \\
\enddata  
\tablecomments{In the H and He burning sections, all the quantities refer to the values at the end of each nuclear burning, the only exception being $\rm M_{\rm CC}$, that is the maximum extension in mass of the convective core during the nuclear burning, and $\rm t$, that is the evolutionary time of each burning stage.}                                          
\end{deluxetable*} 

\newpage 

%\floattable
\startlongtable
\begin{deluxetable*}{lcccccccccc}\label{tab_main_prop2}
\centerwidetable
\tablewidth{0pt}
\tabletypesize{\footnotesize}
%\tabletypesize{\scriptsize}
\movetableright=1cm
\tablecolumns{11}
\tablecaption{Main properties of the computed models}
\tablehead{\colhead{} & \colhead{9.22} & \colhead{9.25} & \colhead{9.30} & \colhead{9.50} & \colhead{9.80} & \colhead{10.00} & \colhead{11.00} & \colhead{12.00} & \colhead{13.00} & \colhead{15.00} } 
\startdata    
\cutinhead{ H burning}                                       
  $\rm M_{CC}$              & 3.33     & 3.34    & 3.36    & 3.57    & 3.68    & 3.76     & 4.18    & 4.75    & 5.28    & 6.45    \\   
  $\rm t$                   & 2.91(7)  & 2.90(7) & 2.87(7) & 2.76(7) & 2.60(7) & 2.51(7)  & 2.11(7) & 1.83(7) & 1.62(7) & 1.31(7) \\   
  $\rm M_{He }$             & 1.66     & 1.66    & 1.72    & 1.76    & 1.84    & 1.88     & 2.19    & 2.56    & 2.91    & 3.71    \\   
  $\rm M_{tot}$             & 9.18     & 9.21    & 9.26    & 9.46    & 9.76    & 9.96     & 10.90   & 11.90   & 12.90   & 14.88   \\   
\cutinhead{ He ignition}                                                               
  $\rm M_{He}$              & 1.73     & 1.74    & 1.77    & 1.81    & 1.89    & 1.94     & 2.24    & 2.57    & 2.91    & 3.66    \\   
  $\rm M_{CE}$              & 2.75     & 2.59    & 2.61    & 2.69    & 2.78    & 2.84     & 3.24    & 3.67    & 4.13    & 5.12    \\   
\cutinhead{ He burning}                                                                 
  $\rm M_{CC}$              & 1.09     & 1.10    & 1.11    & 1.16    & 1.21    & 1.25     & 1.49    & 1.82    & 2.11    & 2.71    \\    
  $\rm t$                   & 2.50(6)  & 2.48(6) & 2.46(6) & 2.36(6) & 2.21(6) & 2.12(6)  & 1.74(6) & 1.51(6) & 1.30(6) & 1.01(6) \\
\enddata  
\tablecomments{Same as in Table \ref{tab_main_prop}}                                          
\end{deluxetable*} 

\newpage 

%\floattable
\startlongtable
\begin{deluxetable*}{lcccccccc}\label{tpprop700}
\centerwidetable
\tablewidth{0pt}
\tabletypesize{\footnotesize}
%\tabletypesize{\scriptsize}
\movetableright=1cm
\tablecolumns{9}
\tablecaption{Main properties of the TP phase of the $\rm 7.00~M_\odot$ model}
\tablehead{ \colhead{\rm TP} & \colhead{$\Delta t_{\rm pulse}$} & \colhead{${\rm Log}\left(L_{\rm He}/L_\odot\right)^{\rm max}$} & \colhead{$\Delta M_{\rm pulse}^{\rm max}$} & \colhead{$T_{\rm He}^{\rm max}$} & \colhead{$\lambda$} & \colhead{$\Delta t_{\rm inter}$} & \colhead{$\Delta M_{\rm He}$} & \colhead{$\Delta M_{\rm CO}$} \\ 
            \colhead{}      & \colhead{(yr)}                   & \colhead{}                                                     & \colhead{$(\rm M_\odot)$}                  & \colhead{$(\rm 10^{8}~K)$}       & \colhead{}          & \colhead{$(yr)$}                 & \colhead{$(\rm M_\odot)$}     & \colhead{$(\rm M_\odot)$}      } 
\startdata    
    1 &   89.82555 &    4.59285 &   0.0000e+00 &    2.24571 &    0.00000 &        0.00000 &   0.0000e+00 &   0.0000e+00 \\
    2 &  111.22892 &    4.96992 &   6.3260e-04 &    2.28490 &    0.00000 &      810.01669 &   3.3420e-04 &   3.4780e-04 \\
    3 &  115.42192 &    5.16239 &   7.6520e-04 &    2.39542 &    0.00000 &      830.90516 &   3.5400e-04 &   3.7560e-04 \\
    4 &  117.34890 &    5.32072 &   8.2270e-04 &    2.46989 &    0.00000 &      864.45584 &   3.7510e-04 &   3.9770e-04 \\
    5 &  121.55523 &    5.45162 &   8.3630e-04 &    2.51289 &    0.00000 &      893.41053 &   3.9600e-04 &   4.1390e-04 \\
    6 &  111.21537 &    5.56693 &   8.6150e-04 &    2.55107 &    0.07813 &      913.12543 &   4.1340e-04 &   4.2040e-04 \\
    7 &  101.03505 &    5.67624 &   8.7040e-04 &    2.60405 &    0.16083 &      940.51568 &   4.3400e-04 &   4.0980e-04 \\
    8 &   88.24332 &    5.77267 &   2.5130e-04 &    2.62902 &    0.23434 &      970.62890 &   4.5490e-04 &   4.4900e-04 \\
\enddata  
\tablecomments{The meaning of the columns are the following: TP is the pulse number; $\Delta t_{\rm pulse}$ is the duration of the pulse in years; ${\rm Log}\left(L_{\rm He}/L_\odot\right)^{\rm max}$ is the maximum luminosity provided by the He burning reactions during the pulse in solar luminosity; $\Delta M_{\rm pulse}^{\rm max}$ is the maximum extension of the He convective shell during the pulse in solar masses; $T_{\rm He}^{\rm max}$ is the maximum temperature of the He burning shell during the pulse in Kelvin degree; $\lambda$ is the measure of the efficiency of the third dredge up and is defined in section \ref{sec:agb}; $\Delta t_{\rm inter}$ is the time elapsed between two consecutive thermal pulses in year; $\Delta M_{\rm He}$ and $\Delta M_{\rm CO}$ are the increase of the He and CO core, respectively, in solar masses, between two consecutive thermal pulses.}                                        
\end{deluxetable*} 

\newpage 

%\floattable
\startlongtable
\begin{deluxetable*}{lcccccccc}\label{tpprop750}
\centerwidetable
\tablewidth{0pt}
\tabletypesize{\footnotesize}
%\tabletypesize{\scriptsize}
\movetableright=1cm
\tablecolumns{9}
\tablecaption{Main properties of the TP phase of the $\rm 7.50~M_\odot$ model}
\tablehead{ \colhead{\rm TP} & \colhead{$\Delta t_{\rm pulse}$} & \colhead{${\rm Log}\left(L_{\rm He}/L_\odot\right)^{\rm max}$} & \colhead{$\Delta M_{\rm pulse}^{\rm max}$} & \colhead{$T_{\rm He}^{\rm max}$} & \colhead{$\lambda$} & \colhead{$\Delta t_{\rm inter}$} & \colhead{$\Delta M_{\rm He}$} & \colhead{$\Delta M_{\rm CO}$} \\ 
            \colhead{}      & \colhead{(yr)}                   & \colhead{}                                                     & \colhead{$(\rm M_\odot)$}                  & \colhead{$(\rm 10^{8}~K)$}       & \colhead{}          & \colhead{$(yr)$}                 & \colhead{$(\rm M_\odot)$}     & \colhead{$(\rm M_\odot)$}      } 
\startdata    
    1 &   44.37918 &    4.70594 &   0.0000e+00 &    2.38536 &    0.00000 &        0.00000 &   0.0000e+00 &   0.0000e+00 \\
    2 &   52.11441 &    5.04933 &   3.2920e-04 &    2.39576 &    0.00000 &      419.64009 &   2.0260e-04 &   2.0100e-04 \\
    3 &   53.95405 &    5.18454 &   3.9090e-04 &    2.47251 &    0.00000 &      432.30629 &   2.1320e-04 &   2.1680e-04 \\
    4 &   56.06827 &    5.29280 &   3.9980e-04 &    2.52862 &    0.00000 &      447.35448 &   2.2440e-04 &   2.2900e-04 \\
    5 &   56.34100 &    5.39051 &   4.2360e-04 &    2.56485 &    0.00000 &      463.38110 &   2.3510e-04 &   2.3760e-04 \\
    6 &   52.76081 &    5.48184 &   4.4310e-04 &    2.59733 &    0.05947 &      476.65252 &   2.4550e-04 &   2.4740e-04 \\
    7 &   50.08207 &    5.56786 &   3.9450e-04 &    2.64991 &    0.12070 &      491.06480 &   2.5600e-04 &   2.9240e-04 \\
    8 &   47.81777 &    5.64817 &   3.8670e-04 &    2.68787 &    0.16448 &      504.01888 &   2.6630e-04 &   2.2620e-04 \\
\enddata  
\tablecomments{Same as Table \ref{tpprop700}}                                        
\end{deluxetable*} 

\newpage 

%\floattable
\startlongtable
\begin{deluxetable*}{lcccccccc}\label{tpprop800}
\centerwidetable
\tablewidth{0pt}
\tabletypesize{\footnotesize}
%\tabletypesize{\scriptsize}
\movetableright=1cm
\tablecolumns{9}
\tablecaption{Main properties of the TP phase of the $\rm 8.00~M_\odot$ model}
\tablehead{ \colhead{\rm TP} & \colhead{$\Delta t_{\rm pulse}$} & \colhead{${\rm Log}\left(L_{\rm He}/L_\odot\right)^{\rm max}$} & \colhead{$\Delta M_{\rm pulse}^{\rm max}$} & \colhead{$T_{\rm He}^{\rm max}$} & \colhead{$\lambda$} & \colhead{$\Delta t_{\rm inter}$} & \colhead{$\Delta M_{\rm He}$} & \colhead{$\Delta M_{\rm CO}$} \\ 
            \colhead{}      & \colhead{(yr)}                   & \colhead{}                                                     & \colhead{$(\rm M_\odot)$}                  & \colhead{$(\rm 10^{8}~K)$}       & \colhead{}          & \colhead{$(yr)$}                 & \colhead{$(\rm M_\odot)$}     & \colhead{$(\rm M_\odot)$}      } 
\startdata    
    1 &   18.29010 &    4.86701 &   0.0000e+00 &    2.54921 &    0.00000 &        0.00000 &   0.0000e+00 &   0.0000e+00 \\
    2 &   19.71649 &    4.99491 &   9.6200e-05 &    2.46998 &    0.00000 &      167.70801 &   9.2800e-05 &   9.2900e-05 \\
    3 &   21.16711 &    5.10723 &   1.1190e-04 &    2.54650 &    0.00000 &      172.35960 &   9.6500e-05 &   9.9400e-05 \\
    4 &   21.81918 &    5.18338 &   1.3880e-04 &    2.60576 &    0.00000 &      177.92965 &   1.0070e-04 &   1.0160e-04 \\
    5 &   22.31795 &    5.27167 &   1.4490e-04 &    2.64632 &    0.00000 &      182.94191 &   1.0450e-04 &   1.0530e-04 \\
    6 &   22.23202 &    5.34037 &   1.7580e-04 &    2.67582 &    0.01196 &      188.46944 &   1.0870e-04 &   1.0660e-04 \\
    7 &   21.95982 &    5.42845 &   1.9100e-04 &    2.73056 &    0.04178 &      193.68099 &   1.1250e-04 &   1.1000e-04 \\
    8 &   21.69375 &    5.51369 &   1.8770e-04 &    2.78321 &    0.07198 &      199.54792 &   1.1670e-04 &   1.0990e-04 \\
\enddata  
\tablecomments{Same as Table \ref{tpprop700}}                                        
\end{deluxetable*}

\newpage 

%\floattable
\startlongtable
\begin{deluxetable*}{lcccccccc}\label{tpprop850}
\centerwidetable
\tablewidth{0pt}
\tabletypesize{\footnotesize}
%\tabletypesize{\scriptsize}
\movetableright=1cm
\tablecolumns{9}
\tablecaption{Main properties of the TP phase of the $\rm 8.50~M_\odot$ model}
\tablehead{ \colhead{\rm TP} & \colhead{$\Delta t_{\rm pulse}$} & \colhead{${\rm Log}\left(L_{\rm He}/L_\odot\right)^{\rm max}$} & \colhead{$\Delta M_{\rm pulse}^{\rm max}$} & \colhead{$T_{\rm He}^{\rm max}$} & \colhead{$\lambda$} & \colhead{$\Delta t_{\rm inter}$} & \colhead{$\Delta M_{\rm He}$} & \colhead{$\Delta M_{\rm CO}$} \\ 
            \colhead{}      & \colhead{(yr)}                   & \colhead{}                                                     & \colhead{$(\rm M_\odot)$}                  & \colhead{$(\rm 10^{8}~K)$}       & \colhead{}          & \colhead{$(yr)$}                 & \colhead{$(\rm M_\odot)$}     & \colhead{$(\rm M_\odot)$}      } 
\startdata    
    1 &    8.15592 &    4.90472 &   0.0000e+00 &    2.66934 &    0.00000 &        0.00000 &   0.0000e+00 &   0.0000e+00 \\
    2 &    8.45305 &    4.98826 &   1.6500e-05 &    2.53505 &    0.00000 &       74.46446 &   4.6800e-05 &   4.7100e-05 \\
    3 &    8.75214 &    5.05037 &   2.0800e-05 &    2.57771 &    0.00000 &       75.69203 &   4.7900e-05 &   4.8700e-05 \\
    4 &    9.00542 &    5.10690 &   4.0300e-05 &    2.64540 &    0.00000 &       76.95628 &   4.9000e-05 &   4.9900e-05 \\
    5 &    9.20708 &    5.16083 &   5.9400e-05 &    2.68255 &    0.00000 &       78.46501 &   5.0200e-05 &   5.1100e-05 \\
    6 &    9.62853 &    5.21936 &   4.7700e-05 &    2.69753 &    0.00000 &       79.89779 &   5.1300e-05 &   5.1700e-05 \\
    7 &    9.65059 &    5.26992 &   6.8300e-05 &    2.72114 &    0.00000 &       81.18954 &   5.2600e-05 &   5.2600e-05 \\
    8 &    9.65526 &    5.32230 &   6.2300e-05 &    2.74320 &    0.00370 &       83.15922 &   5.4000e-05 &   5.4200e-05 \\
\enddata  
\tablecomments{Same as Table \ref{tpprop700}}                                        
\end{deluxetable*}

\newpage 

%\floattable
\startlongtable
\begin{deluxetable*}{lcccccccc}\label{tpprop880}
\centerwidetable
\tablewidth{0pt}
\tabletypesize{\footnotesize}
%\tabletypesize{\scriptsize}
\movetableright=1cm
\tablecolumns{9}
\tablecaption{Main properties of the TP phase of the $\rm 8.80~M_\odot$ model}
\tablehead{ \colhead{\rm TP} & \colhead{$\Delta t_{\rm pulse}$} & \colhead{${\rm Log}\left(L_{\rm He}/L_\odot\right)^{\rm max}$} & \colhead{$\Delta M_{\rm pulse}^{\rm max}$} & \colhead{$T_{\rm He}^{\rm max}$} & \colhead{$\lambda$} & \colhead{$\Delta t_{\rm inter}$} & \colhead{$\Delta M_{\rm He}$} & \colhead{$\Delta M_{\rm CO}$} \\ 
            \colhead{}      & \colhead{(yr)}                   & \colhead{}                                                     & \colhead{$(\rm M_\odot)$}                  & \colhead{$(\rm 10^{8}~K)$}       & \colhead{}          & \colhead{$(yr)$}                 & \colhead{$(\rm M_\odot)$}     & \colhead{$(\rm M_\odot)$}      } 
\startdata    
    1 &    3.43398 &    4.81971 &   0.0000e+00 &    2.75444 &    0.00000 &        0.00000 &   0.0000e+00 &   0.0000e+00 \\
    2 &    3.66142 &    4.86343 &   0.0000e+00 &    2.78394 &    0.00000 &       35.35383 &   2.3700e-05 &   2.4300e-05 \\
    3 &    3.80866 &    4.91269 &   8.3000e-06 &    2.78179 &    0.00000 &       35.37903 &   2.3900e-05 &   2.4500e-05 \\
    4 &    3.92747 &    4.95956 &   9.2000e-06 &    2.78450 &    0.00000 &       35.78979 &   2.4200e-05 &   2.4800e-05 \\
    5 &    3.98919 &    4.99723 &   0.0000e+00 &    2.78093 &    0.00000 &       36.10830 &   2.4600e-05 &   2.7400e-05 \\
    6 &    4.04114 &    5.03525 &   1.7600e-05 &    2.67570 &    0.00000 &       36.40138 &   2.4900e-05 &   2.3200e-05 \\
    7 &    4.10827 &    5.07228 &   1.1600e-05 &    2.70671 &    0.00000 &       36.87323 &   2.5200e-05 &   2.5400e-05 \\
    8 &    4.21462 &    5.10703 &   1.6600e-05 &    2.73201 &    0.00000 &       37.44727 &   2.5700e-05 &   2.6200e-05 \\
\enddata  
\tablecomments{Same as Table \ref{tpprop700}}                                        
\end{deluxetable*}

\newpage 

%\floattable
\startlongtable
\begin{deluxetable*}{lcccccccc}\label{tpprop900}
\centerwidetable
\tablewidth{0pt}
\tabletypesize{\footnotesize}
%\tabletypesize{\scriptsize}
\movetableright=1cm
\tablecolumns{9}
\tablecaption{Main properties of the TP phase of the $\rm 9.00~M_\odot$ model}
\tablehead{ \colhead{\rm TP} & \colhead{$\Delta t_{\rm pulse}$} & \colhead{${\rm Log}\left(L_{\rm He}/L_\odot\right)^{\rm max}$} & \colhead{$\Delta M_{\rm pulse}^{\rm max}$} & \colhead{$T_{\rm He}^{\rm max}$} & \colhead{$\lambda$} & \colhead{$\Delta t_{\rm inter}$} & \colhead{$\Delta M_{\rm He}$} & \colhead{$\Delta M_{\rm CO}$} \\ 
            \colhead{}      & \colhead{(yr)}                   & \colhead{}                                                     & \colhead{$(\rm M_\odot)$}                  & \colhead{$(\rm 10^{8}~K)$}       & \colhead{}          & \colhead{$(yr)$}                 & \colhead{$(\rm M_\odot)$}     & \colhead{$(\rm M_\odot)$}      } 
\startdata    
    1 &    1.65055 &    4.79497 &   0.0000e+00 &    2.83896 &    0.00000 &        0.00000 &   0.0000e+00 &   0.0000e+00 \\
    2 &    1.70540 &    4.82654 &   0.0000e+00 &    2.85654 &    0.00000 &       18.10599 &   1.2500e-05 &   1.2800e-05 \\
    3 &    1.81595 &    4.85967 &   0.0000e+00 &    2.86775 &    0.00000 &       18.32212 &   1.2600e-05 &   1.3000e-05 \\
    4 &    1.85047 &    4.89298 &   0.0000e+00 &    2.89756 &    0.00000 &       18.37811 &   1.2800e-05 &   1.3400e-05 \\
    5 &    1.92638 &    4.92275 &   0.0000e+00 &    2.92789 &    0.00000 &       18.49531 &   1.2900e-05 &   1.3000e-05 \\
    6 &    1.95988 &    4.95348 &   1.8000e-06 &    2.93369 &    0.00000 &       18.55105 &   1.2900e-05 &   1.3600e-05 \\
    7 &    1.97685 &    4.98545 &   0.0000e+00 &    2.95065 &    0.00000 &       18.71940 &   1.3100e-05 &   1.3400e-05 \\
    8 &    2.02927 &    5.01602 &   0.0000e+00 &    2.92993 &    0.00000 &       18.92700 &   1.3200e-05 &   1.3600e-05 \\
\enddata  
\tablecomments{Same as Table \ref{tpprop700}}                                        
\end{deluxetable*} 

\newpage 

%\floattable
\startlongtable
\begin{deluxetable*}{lcccccccc}\label{tpprop905}
\centerwidetable
\tablewidth{0pt}
\tabletypesize{\footnotesize}
%\tabletypesize{\scriptsize}
\movetableright=1cm
\tablecolumns{9}
\tablecaption{Main properties of the TP phase of the $\rm 9.05~M_\odot$ model}
\tablehead{ \colhead{\rm TP} & \colhead{$\Delta t_{\rm pulse}$} & \colhead{${\rm Log}\left(L_{\rm He}/L_\odot\right)^{\rm max}$} & \colhead{$\Delta M_{\rm pulse}^{\rm max}$} & \colhead{$T_{\rm He}^{\rm max}$} & \colhead{$\lambda$} & \colhead{$\Delta t_{\rm inter}$} & \colhead{$\Delta M_{\rm He}$} & \colhead{$\Delta M_{\rm CO}$} \\ 
            \colhead{}      & \colhead{(yr)}                   & \colhead{}                                                     & \colhead{$(\rm M_\odot)$}                  & \colhead{$(\rm 10^{8}~K)$}       & \colhead{}          & \colhead{$(yr)$}                 & \colhead{$(\rm M_\odot)$}     & \colhead{$(\rm M_\odot)$}      } 
\startdata    
    1 &    1.15226 &    4.79895 &   0.0000e+00 &    2.83820 &    0.00000 &        0.00000 &   0.0000e+00 &   0.0000e+00 \\
    2 &    1.18284 &    4.82525 &   0.0000e+00 &    2.87107 &    0.00000 &       13.33709 &   9.6000e-06 &   9.9000e-06 \\
    3 &    1.23655 &    4.85639 &   0.0000e+00 &    2.87721 &    0.00000 &       13.41713 &   9.7000e-06 &   9.9000e-06 \\
    4 &    1.26437 &    4.88084 &   0.0000e+00 &    2.90764 &    0.00000 &       13.51299 &   9.7000e-06 &   1.0100e-05 \\
    5 &    1.28886 &    4.90511 &   0.0000e+00 &    2.93080 &    0.00000 &       13.47634 &   9.8000e-06 &   1.0300e-05 \\
    6 &    1.30935 &    4.93650 &   0.0000e+00 &    2.94823 &    0.00000 &       13.65281 &   9.9000e-06 &   1.0200e-05 \\
    7 &    1.29946 &    4.95740 &   0.0000e+00 &    2.95029 &    0.00000 &       13.63861 &   9.9000e-06 &   1.0200e-05 \\
    8 &    1.35198 &    4.97593 &   0.0000e+00 &    2.99020 &    0.00000 &       13.70251 &   1.0000e-05 &   1.0300e-05 \\
\enddata  
\tablecomments{Same as Table \ref{tpprop700}}                                        
\end{deluxetable*}

\newpage 

%\floattable
\startlongtable
\begin{deluxetable*}{lcccccccc}\label{tpprop910}
\centerwidetable
\tablewidth{0pt}
\tabletypesize{\footnotesize}
%\tabletypesize{\scriptsize}
\movetableright=1cm
\tablecolumns{9}
\tablecaption{Main properties of the TP phase of the $\rm 9.10~M_\odot$ model}
\tablehead{ \colhead{\rm TP} & \colhead{$\Delta t_{\rm pulse}$} & \colhead{${\rm Log}\left(L_{\rm He}/L_\odot\right)^{\rm max}$} & \colhead{$\Delta M_{\rm pulse}^{\rm max}$} & \colhead{$T_{\rm He}^{\rm max}$} & \colhead{$\lambda$} & \colhead{$\Delta t_{\rm inter}$} & \colhead{$\Delta M_{\rm He}$} & \colhead{$\Delta M_{\rm CO}$} \\ 
            \colhead{}      & \colhead{(yr)}                   & \colhead{}                                                     & \colhead{$(\rm M_\odot)$}                  & \colhead{$(\rm 10^{8}~K)$}       & \colhead{}          & \colhead{$(yr)$}                 & \colhead{$(\rm M_\odot)$}     & \colhead{$(\rm M_\odot)$}      } 
\startdata    
    1 &    0.86314 &    4.78280 &   0.0000e+00 &    2.87263 &    0.00000 &        0.00000 &   0.0000e+00 &   0.0000e+00 \\
    2 &    0.88962 &    4.80465 &   0.0000e+00 &    2.88928 &    0.00000 &       10.52799 &   7.6000e-06 &   7.7000e-06 \\
    3 &    0.94144 &    4.83235 &   0.0000e+00 &    2.91490 &    0.00000 &       10.68180 &   7.7000e-06 &   7.9000e-06 \\
    4 &    0.97891 &    4.85613 &   0.0000e+00 &    2.93082 &    0.00000 &       10.64835 &   7.7000e-06 &   7.8000e-06 \\
    5 &    1.01126 &    4.88220 &   0.0000e+00 &    2.95641 &    0.00000 &       10.76478 &   7.8000e-06 &   8.2000e-06 \\
    6 &    1.01921 &    4.90454 &   0.0000e+00 &    2.95521 &    0.00000 &       10.76601 &   7.8000e-06 &   8.0000e-06 \\
    7 &    1.02637 &    4.92638 &   0.0000e+00 &    2.96556 &    0.00000 &       10.84637 &   7.9000e-06 &   8.3000e-06 \\
    8 &    1.04040 &    4.94952 &   0.0000e+00 &    2.99702 &    0.00000 &       10.90800 &   8.0000e-06 &   8.1000e-06 \\
\enddata  
\tablecomments{Same as Table \ref{tpprop700}}                                        
\end{deluxetable*} 

\newpage 

%\floattable
\startlongtable
\begin{deluxetable*}{lcccccccc}\label{tpprop915}
\centerwidetable
\tablewidth{0pt}
\tabletypesize{\footnotesize}
%\tabletypesize{\scriptsize}
\movetableright=1cm
\tablecolumns{9}
\tablecaption{Main properties of the TP phase of the $\rm 9.15~M_\odot$ model}
\tablehead{ \colhead{\rm TP} & \colhead{$\Delta t_{\rm pulse}$} & \colhead{${\rm Log}\left(L_{\rm He}/L_\odot\right)^{\rm max}$} & \colhead{$\Delta M_{\rm pulse}^{\rm max}$} & \colhead{$T_{\rm He}^{\rm max}$} & \colhead{$\lambda$} & \colhead{$\Delta t_{\rm inter}$} & \colhead{$\Delta M_{\rm He}$} & \colhead{$\Delta M_{\rm CO}$} \\ 
            \colhead{}      & \colhead{(yr)}                   & \colhead{}                                                     & \colhead{$(\rm M_\odot)$}                  & \colhead{$(\rm 10^{8}~K)$}       & \colhead{}          & \colhead{$(yr)$}                 & \colhead{$(\rm M_\odot)$}     & \colhead{$(\rm M_\odot)$}      } 
\startdata    
    1 &    0.75097 &    4.80559 &   0.0000e+00 &    2.85912 &    0.00000 &        0.00000 &   0.0000e+00 &   0.0000e+00 \\
    2 &    0.75269 &    4.80162 &   0.0000e+00 &    2.86844 &    0.00000 &        7.90348 &   5.7000e-06 &   5.9000e-06 \\
    3 &    0.75519 &    4.82036 &   0.0000e+00 &    2.87524 &    0.00000 &        7.94354 &   5.7000e-06 &   6.0000e-06 \\
    4 &    0.78747 &    4.83328 &   0.0000e+00 &    2.89672 &    0.00000 &        8.01730 &   5.8000e-06 &   6.1000e-06 \\
    5 &    0.83944 &    4.85992 &   0.0000e+00 &    2.93044 &    0.00000 &        8.12756 &   5.8000e-06 &   6.0000e-06 \\
    6 &    0.82561 &    4.87533 &   0.0000e+00 &    2.93609 &    0.00000 &        8.07005 &   5.9000e-06 &   6.2000e-06 \\
    7 &    0.87714 &    4.90579 &   0.0000e+00 &    2.97745 &    0.00000 &        8.12719 &   5.9000e-06 &   6.1000e-06 \\
    8 &    0.88349 &    4.93938 &   0.0000e+00 &    2.99234 &    0.00000 &        8.18389 &   6.0000e-06 &   6.3000e-06 \\
\enddata  
\tablecomments{Same as Table \ref{tpprop700}}                                        
\end{deluxetable*}

\newpage 

%\floattable
\startlongtable
\begin{deluxetable*}{lcccccccc}\label{tpprop920}
\centerwidetable
\tablewidth{0pt}
\tabletypesize{\footnotesize}
%\tabletypesize{\scriptsize}
\movetableright=1cm
\tablecolumns{9}
\tablecaption{Main properties of the TP phase of the $\rm 9.20~M_\odot$ model}
\tablehead{ \colhead{\rm TP} & \colhead{$\Delta t_{\rm pulse}$} & \colhead{${\rm Log}\left(L_{\rm He}/L_\odot\right)^{\rm max}$} & \colhead{$\Delta M_{\rm pulse}^{\rm max}$} & \colhead{$T_{\rm He}^{\rm max}$} & \colhead{$\lambda$} & \colhead{$\Delta t_{\rm inter}$} & \colhead{$\Delta M_{\rm He}$} & \colhead{$\Delta M_{\rm CO}$} \\ 
            \colhead{}      & \colhead{(yr)}                   & \colhead{}                                                     & \colhead{$(\rm M_\odot)$}                  & \colhead{$(\rm 10^{8}~K)$}       & \colhead{}          & \colhead{$(yr)$}                 & \colhead{$(\rm M_\odot)$}     & \colhead{$(\rm M_\odot)$}      } 
\startdata    
    1 &    0.49000 &    4.79053 &   0.0000e+00 &    2.96686 &    0.00000 &        0.00000 &   0.0000e+00 &   0.0000e+00 \\
    2 &    0.51123 &    4.81073 &   0.0000e+00 &    2.97870 &    0.00000 &        6.31323 &   4.6000e-06 &   4.7000e-06 \\
    3 &    0.54396 &    4.84059 &   0.0000e+00 &    3.01213 &    0.00000 &        6.35516 &   4.5000e-06 &   4.7000e-06 \\
    4 &    0.55834 &    4.85717 &   0.0000e+00 &    3.01421 &    0.00000 &        6.48414 &   4.7000e-06 &   4.8000e-06 \\
    5 &    0.62452 &    4.94737 &   0.0000e+00 &    3.08615 &    0.00000 &        6.76557 &   4.9000e-06 &   4.8000e-06 \\
    6 &    0.62582 &    4.94290 &   0.0000e+00 &    3.07472 &    0.00000 &        6.72019 &   4.9000e-06 &   5.0000e-06 \\
    7 &    0.62390 &    4.95652 &   0.0000e+00 &    3.09052 &    0.00000 &        6.71941 &   4.9000e-06 &   5.1000e-06 \\
    8 &    0.63878 &    4.97361 &   0.0000e+00 &    3.08619 &    0.00000 &        6.80544 &   4.9000e-06 &   5.1000e-06 \\
\enddata  
\tablecomments{Same as Table \ref{tpprop700}}                                        
\end{deluxetable*}

\end{document}